# Space Station Rotational Stability

David W. Jensen, Ph.D.
*Technical Fellow, Retired*
*Rockwell Collins / Collins Aerospace*
*Cedar Rapids, IA 52302*
*david.jensen@alumni.iastate.edu*

## Abstract

*Designing for rotational stability can dramatically affect the geometry of a space station. If improperly designed, the rotating station could end up catastrophically tumbling end-over-end. Active stabilization can address this problem; however, designing the station with passive rotation stability provides a lower-cost solution. This paper presents passive rotational stability guidelines for four space station geometries. Station stability is first analyzed with thin-shell and thick-shell models. Stability is also analyzed with models of the station's major constituent parts, including outer shells, spokes, floors, air, and shuttle bays.*

**Keywords:** Rotational stability, space station, station size, station geometry, human limits, gravity, air pressure

## 1 Introduction

Rotational stability is the property of a body for which a small angular displacement sets up a restoring torque that tends to return the body to its original position [McGraw-Hill 2023]. Recently, a research team considered this stability for a rotating cylindrical space station [Globus et al. 2007]. They analyzed a thin shell model of a rotating cylindrical space station. There is much more to space stations than a thin shell cylinder model. Our Asteroid Restructuring paper introduced station details such as centripetal gravity, air pressure, and multiple floors for four station geometries [Jensen 2023]. This paper focuses on the rotational stability of the four main types of station geometries. Not only does it cover hollow thin-shell models, but it also evaluates rotational stability for stations with their constituent components such as spokes, shuttle bays, air, and floors. Figure 1-1 shows an external view of a rotating station with many of those components.

Section 2 briefly reviews the four station geometries, including refinements for stability and gravity constraints. Other constraints are identified, and renderings are provided to help visualize these large space stations.

Section 3 provides more details on space station limitations. Human frailty imposes many limits on the design of a space station. These limits include gravity, air pressure, rotational stability, population, and radiation. These limits and key characteristics are reviewed for use in stability modeling.

Section 4 defines rotational stability, explores its background, and provides alternative approaches. It provides a set of three rules required for passive rotational stability. Stability analysis is the purpose of this paper, and this section provides background for this analysis.

Section 5 introduces the analysis of rotational stability. For this analysis, stations are decomposed into constituent components. The components are analyzed separately and then combined to represent the station. This section defines those components and provides the analytic equations used to model them and the rotational stability.

Section 6 covers the single-floor rotational stability. This extends the original work of [Globus et al. 2007] and our own work in [Jensen 2023]. It focuses on the rotational stability of a projected single floor on only the outer shell. All four geometries are evaluated with thin and thick shells.

Section 7 evaluates the stability of multiple-component stations. This evaluation includes the rotational stability of all the major components of the stations. These components include the shell, spokes, main floor, multiple floors, air, shuttle bay, and dividers. The station stability is found by summing the moments of inertia of all the components.

Section 8 summarizes the station stability results. This paper details the rotational stability of large, complex space stations. This section summarizes the station constraints from this analysis and presents the stability results using the single-floor and multiple-component models.

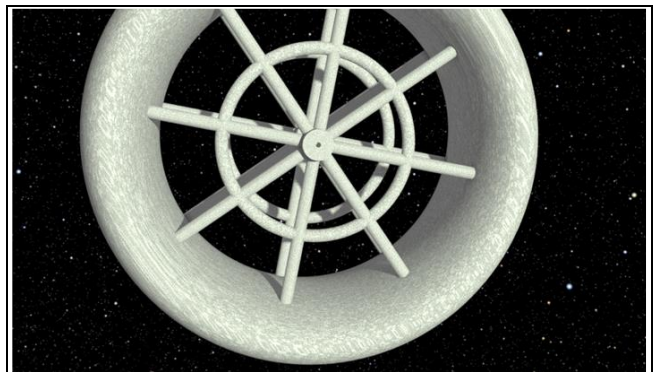

*Credit: Self-produced using Blender. Background Milky Way from NASA SVS Deep Star Maps [NASA Image Public Domain]*

**Figure 1-1 – Example Rotating Torus Space Station**

This paper details the rotational stability for rotating space stations. We review and extend results from our Asteroid Restructuring paper [Jensen 2023]. We use details from our *Design Limits on Large Space Stations* paper [Jensen 2024d]. Because this paper is a continuation of those previous papers, their details are often cited for context. We also liberally re-use charts, images, and text from those publications.

## 2 Space Stations

This section introduces the space station geometries and concepts used in this paper. The concepts include summaries of the geometries and assessment approaches. It also introduces the refinement of these geometries to address gravity and stability constraints. Other constraints are briefly introduced. This section also provides renderings to help visualize these large space stations.

### 2.1 Geometries

There are four common space station geometries: Sphere, Dumbbell, Torus, and Cylinder [Johnson and Holbrow 1977]. All four of these geometries have symmetry to support spinning and the production of centripetal gravity. They also have hollow regions to hold the atmosphere.

### 2.2 Station Assessments

Studies have used multiple approaches to assess and select a station geometry. These approaches include maximizing volume, and O'Neill found that the sphere geometry was superior [O'Neill 1976]. Another was to minimize the mass for a given population, and the cylinder geometry was superior [O'Neill 1976] [Globus et al. 2007]. A third example was to minimize the station mass for a given rotation radius, and the torus geometry was superior [Johnson and Holbrow 1977] [Misra 2010]. Providing passive rotational balance imposes a geometry constraint on space station designs [Brown 2002]. This constraint dramatically changes the shape of cylinder space stations.

These historical assessments and selections were typically based on only thin space station shells and a single projected floor. Because this study uses thick shells and multiple floors, refinement of those earlier geometries is explored.

### 2.3 Refined Space Station Geometries

This paper presents a refined set of space station geometries to address rotational stability and gravity constraints. Figure 2-1 introduces those refined geometries with line drawings. Long cylindrical stations become short hatbox stations; spherical stations become ellipsoidal stations; circular cross-section torus stations become elliptical cross-section torus stations; and dumbbell structures are doubled. This paper introduces other constraints from multiple floors, air pressure, and stability. Our geometry designs are ultimately modeled with multiple components to reflect these constraints and to evaluate the station's stability.

### 2.4 Station Details

The rendering of a large elliptical torus space station helps to visualize our goal. Figure 1-1 shows an exterior view. Figure 2-2 includes a cutaway rendering and shows an interior view. These renderings show many of the station's features. The interior view shows a central floor curving upward, providing open space and an excellent vista for the residents. Below that central floor are multiple lower floors that can be seen in the cutaway drawing. These floors provide space for agriculture, industry, commercial, residential, and services. The station is designed to provide an Earthlike gravity range from the main floor to the bottom outer rim. The station is also designed to provide passive rotational balance. The station uses high-tensile strength structures for its construction.

### 2.5 Multiple Floors

Adding multiple floors to a space station greatly increases the available floor space [Jensen 2023]. Examples of terrestrial multiple-floors structures include skyscrapers, underground cities, submarines, and cruise ships. Limited space and costs in urban environments promote high-rise living. Adding floors to a structure greatly increases the available floor space. Historic and modern underground cities exist [Garrett 2019]. Entrepreneurs have begun to convert abandoned military missile silos into multiple-floors homes and underground cities [Garrett 2019]. Our papers [Jensen 2023] and [Jensen 2024d] provide more information on multiple floors in rotating space stations.

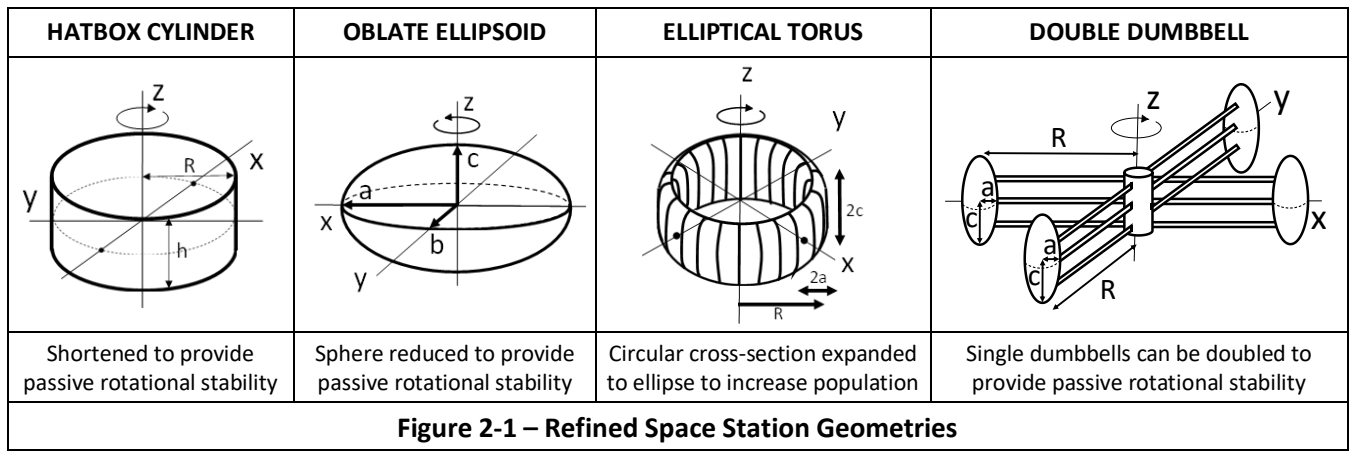

| HATBOX CYLINDER | OBLATE ELLIPSOID | ELLIPTICAL TORUS | DOUBLE DUMBBELL |
|---|---|---|---|
| Shortened to provide passive rotational stability | Sphere reduced to provide passive rotational stability | Circular cross-section expanded to ellipse to increase population | Single dumbbells can be doubled to provide passive rotational stability |

**Figure 2-1 – Refined Space Station Geometries**

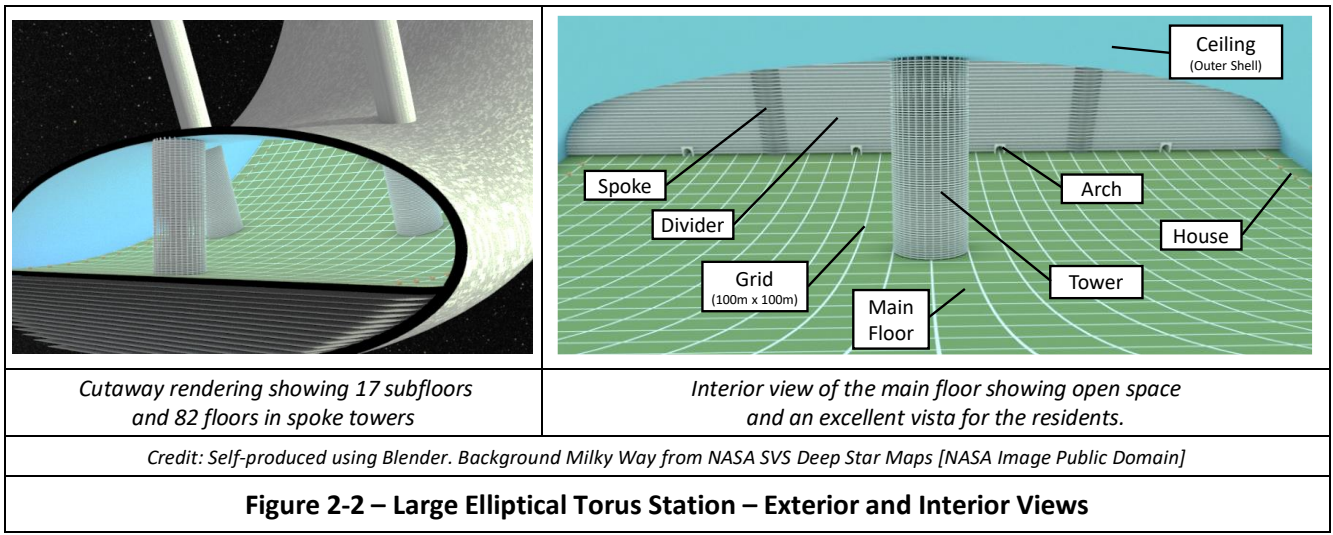

*Cutaway rendering showing 17 subfloors and 82 floors in spoke towers*

*Interior view of the main floor showing open space and an excellent vista for the residents.*

Credit: Self-produced using Blender. Background Milky Way from NASA SVS Deep Star Maps [NASA Image Public Domain]

**Figure 2-2 – Large Elliptical Torus Station – Exterior and Interior Views**

Figure 2-2 also helps to visualize the multiple components in these stations. This torus station has a major radius of 2300 meters. The torus has an elliptic cross-section with minor radii of 400 and 1150 meters. The interior view rendering labels many of the components. This station provides an open vista like a long valley of 6 kilometers over the curved top floor. The cutaway rendering shows the multiple floors beneath the main floor in the torus station. More details on this torus are in [Jensen 2023].

## 3  Space Station Limitations

Space stations must be designed to support personnel. The first set of constraints considered are those affecting the welfare of the occupants. These limits include gravity, air pressure, material, top floor, population, and radiation. Human frailty imposes many limits on the design of a space station. These limits constrain the station design and, ultimately, the station's stability. The risk from rotational instabilities in space stations is feedback that could cause an abrupt change in orientation in the rotating space station [Globus et al. 2007].

### 3.1  Gravity Limits

Gravity plays an important role in the regions where humans will live long-term. Our human bodies have evolved in a very narrow range of gravity. Centripetal gravity, with its benefits and detriments, was covered in our previous paper [Jensen 2023]. Rotating space stations can be designed to create Earth-like centripetal gravity in the most inhabited regions. Keeping the rotation less than 1 rpm will help prevent motion sickness. This rotation rate and gravity imply a station radius greater than 900 meters. The gravity of an example large station is reviewed in this section. The section also defines the range of acceptable gravity and the gravity range effect on the station design.

In a rotating station, the centripetal gravity is always directed radially towards the center of rotation. In the rotating station, this force is $\omega^2 R$, where $\omega$ is the station rotation speed, and R is the radius to the center. The variable $g_0$ is the centripetal gravity acceleration at the station's outer rim. The station's centripetal gravity is $g_h = g_0 (R-h)/R$, where R is the rotating station's radius and h is the height above the outer rim. Figure 3-1 shows this centripetal gravity for several station sizes. These stations rotate at a speed to produce the gravity $g_0$ equal to 1.05g at the outer rim (h=0). The centripetal gravity in small stations reduces quickly with increasing height. The centripetal gravity in large stations is more similar to Earth's gravity with height. Earth's gravity is $g_h = g_0 (R/(R+h))^2$ where R is the Earth's radius, and h is the height above Earth's surface.

With an outer torus major radius of 2400 meters, rotating the station once every 1.6 minutes produces a sensation of Earth-like gravity on the main floor of the large outer elliptic torus. The station used in Figure 1-1 has a gravity range of 1.0g to 1.12g on the floors of the large torus. Stations can be designed with different radii and rotation rates to vary the gravity values over the station floors. Historically, organizations have used different ranges in their designs. Most are designed using a single floor at 1.0g. A 1977 NASA study used a gravity range of 0.9g to 1.0g [Johnson and Holbrow 1977]. They also considered a relaxed constraint range of 0.7g to 1.0g. A decade later, NASA authors designed a torus station with a gravity range from 0.97g to 1.03g [Queijo et al. 1988]. They felt this range would not have any significant influence on human physiology and performance.

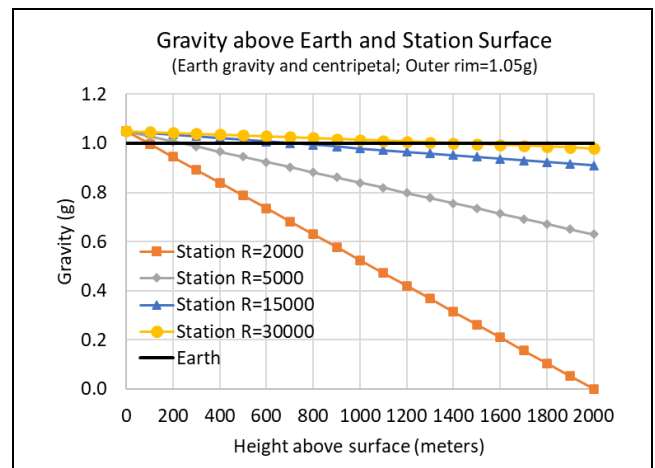

**Figure 3-1 – Earth and Centripetal Gravity**

We usually design the multiple-component stations with a gravity range of 0.95g to 1.1g. The inhabitants will spend most of their time on floors with Earth-like gravity. The gravity range of 0.95g to 1.05g over those floors would minimize the health risk to the residents.

Gravity range details in this subsection were introduced in [Jensen 2023] and considered two classes of rotating stations. One class rotates about an axis outside the enclosed habitable region and includes the torus and dumbbells. The other class rotates about an axis inside the enclosed region and includes the cylinder and ellipsoid. The number of floors in the station volume can be varied in our analysis. So can the floor locations for the minimum and maximum gravities.

Table 3-1 includes equations that define the relationships between the gravities, the radius, and the multiple-floors height. This analysis introduces a scale variable m that defines R as equal to m times a. The scale m is equal to the minimum gravity over the difference of the maximum and minimum gravity. The same table works for a cylinder or ellipsoid. Using the outer radius Ro to analyze those geometries is often convenient. Table 3-1 provides a small table of the habitability scale metrics for a range of minimum gravities (gmin) and maximum gravities (gmax). Broader ranges of gravities support more floors and greater populations in the station. Two values are highlighted in the table: 9.5 and 6.33. These represent a minimum of 0.95g on the top floor and a maximum of 1.05g or 1.1g on the outer rim. Those values and gravity ranges are used extensively throughout this paper and our earlier research [Jensen 2023].

## 3.2  Air Pressure Limits

Air is obviously required to support our station's population. Early space missions used low-pressure, pure oxygen to reduce weight and still provide a breathable atmosphere. We assume the station will have an Earth-like mixture of nitrogen, oxygen, and carbon dioxide. This section provides a brief overview of the limits from air pressure.

Human bodies have adapted to a range of air pressures on Earth's surface. The highest air pressures are at low elevations. The Dead Sea is 430 meters below sea level and has an air pressure of 106,624.5 Pascals. Sea level has an air pressure of 101,325 Pascals. The highest permanent settlement, La Rinconada, is 5100 meters above sea level and has a low air pressure of 53,500 Pascals. This is in the range where severe altitude sickness can begin to occur (above 3,500 meters). Mountain climbers call the region above 8,000 meters the death zone with less than 35,600 Pascals air pressure. Mount Everest has an altitude of 8,850 meters and air pressure of 33,700 Pascals.

This study uses a maximum air pressure of sea level and a minimum of the air pressure at an altitude of Denver. Sea level is 101,325 Pascals. The Denver air pressure is 83,728

| Table 3-1 – Half-Filled Floor Scaling Factor for Various Centripetal Gravity Ranges |||||
|---|---|---|---|---|
| Scale (m) | gmin | | | |
| gmax | 0.85g | 0.9g | 0.95g | 1.0g |
| 1.05g | 4.25 | 6.00 | 9.50 | 20.0 |
| 1.1g | 3.40 | 4.50 | 6.33 | 10.0 |
| 1.15g | 2.83 | 3.60 | 4.75 | 6.67 |
| 1.2g | 2.43 | 3.00 | 3.80 | 5.00 |

*Torus (and Dumbbell) Major Radius=R; Minor radius=a; Outer Radius=R+a; R=m×a; m=gmin/(gmax-gmin)*

*Ellipsoid (and Cylinder) Outer Radius=$R_o$; Top Floor Height=h; $R_o$ = (m+1) h = m̂×h; Top Floor Radius R=$R_o$-h; m̂ = gmax / (gmax-gmin)*

Pascals at 1609 meters above sea level. On rotating stations, this implies a similar maximum habitable altitude. These limits are conservative. Nearly all individuals would breathe comfortably at these air pressures. Large air pressures are denser and would use more air to fill the station. Smaller air pressures would be uncomfortable for some colonists.

This study uses the air pressure analysis and equations from [Jensen 2024d]. That analysis is built on works by [Lente and Ösz 2020] and [O'Neill 1974]. As a summary, we include the resulting equations from that effort. The following equation is used to calculate the Earth's air pressure at altitudes above its surface:

$$P_h = P_0 \, exp\left(-\frac{Mg_0}{kT}h\right)$$

Where $P_0$ is the air pressure at the outer radius and h is the height above Earth's surface. The variable M is the air molar mass constant, $g_0$ is the gravity constant, k is the gas constant, and T is temperature. In a rotating cylinder (and ellipsoid) space station, the following equation computes the air pressure above the outer rotating rim:

$$P_h = P_0 \, exp\left(-\frac{Mg_0}{kT}\left(h - \frac{h^2}{2R}\right)\right)$$

Constants are the same as with the Earth air pressure equation. For the rotating station, R is the outer radius of the station, and h is the height above that outer rim. Similarly, for the torus (and dumbbells), the radius R is typically the major axis radius, and the outer rotating rim is at that radius plus the minor axis, R+r. With this change, the equation becomes:

$$P_h = P_0 \, exp\left(-\frac{Mg_0}{kT}\left(h - \frac{h^2}{2(R+r)}\right)\right)$$

Assuming constant temperature, the air density analysis in the station produces the same equation. Additional details and the values of these constants are in the air pressure constraints section of [Jensen 2024d]. All stations have sea-level air pressure at the outer rim by design. The air pressure decreases with height, just like the pressure on Earth. The air pressure in rotating stations decreases with height but not as much as on the Earth. Air density becomes important because of the air mass and its influence on the station's stability.

## 3.3 Station Material Limits

The materials used in the construction of a rotating space station must be able to withstand a variety of stresses and forces. The materials must be strong enough to support the weight of the station and its inhabitants, as well as withstand the centrifugal forces generated by the rotation. The stresses that space station materials must endure include mechanical, thermal, and radiation stresses. Mechanical stresses are caused by the station's weight and the forces that act upon it, such as when the station is maneuvered or docked with another spacecraft. Thermal stresses result from the extreme temperature variations that the station could experience. Radiation stresses come from the station's exposure to cosmic radiation and solar flares.

A report on Space Structures and Support Systems [Bell and Hines 2012] focused on glass, basalt, metals, concrete, and anhydrous glass. They presented that glass, basalt, and metals have high compressive strength [Bell and Hines 2012]. Of those materials, anhydrous glass had the greatest tensile strength. Our previous paper [Jensen 2023] covered stresses produced in the rotating space station. Newer details in [Jensen 2024d] provide maximum station radius values. Table 3-2 replicates a subset of those results. These results use the hoop stress from the rotating station. Steel could be used to create the exterior station shell, which could be almost 4 kilometers in radius. The tensile strength of anhydrous glass suggests that a station of over 25 km could be feasible. Table 3-2 includes a filled shell structure using anhydrous trusses and processed regolith fill. The structure does have lower tensile strength; however, the low density keeps the station radius close to 20 kilometers. Conservatively, a much smaller radius closer to 3 or 4 kilometers seems more appropriate. Such stations may seem absurd today; however, in the future, such sizes may be possible using in-situ extraterrestrial materials, robotic automation, self-replication, and artificial intelligence. Asteroids, moons, and dwarf planets can provide nearly unlimited amounts of building material. Our study has shown that automation can provide nearly unlimited amounts of labor to build these rotating space stations [Jensen 2023].

Figure 3-2 presents the working stress for various materials. Again, these stresses were covered in [Jensen 2023]. The working stress is the sum of the stresses from the air pressure, the centripetal forces on the shell, and the centripetal forces on the internal structures and furnishings. The chart shows the working stresses in megapascals ranging from 1 to 20,000 on the logarithmic y-axis. The x-axis shows the outer rim radius of the rotating station and ranges from 100 to 100,000 meters. The station is rotating at a speed to produce 1g at the outer rim. The chart includes the material stresses for four materials. The anhydrous glass has the largest tensile strength, and aluminum has the smallest. Steel, with the highest density, creates the largest working stress. The filled structure, with the lowest density, creates the smallest working stress. All the materials support their working stress below the rotation radius of 10,000 meters. Because of this material strength limit, we use this radius as a maximum value. Note that filled structures have the potential to increase this rotation radius to over 40,000 meters, and anhydrous glass could increase the radius to over 100,000 meters.

A simple analysis of the strength of materials shows that large stations can be built. The desired construction materials for rotating space stations must be strong, lightweight, durable, and resistant to the harsh conditions of space. We advocate that anhydrous glass from asteroid oxides is a valid choice for constructing large stations.

## 3.4 Multiple Floors Limits

There are issues with creating and using many floors. Studies have found that people living in highrises on the Earth suffer from issues such as greater mental health problems, higher fear of crime, fewer positive social interactions, and more difficulty with child-rearing [Barr 2018]. NASA studies have considered the potential oppressive closed-quarters ambience to be a risk to the colonists' psychological well-being [Johnson and Holbrow 1977] [Keeter 2020]. Fortunately, researchers offer approaches to address these issues and risks using proper planning and space allocation. As such, we

| Table 3-2 – Materials and Space Habitat Radius | | | |
|---|---|---|---|
| Material | Tensile Strength (MPa) | Density (g/cm3) | Radius (km) |
| Anhydrous Glass (max) | 13,800 | 2.70 | 123.4 |
| Anhydrous Glass | 3,000 | 2.70 | 26.8 |
| Filled Structure | 1,500 | 1.72 | 21.1 |
| Steel | 1,240 | 7.80 | 3.8 |
| Aluminum | 352 | 2.65 | 3.2 |
| Credit: Self-produced in [Jensen 2023] using data from [O'Neill 1974] [McKendree 1995] [Bell and Hines 2012]; [Facts]. | | | |

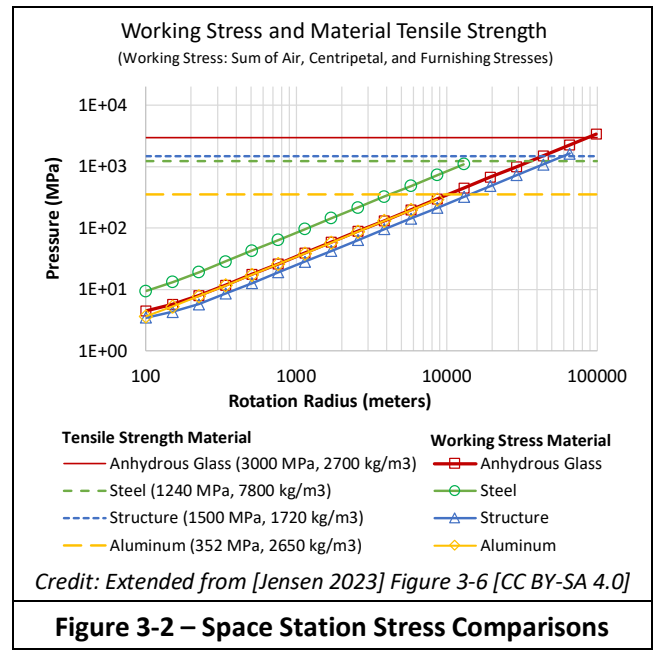

Credit: Extended from [Jensen 2023] Figure 3-6 [CC BY-SA 4.0]

**Figure 3-2 – Space Station Stress Comparisons**

believe using many floors is acceptable for space station habitation.

## 3.5 Top Floor Limits

Historically, most space stations have a single floor on the interior of their outer shell. Our research has introduced and analyzed multiple floors in the rotating station geometries. The entire station could be filled with floors; however, leaving open space above the floors produces an aesthetically pleasing region for the inhabitants. The top floor could be anywhere from the outer rim, to the center, or to the top rim. Figure 2-2 provides a rendering of the multiple floors and the open space.

There are limits on the number of floors in our multiple-floors station designs. The viable gravity range is one of those limits. The floors typically range from 0.95g to 1.05g. A viable air pressure is another limit in the design. The air pressure in the habitable portion of the stations ranges from sea level to at least Denver air pressure. Like on the Earth, the air density in the rotating station varies with height. The station size and centripetal gravity affect the station air pressure. In our large rotating stations, the air pressure limits the top floor height to about 1600 meters. The station could have a maximum of over 300 floors with a 5-meter spacing between the floors. Generally, the minimum gravity and air pressure limits represent the habitable top floor height for the station designs. In small stations, the gravity limits the height of the top floor. In large stations, the air pressure limits the position of the top floor.

This top floor location and limits are detailed in our original *Asteroid Restructuring* paper [Jensen 2023] and our *Large Space Station*s paper [Jensen 2024d]. The large space stations paper introduces a straightforward approach to increase the top floor in large stations above the minimum air pressure limit. Consider a floor at the air pressure height limit. That floor could be made airtight. Immediately below that floor the air pressure would be at the Denver limit. Immediately above that top floor, the air pressure could be raised to the sea level maximum. Additional floors could be constructed above that original top floor. Another 1600 meters of floors could be added before reaching the Denver air pressure limit again.

As an example of this air pressure and the use of airtight floors, we include Figure 3-3. The chart shows the air density in a 35,000-meter radius station along the left y-axis and ranges from 0 to 1.3 kilograms per meter cubed. The x-axis shows the height in the station. For reference, the right y-axis shows the gravity in the station and ranges from 0g to 1.3g. The chart shows the air density at station heights with the air pressure limit (open floors), the gravity limit (airtight layers), and a non-rotating uniform density. Without airtight layers (open floor), the red line shows the air density monotonically decreasing (using air density equations).

The stability analysis considers stations with the air pressure limit (open floors) and the gravity limit (airtight layers). The mass of the air and the number of floors are considerably different between the two designs. The air-pressure top floor design has less air, fewer floors, and less weight than the gravity-limited design. Our Large Station paper includes additional details and mass examples [Jensen 2024d].

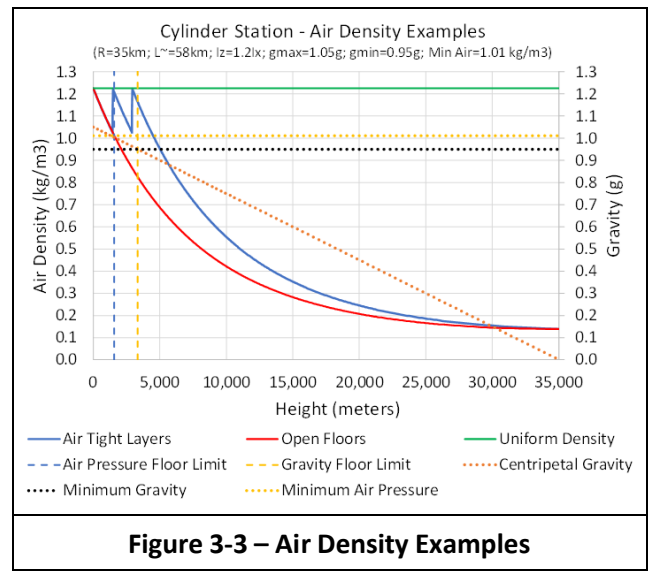

**Figure 3-3 – Air Density Examples**

## 3.6 Population Limits

This section briefly reviews the population limits for our rotating space station. These concepts were introduced and detailed in our Restructuring Asteroid paper [Jensen 2023]. Some of the graphs and text are reused to review details. This section briefly overviews two topics from those papers for context. First is the required allocation of floor space. Second is a review of the effects from single and multiple floors in the design.

### 3.6.1 Floor Allocation Usage

The allocation of floor space to categories such as living, agriculture, industry, and openness creates limits for the population. Everyone in a space station needs space to support their living, working, industry, and agriculture needs. As an example, a 1970s NASA design study created a set of surface usage metrics [Johnson and Holbrow 1977]. These included residential, open spaces, transportation, and agriculture. These totaled to 155.2 square meters per person. Using multiple-story buildings, the projected floor space on the single floor became 67.0 square meters per person with an average height of 11.2 meters.

The stations in our studies have multiple floors and consider floor spacings of 5, 10, and 15 meters, and they use slightly different categories (open areas, support, agriculture, industry, and residential). Usage expectations and requirements have changed in the 50 years since the NASA study. With a fixed floor spacing of 15 meters, the updated space allocation provides each individual 65.5 square meters per person, which is close to the 1970s NASA metric. With the 5-meter floor spacing, each person is allocated 144.2 square meters. Details are available for these different usage metrics [Jensen 2023].

### 3.6.2 Single and Multiple Floors Examples

Figure 3-4 shows populations in a torus space habitat. These estimates use the original NASA estimate of 67 square meters per person. The population is based on the available surface area of the floors. The vertical axis shows the population and ranges from 10 to 1 billion on a logarithmic scale. The horizontal axis shows the major radius of the torus and ranges from 100 to 40,000 on a logarithmic scale.

Figure 3-4 shows three population allocations for a torus surface area. The lowest population results from using only a single floor projected across the minor diameter. This living area is twice the minor radius times the major radius circumference. This population is proportional to the radius squared. Larger populations are supported with the multiple floors. Those multiple floors are under the center main floor; see Figure 2-2. As the minor radius increases, the space under the center diameter floor increases. This space provides additional floors. For stations with a small radius, Figure 3-4 shows little increase in population by adding floors. As the radius increases, the benefit from adding floors appears. A torus station with a major radius of 1000 meters and a minor radius of 100 meters would have about 20 floors under the main floor. Each floor extends entirely around the major radius of the torus. The population with multiple floors is nearly proportional to the radius cubed.

There will be open vistas on the main floor for colonists to enjoy. Unfortunately, with large habitats using multiple floors, most of the square footage is on the lower levels. We tend to think the lowest floors would be undesirable as living quarters. To compensate, Figure 3-4 also includes a population estimate using only the top 20 floors. The lower floors are not used in our population estimate but could support additional industrial, research, ventilation, transportation, storage, and agriculture. Figure 3-4 shows population results that are first proportional to the radius cubed. Once there are more than 20 floors (at a radius of about 1000 meters), the population is only proportional to the radius squared. This 20-floor allocation is only relevant for large stations.

Figure 3-4 includes several tori at their published major radii and population. These include Tiny Torus [Johnson and Holbrow 1977], Stanford Torus [O'Neill et al. 1979], Atira Torus [Jensen 2023], and Elysium Torus [Brody 2013].

### 3.6.3 Population Summary

More details on the single and multiple-floors populations are in [Jensen 2023] and [Jensen 2024d]. For all four geometries, using multiple floors significantly improved the supported population. For a single floor, the population is proportional to the radius squared. For geometries with multiple floors, the population is nearly proportional to the radius cubed. Multiple floors take advantage of the internal station volume.

## 3.7 Radiation

There is sufficient radiation in space to cause sickness (e.g., damage to DNA and cancer). The radiation comes from galactic cosmic rays (GCR) and from solar particle events (SPE) [Dunbar 2019]. Earth's magnetic field and thick atmosphere protect people on the surface from this radiation.

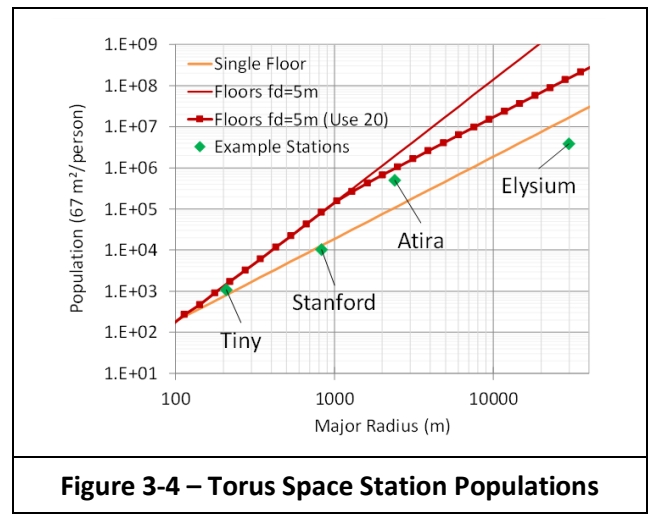

**Figure 3-4 – Torus Space Station Populations**

Various studies suggest using a layer of regolith to provide protection from cosmic rays. The depth of the regolith ranges from 2.5 meters [O'Neill 2008] to more than 8 meters [Turner and Kunkel 2017] to reduce the radiation to an Earth background level. In our station design, the outer shell has a thick regolith layer to protect our station's residents. A truss framework used as the shell provides most of the structural integrity. Ten-meter walls would be sufficient to provide radiation protection. We prefer solutions with greater thickness on the outer walls. The extra thickness provides additional integrity for potential debris, small meteoroids, and ship collisions. Our designs typically use a shell thickness of 20 meters in our analysis.

Historically, researchers have been concerned with the substantial structural shell weight penalty from an attached shielding. As a reference, a 1979 NASA paper considered radiation shielding and offered requirements and analysis for the various geometries [Bock, Lambrou, and Simon 1979]. They considered an attached shield integral to the design and an unattached stationary shielding with a separate living area rotating inside. As another reference, a different author stated that the material strength of attached shielding would not support a rotation rate greater than 1 revolution per minute [Graem 2006]. This implies stations must be larger than a 900-meter radius to have an integrated attached shield. Even for thin-shell single-floor designs, one quickly finds that shielding mass dominates the station structural mass [Bock, Lambrou, and Simon 1979].

The mass and inertia of this shield will affect the station's inertia. For smaller stations, the shielding is 100 to 1000 times greater than the structural mass of the habitat interior. For large stations, the shielding is only 10 times greater than the structural mass of the habitat interior. The external shell and shielding still dominate the mass and inertia for multiple-floors stations. With very large multiple-floors stations, other components of the station begin to equal or exceed the outer shell (shield) mass and inertia.

## 4 Rotational Stability

This section provides background and definitions to introduce rotational stability. It also covers alternative approaches to stabilize the rotating space station. A preview of the rotational stability of a cylinder is included.

### 4.1 Rotational Stability Background

Newton's first law of motion is a body in motion tends to remain in motion. The space station will rotate about a central axis to produce centrifugal force as a psuedogravity. In the vacuum of space, once the space station is spinning, it will tend to remain spinning. Other forces on a space station could potentially disrupt the spinning stability. Examples of these forces include crafts arriving and leaving the station, movement of objects internally, solar wind, and micrometeors.

The stability criterion was recognized and addressed in a recent space station design [Globus et al. 2007]. The team described in the Kalpana One Orbital Station paper:

*"In an ideal space environment, any cylinder rotating about its longitudinal axis will continue to do so forever; but in the real space environment perturbations cause rotating systems to eventually rotate about the axis with the greatest angular moment of inertia. If that axis is not along the cylinder length, this introduces a catastrophic failure mode where the settlement gradually changes its rotational axis until it is tumbling end-over-end."* [Globus et al. 2007]

Obviously, it is essential to prevent a catastrophic instability that results in a settlement tumbling end-over-end.

### 4.2 Spin Stabilized System

There is a risk for rotation instabilities with asymmetric geometries. This instability was named after the Soviet cosmonaut Vladimir Dzhanibekov, who discovered this effect on the MIR space station in 1985. The cosmonaut spun a T-shaped handle and discovered bi-stable rotation states. A Berkeley website [O'Reilly et al. 2021] provides an analysis of this rotational behavior; see Figure 4-1. This site also includes a link to a video showing a rotating T-handle in zero gravity on the ISS (International Space Station). The abrupt

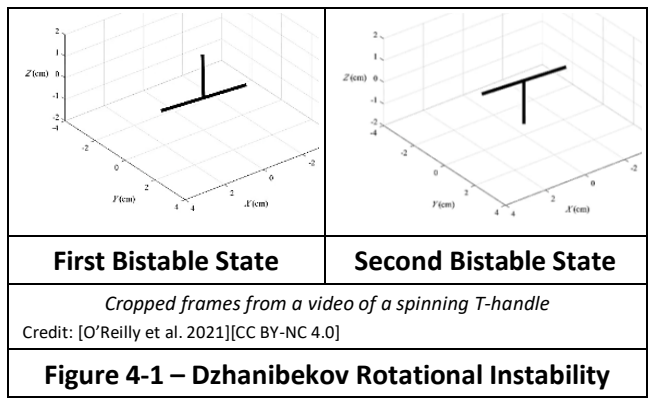

| First Bistable State | Second Bistable State |
|---|---|

*Cropped frames from a video of a spinning T-handle*
Credit: [O'Reilly et al. 2021][CC BY-NC 4.0]

**Figure 4-1 – Dzhanibekov Rotational Instability**

changes in orientation between two states would not be desirable in a space station.

Even though our space stations are symmetrical about the rotation axis, one must still design the geometry to maintain stability and avoid this behavior. In the *Elements of Spacecraft Design*, Charles Brown recommends that the desired axis of rotation should have an angular moment of inertia (MOI) at least 1.2 times greater than any other axis to provide rotational stability [Brown 2002]. Moments of inertia can be computed for each of the three rotation axes of a space station. The following quotes from the Globus paper and a more recent Fitzpatrick paper serve as our rules to provide passive rotational stability:

*"Experience with spin-stabilized spacecraft suggests that the desired axis of rotation should have an angular moment of inertia at least 1.2 times greater than any other axis."* [Globus et al. 2007]

*"Finally, if two of the principal moments are the same then it can be shown that the body is only stable to small perturbations when rotating about the principal axis whose moment is distinct from the other two."* [Fitzpatrick 2011]

*"In conclusion, a rigid body with three distinct principal moments of inertia is stable to small perturbations when rotating about the principal axes with the largest and smallest moments, but is unstable when rotating about the axis with the intermediate moment."* [Fitzpatrick 2011]

Not meeting these guidelines could lead to instabilities. Historic station designs did not appear to use these stability rules. The paper on the Kalpana cylinder space station investigated this inertial stability [Globus et al. 2007]. The researchers noted that the Bernal Sphere and the O'Neill Cylinder would be unstable because their desired axis of rotation was not 1.2 times greater than the other axes. They applied this guideline to design the Kalpana cylinder design space station. Applying this stability criterion dramatically changed the original long-and-narrow O'Neill Cylinders [O'Neill 1976] to the short-and-squatty Kalpana One Station [Globus et al. 2007].

### 4.3 Rotational Stability Alternatives

There are multiple ways to maintain the rotational stability. The paper [Jensen 2024d] covers three techniques: passive spin stabilization, three-axis stability with thrusters, and three-axis stabilization with momentum wheels. That paper covers the advantages and disadvantages of those three techniques.

This paper focuses on the passive spin stabilization for our designs. The space stations of this study are rotating to provide centripetal gravity to the residents. This rotation provides spin stabilization; however, this stabilization will not be perfect. The mass distribution will not be uniform about the rotation axis. Shuttles, material, equipment, and people will be in motion in the station and possibly destabilize the station. Large space stations will have large angular momentum, and most movements will have little impact on the stability. Ultimately, station designs will use computers and accelerometers to provide active control and counteract such instabilities. Authors have suggested active control techniques using motors to move large weights on cables [Globus et al. 2007] or using pumps to move water [Lipsett 2005].

## 4.4 Rotational Stability Preview

Our study follows guidelines and rules based on [Fitzpatrick 2011] [Globus et al. 2007] [Brown 2002] and [Birse 2000]. A key stability rule is that the moment of inertia (MOI) about the rotation axis should be greater than 1.2 times the other axes [Brown 2002]. With the rotation axis assigned as the z-axis, the system would be stable when Iz≥1.2 Ix.

Consider the cylinder geometry as an example of this relationship. In *§6 Single Floor Rotational Stability*, we provide the thin shell analysis for the cylinder. That analysis finds the cylinder length over the cylinder radius (L/R) should be 1.3 to be stable (Iz=1.2Ix). This analysis and result match the Globus study [Globus et al. 2007].

The stability of the cylinder geometry includes multiple station components (outer shell, endcaps, spokes, multiple floors, air, and a center shuttle bay). In *§7 Multiple Component Rotational Stability,* we detail these station components. Figure 4-2 shows the stability when including the multiple components and varying the cylinder geometry. The x-axis shows the cylinder rotation radius on a logarithmic scale ranging from 100 to over 100,000 meters. The y-axis shows the stability as the ratio of the z moment of inertia (Iz) over the x moment of inertia (Ix). The minimum stability limit is a dotted black line at 1.2 on the stability axis. Geometries with stability values greater than 1.2 are rotationally stable, and values less than 1.2 are unstable.

Using the thin shell L/R=1.3 and multiple components, the chart data shows the station stability crosses the 1.2 limit. This shows instability with the multiple components when the radius is less than 1300 meters and stable for larger radii. Figure 4-2 shows geometries that are longer in length tend to be more rotationally unstable; see L/R=2.0. It also shows shorter geometries (cylinder hatbox) are more rotationally stable; see the L/R=0.5. In our analysis, the stability ratio of 1.2 is attained by varying the L/R with the changing radius. Most of the following sections use this stability ratio.

## 5 Station Component Analysis

This section introduces the components used to model our stations. One can model a space station as its outer shell. The outer shell is often the highest mass component in the station. The mass and geometry are combined to determine moments of inertia (MOIs) and the station stability. Decomposing the station into its constituent pieces improves the model results. The pieces are analyzed separately and combined to represent the components and the station. Those components include the outer shell, spokes, main floor, multiple floors, air, shuttle bay, and dividers. This section defines those components for each station geometry and provides analytic equations that model their volume and mass.

### 5.1 Station Component Equations

Figure 2-1 illustrated the four space station geometries. This subsection describes the major components used in our analysis for each of the four refined space station geometries. Different sizes, densities, and quantities of individual pieces represent the major station components. Equations to compute the volume and the mass of those components follow. Variable subscripts define the component. They also define the inner dimensions with subscript "i". Outer dimensions are labeled with subscript "o". This analysis uses the density values from Table 5-1 with the volumes to compute the mass.

**Hatbox Cylinder**: The hatbox cylinder is comprised of 7 components. It is decomposed into the outer shell, end caps, the air, the floors, the main floor, the spokes, and the shuttle bay. The variables in the equations include subscripts for the inner and outer dimensions. The equations are:

$$m_{shell} = \rho_{fill} V_{shell} = \rho_{fill}\, pi\, L\, (R_o^2 - (R_o + t)^2)$$
$$m_{endcaps} = 2\, \rho_{fill} V_{endcap} = \rho_{fill}\, pi\, t\, R_o^2$$
$$m_{air} = \rho_{air} V_{air} = \rho_{air}\, pi\, L\, R_o^2$$
$$m_{floor5} = \rho_{floor5} V_{floor5} = \rho_{floor5}\, pi\, L\, (R_o^2 - r_m^2)$$
$$m_{main} = \rho_{fill} V_{main} = \rho_{fill}\, pi\, h\, (r_m^2 - (r_m + t)^2)$$
$$m_{spoke} = \rho_{spoke} V_{spoke} = \rho_{spoke}\, \pi\, r_{spoke}^2\, 2\, R_o$$
$$m_{bay} = \rho_{bay}\, \pi\, r_{bay}^2\, 2\, L_{bay} + \rho_{spoke}\, \pi\, r_{spoke}^2\, 2\, \max(c - L_{bay}, 0)$$

Where L=cylinder station length; $R_o$=radius to the outer floor; t=thickness of shell and endcaps. The variable $r_m$

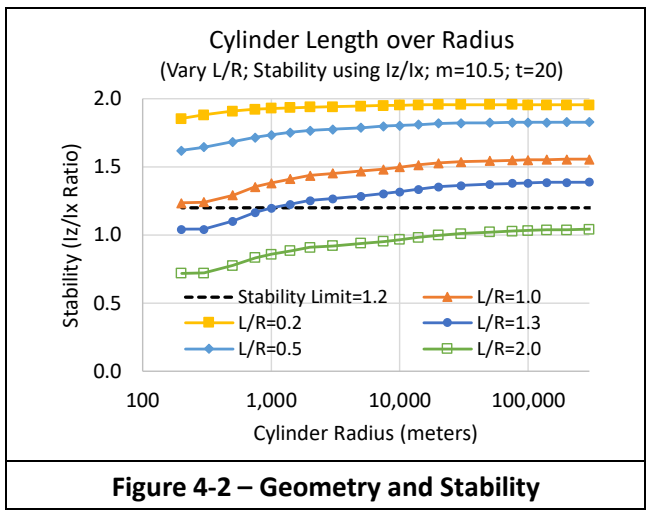

**Figure 4-2 – Geometry and Stability**

| Table 5-1 – Densities of Station Components | | | |
|---|---|---|---|
| **Density** | **Value (kg/m3)** | **Density** | **Value (kg/m3)** |
| $\rho_{steel}$ | 7850 | $\rho_{spoke}$ | 337.4 |
| $\rho_{basalt}$ | 2790 | $\rho_{structure}$ | 64.4 |
| $\rho_{rod}$ | 2790 | $\rho_{floor5}$ | 32.2 |
| $\rho_{panels}$ | 2790 | $\rho_{floor10}$ | 17.0 |
| $\rho_{aluminum}$ | 2650 | $\rho_{floor15}$ | 11.2 |
| $\rho_{bay}$ | 2291 | $\rho_{air}$ | 1.5 |
| $\rho_{fill}$ | 1721 | $\rho_{sealevel}$ | 1.225 |

represents the radius to the main floor where $r_m=R_i-h_m$, $h_m=R_i/m$, and $m=g_{max}/(g_{max}-g_{min})$. $L_{bay}$ is the maximum shuttle bay length. The shuttle bay and spokes are interior to the shell and endcaps. This air mass equation uses a uniform air density in the station. Later analyses use varying air density to account for the centripetal gravity effects.

**Oblate Ellipsoid:** The oblate ellipsoid is comprised of 6 components. It is decomposed into the outer shell, air, floors, main floor, spokes, and shuttle bay. The equations are:

$$V_{i-shell} = \frac{4}{3} pi\ abc;\ V_{o-shell} = \frac{4}{3} pi\ (a+t)(b+t)(c+t)$$

$$V_{shell} = \frac{4}{3} pi\ (t\ (ab + ac + bc) + t^2\ (a+b+c) + t^3)$$

$$V_{shell} = V_{o-shell} - V_{i-shell}$$

$$m_{shell} = \rho_{fill}\ V_{shell}$$

$$m_{air} = \rho_{air}\ V_{air} = \rho_{air}(V_{i-shell}) = \rho_{air}\left(\frac{4}{3} pi\ abc\right)$$

$$m_{main} = \rho_{fill}\ V_{main} = \rho_{fill}\ t\ 2\ pi\ a_m\ 2\ c_m$$

$$m_{spoke} = \rho_{spoke}\ V_{spoke} = \rho_{spoke}\ \pi\ r_{spoke}^2\ L_{spoke}$$

$$m_{bay} = \rho_{bay}\ \pi\ r_{bay}^2\ 2\ L_{bay} + \rho_{spoke}\ \pi\ r_{spoke}^2\ 2\ \max(c - L_{bay}, 0)$$

Where c is the distance to the ellipsoid outer shell on the axis of rotation, and a and b are the other axes and equal distances to the outer shell. The floors in the ellipsoid are cylinders. The variable $a_m$ represents the radius to the main floor where $a_m = R_i - h_m$, $h_m = R_i/m$, and $m = g_{max}/(g_{max}-g_{min})$; the width of the main floor is $2c_m$ where $c_m = c_i\sqrt{1 - a_m^2/a_i^2}$; the variable t is the thickness of the shell and main floor. Like the cylinder, the shuttle bay and spokes are interior to the ellipsoid shell.

The volume of the multiple floors is challenging to visualize in the ellipsoid. Figure 5-1 is included to assist. Our analysis uses the volumes of an ellipsoid, a cylinder at the main floors, and ellipsoid endcaps on the cylinder. The length ci is on the shorter polar axis along the rotation axis. The variable $h_c$ represents the height of the endcap $h_c = (c_i - c_m)$ and $a_i$ is equal to $b_i$ in the oblate ellipsoid. The equations for those volumes are:

$$V_{acylinder} = pi\ a_m^2\ 2\ c_m = 2\ pi\ a_m^2\ c_i\sqrt{1 - a_m^2/a_i^2}$$

$$V_{aendcap} = pi\ a_i^2\left(\frac{h_c^2}{3c_i^2}3c_i - \frac{h^2}{3c_i^2}h_c\right)$$

$$V_{aendcap} = \frac{1}{3}pi\ \frac{a_i^2}{c_i^2}h_c^2(3c_i - h_c)$$

$$V_{floor5} = V_{i-shell} - V_{acylinder} - 2\ V_{aendcap}$$

$$V_{floor5} = \frac{4}{3} pi\ abc - 2\ pi\ a_m^2\ c_i\sqrt{1 - a_m^2/a_i^2} - 2\left(\frac{1}{3}pi\ \frac{a_i^2}{c_i^2}h_c^2(3c_i - h_c)\right)$$

$$V_{floor5} = \frac{4}{3} pi\left(abc - \frac{3}{2}a_m^2\ c_i\sqrt{1 - a_m^2/a_i^2} - \frac{1}{2}\frac{a_i^2}{c_i^2}h_c^2(3c_i - h_c)\right)$$

$$m_{floor5} = \rho_{floor5}\ V_{floor5}$$

**Elliptical Torus:** The elliptical torus is comprised of 7 components. It is decomposed into the outer shell, the air, the floors, the main floor, the dividers, the spokes, and the shuttle bay. The main floor of the torus is at the major radius R. The floor is at the center of the torus tube and separates the tube into inner and outer volumes. The outer tube contains the multiple floors. The tube has a minor radius, a, that is coincident with R and a minor radius, c, that is perpendicular

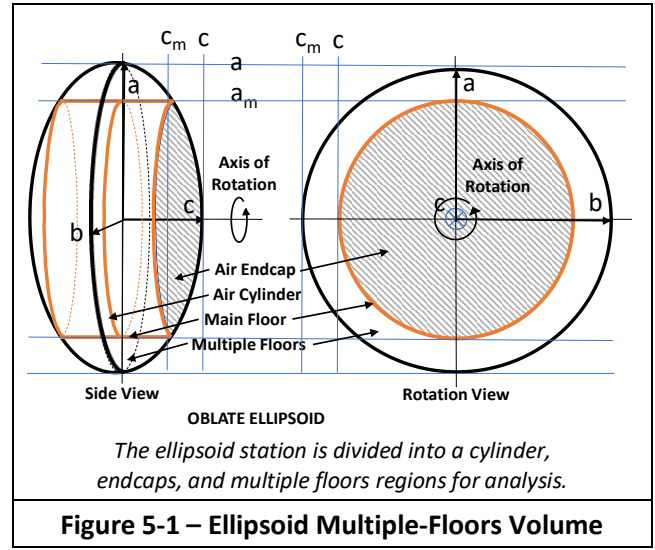

**Figure 5-1 – Ellipsoid Multiple-Floors Volume**

*The ellipsoid station is divided into a cylinder, endcaps, and multiple floors regions for analysis.*

to R. The floors range from radius R to radius R+a. The dividers separate the tube into airtight compartments as part of a fail-safe design. The equations are:

$$V_{itorus} = 2\ pi^2\ a\ c\ R$$

$$V_{ihalf} = \left(\frac{1}{2} - \frac{2}{3pi}\frac{a}{R}\right)V_{itorus};\ V_{ohalf} = \left(\frac{1}{2} + \frac{2}{3pi}\frac{a}{R}\right)V_{itorus}$$

$$m_{air} = \sigma_{air}\ V_{air} = \sigma_{air}\ V_{itorus} = \sigma_{air}\ 2\ pi^2\ a\ c\ R$$

$$m_{floor5} = \rho_{floor5}\ V_{floor5} = \rho_{floor5}\ V_{ohalf} = \rho_{floor5}\left(\frac{1}{2} + \frac{2}{3pi}\frac{a}{R}\right)V_{itorus}$$

$$m_{main} = \rho_{fill}\ V_{main} = \rho_{fill}\ 2\ pi\ R\ 2\ c\ t$$

$$V_{otorus} = 2\ pi^2\ (a+t)\ (c+t)\ R$$

$$V_{shell} = V_{otorus} - V_{itorus} = 2\ pi^2\ (at + ct + t^2)\ R$$

$$m_{shell} = \rho_{fill}\ V_{shell} = \rho_{fill}\ 2\ pi^2\ (at + ct + t^2)\ R$$

$$m_{divider} = \rho_{divider}\ V_{divider} = \rho_{divider}\ t_{divider}\ \pi\ a\ c$$

$$m_{spoke} = \rho_{spoke}\ V_{spoke} = \rho_{spoke}\ \pi\ r_{spoke}^2\ L_{spoke}$$

$$m_{bay} = \rho_{bay}\ \pi\ r_{bay}^2\ 2\ L_{bay} + \rho_{spoke}\ \pi\ r_{spoke}^2\ 2\ \max(c/2 - L_{bay}, 0)$$

Where a=horizontal minor axis; c=vertical minor axis; R=major axis; t=thickness. The divider thickness is about the radius of the spoke. Figure 2-1 does not show the dividers, shuttle bay and spokes.

**Double Dumbbell:** The double dumbbell is comprised 6 components. It is decomposed into the outer shell, air, floors, main floor, spokes, and shuttle bay. The main floor of the dumbbell node is placed at the major radius R. The main floor is at the node's center, and the node is separated into equal inner and outer volumes. The outer volume contains the multiple floors. The equations for single components in the dumbbell follow. There are multiple nodes, spokes, air, main floor, and floors in both the dumbbell and double dumbbell. For one of those nodes:

$$m_{node} = 2\ \rho_{fill}\ V_{node}$$

$$V_{node} = V_{o-node} - V_{i-node}$$

$$V_{i-node} = \frac{4}{3} pi\ abc;\ V_{o-node} = \frac{4}{3} pi\ (a+t)(b+t)(c+t)$$

$$V_{node} = \frac{4}{3} pi\ (t\ (ab + ac + bc) + t^2\ (a+b+c) + t^3)$$

$$m_{spoke} = \rho_{spoke}\ V_{spoke} = \rho_{bay}\ \pi\ r_{spoke}^2\ 2\ R$$

$$m_{bay} = \rho_{bay} V_{bay} = \rho_{bay} \pi r_{bay}^2 L_{bay}$$
$$m_{air} = \rho_{air} V_{air} = \rho_{air} V_{i-node} = \rho_{air} \frac{4}{3} pi\ abc$$
$$m_{floor5} = \rho_{floor5} \frac{1}{2} V_{i-node} = \rho_{floor5} \frac{2}{3} pi\ abc$$
$$m_{main} = \rho_{fill} V_{main} = \rho_{fill}\ t\ pi\ ab$$

Where R is the distance to the center of the dumbbell node. For the prolate ellipsoid nodes, the longer axis c is parallel to the axis of rotation. The other axes, a and b, are equal in length and perpendicular to the axis of rotation. The double dumbbell has four nodes and two sets of spokes.

## 5.2  Ellipsoid Analysis

Several of our station designs use ellipsoids and elliptical cross-sections. Spherical designs offer more structural strength. Spherical designs inefficient space usage. Angled polygons offer simpler and more efficient designs but have higher stress at the angled joints. Elliptical shapes appear to be a good compromise between the spherical and polygon shapes. The elliptical shapes in our stations and components produce cumbersome mass and inertia equations.

The ellipsoid shell can be modeled with two concentric ellipsoids. We design a uniform thickness shell. Figure 5-2 illustrates this using two concentric ellipsoids. As shown, the inner and outer ellipsoids produce a shell of nearly uniform thickness for the space station designs.

Figure 5-3 shows the volume of the shell between two concentric ellipsoids using four models. Two models use the uniform thickness model shown in Figure 5-2. The other two models scale all the ellipsoid dimensions. This approach has historically been used to simplify analytic analysis [Tatum 2017] [Routh 1877]. One clear result is that the scaled axes volume version is significantly greater than the other estimates. To address this variation, it is wise to follow the advice of Tatum regarding ellipsoids: *"At first this looks easy, but I do not think you can do it in terms of elementary functions. No problem, then – just integrate it numerically."* [Tatum 2017].

Two of the models in Figure 5-3 represent the results from such a numeric approach. That approach divides the ellipsoid shell into many thin disks around the rotation axis. The shell totals are found by summing the mass and moments of inertia of those disks. Figure 5-3 provides the results from the two analytic and two numeric approaches. The volumes are a function of the ellipsoid radii a, b, and c. The x-axis shows the rotation radius, a, ranging from 100 to 1 million meters. The radii b and c are equal, and c=3a. The y-axis shows a normalized volume and ranges from 0.8 to 2.0. The volume results are normalized to the disk summation model. The disk summation results matched the corresponding thick shell equation results of Table 5-2. We also found that the disk summation matched the results of the thin shell equation.

The numeric and analytic analysis results matched for a broad set of ellipsoid and shell dimensions within 99.9%. The station shell will typically be created with 20-meter straight trusses and flat panels. It may be more realistic and

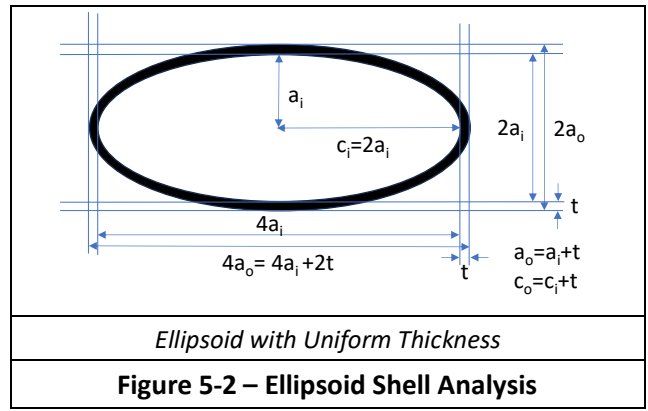

*Ellipsoid with Uniform Thickness*

**Figure 5-2 – Ellipsoid Shell Analysis**

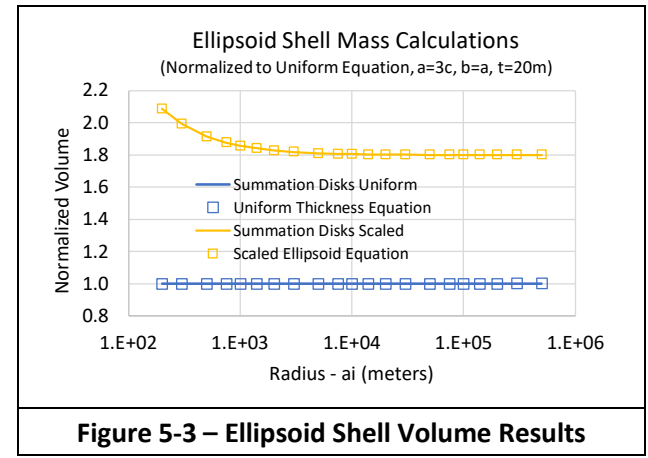

**Figure 5-3 – Ellipsoid Shell Volume Results**

accurate to use these disk elements and create a polygon shape to fit to the ellipsoid shell.

## 5.3  Dumbbell Modeling Details

In a torus geometry space station, the floor curves inside the rotating tube. Visually, an inhabitant would see the floor curving upwards like inside a valley. The floor would be flat across the valley. Cylinders and ellipsoids also have this same floor characteristics of being flat in one direction and curved (upwards) in the rotation direction. In those three geometries, the floor curvature would be constructed to match the geometry curvature. A curved floor in the rotation direction has the advantage that the centripetal gravity is consistently directed perpendicular to the curve. The curved floor would feel flat to an inhabitant walking on the rotating curved floor.

The floor inside a dumbbell only extends across the node. The edges of the curved floor would be bound by the node shell. A flat floor is easier to analyze than a curved floor. It could be easier to construct in small dumbbell nodes. Unfortunately, one would experience a lean when walking on the rotating flat disk. Technically speaking, *"The nonaligned Coriolis component skews the total apparent gravity and alters the apparent slope of surfaces." [Hall 1999]*.

Newton's laws can be transformed into a rotating frame of reference [Hand and Finch 1998]. Those laws of motion take this form to model a rotating body:

$$\underbrace{\boldsymbol{F}}_{\substack{Apparent\\Forces}} = \underbrace{m\boldsymbol{a}}_{\substack{Motion\\Force}} + \underbrace{m\frac{d\boldsymbol{\omega}}{dt}\times\boldsymbol{r}}_{\substack{Euler\\Force}} + \underbrace{2m\boldsymbol{\omega}\times\boldsymbol{v}}_{\substack{Coriolis\\Force}} + \underbrace{m\boldsymbol{\omega}\times(\boldsymbol{\omega}\times\boldsymbol{r})}_{\substack{Centrifugal\\Force}}.$$

This equation is used to evaluate the floors in the dumbbell. More details on this equation, centripetal gravity, and the Coriolis effect are provided in [Jensen 2023]. Flat floors are used for most of our dumbbell analysis. The following subsections offer a brief comparison of the flat and curved floors.

### 5.3.1 Flat Floors

Figure 5-4 illustrates flat floors in a rotating dumbbell node. The second node would be mirrored across the center of rotation, but it is not shown. Because the floor is rotating, the centripetal gravity acceleration is proportional to the distance to the center of rotation. Different locations on the flat floor would have different centripetal gravity. As one moves away from the center of the node in the rotation direction, the distance to the center increases. With the increase in distance, the centripetal gravity increases. The distance also increases with deeper floors in the station. The increasing distance is computed using Pythagorian's formula. It uses the station radius and the position on the floor $\sqrt{(R+d)^2 + f^2}$ where f is the distance from the center in the Y-axis direction and d is the depth below the top floor.

The three charts in Figure 5-5 illustrate the perceived effects from this changing centripetal gravity. Two of the charts show the perceived gravity and slope at the edges of the rotating flat floor. The third shows the perceived gravity as one traverses the floor. These include data from walking on the floor in the spinward and antispinward directions.

The x-axes in two of the charts show the station radius. This is the distance to the center floor in the sphere or ellipsoid. The node in the rotation direction has a width of R/6.33 (same as the sphere radius). The radius ranges from 10 to 100,000 meters on a logarithmic scale. The y-axis in the top chart shows the perceived gravity ranging from 0 to 14 meters per second squared. The y-axis in the middle chart shows the gravity range from 9 to 12 meters per second squared. The y-axis in the bottom chart shows the perceived slope of the floor ranging from 0 to 12 degrees. All the charts show a color-coded set of data for stationary or walking at 1 meter per second in the spinward or antispinward direction.

Figure 5-5a shows there will be a noticeable effect on the gravity in small stations. At a radius of 1000 meters, this effect becomes negligible, where the gravity ranges from 9.73 to 10.12 meters per second squared (0.99g to 1.03g) in the antispinward and spinward direction.

Figure 5-5b shows the perceived gravity increases when walking in the rotation direction from the center to the edge of the floor. The data in this chart uses R equal to 200 meters,

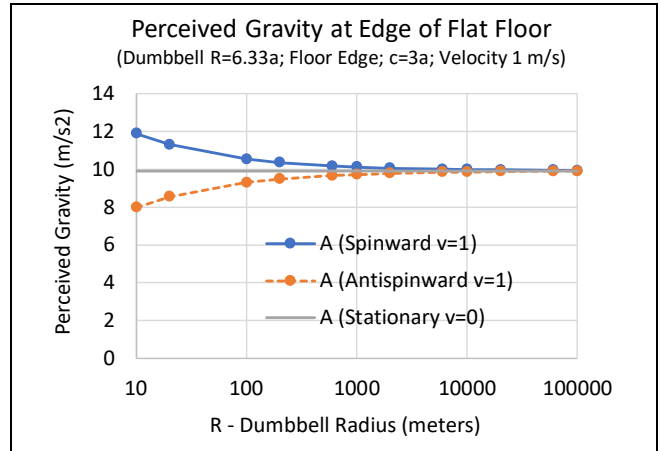

*a) Perceived Gravity at Edge of Dumbbell Flat Floor*

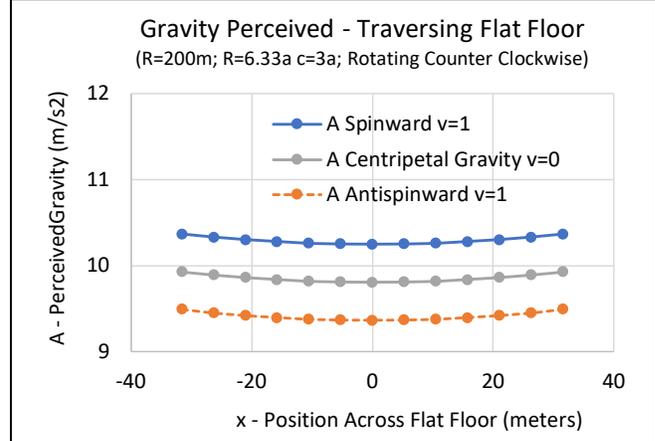

*b) Perceived Gravity Across Flat Floor in Dumbbell*

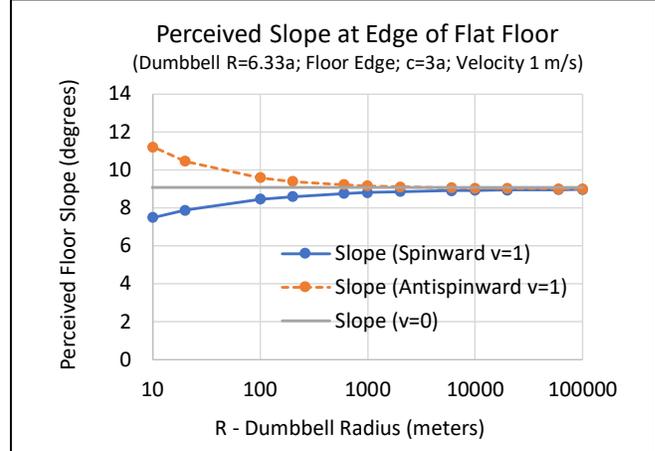

*c) Perceived Slope at Edge of Flat Floor in Dumbbell*

**Figure 5-5 – Perceived Acceleration and Slope with Flat Floor Rotating About Center of Station**

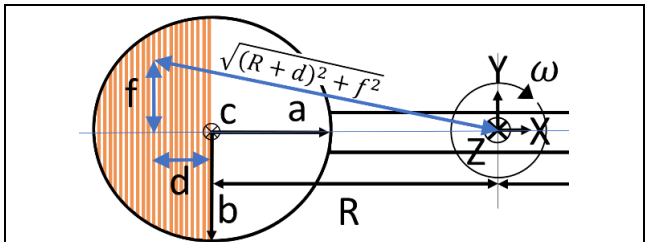

**Figure 5-4 – Flat Dumbbell Floor**

and the floor position ranges from 0 to 31.6 meters. In small stations, there will be a more noticeable increase (or decrease) in gravity as one reverses direction on the flat floor. In this 200-meter case, the gravity will change from 0.97g to 1.06g.

One would also perceive a sloped surface standing at the outer edge of the rotating node's flat floor. Figure 5-5c shows a slope chart. The slope changes are opposite to those covered in the gravity charts. Walking in the spinward direction decreases the perceived slope. The increased effective radius reduces the perceived slope [Hall 1991]. Walking in the antispinward direction causes the perceived slope to increase.

The dumbbell results are above and below a stationary slope of about 17% or 9.6 degrees. This slope would exceed specifications for terrestrial ramps. The ADA and ANSI impose a maximum 2% slope for in-door maneuvering clearances, accessible parking stalls, and other spaces. They recommend that a path can have up to 5% slope before it requires handrails. The 2010 ADA cross-slope limit was 2.1% [Tessmer 2014]. A thorough analysis of footway crossfall gradients for wheelchair accessibility found a guideline of 2.5% to be reasonable [Holloway 2011]. Flat floors need to be designed carefully to avoid slope issues.

The charts show that rotating floors on stations larger than 1000 meters in radius exhibit little difference between walking in the spinward and antispinward directions. Designing for acceptable gravity will not be as constraining as designing for an acceptable slope for these rotating flat floors. Designs will need proportionately much smaller floor lengths to reduce the slope at the edge of the floor. A curved floor or multiple narrow flat floors are two approaches to address this slope issue.

### 5.3.2 Curved Floors

An alternative to the flat dumbbell floor is to construct a curved floor. Figure 5-6 shows an example of the single floor in the rotating dumbbell. The figure shows a single curved floor as a dark grey line in one of the nodes. The floor would be a curve that is equidistance to the center. This is perhaps the most complex floor surface for our surface area analysis. The floor is cylindrical about the rotation axis and is cut from the spherical or ellipsoidal node. A three-dimensional view of the floor is shown to assist the visualization of this floor. This illustration shows the surface area shape of the circle cut from a cylinder. The vertical height is exaggerated with the chart scales. This particular floor is located at radius, R, and depth, d, in the node.

The floor curved radius is dependent primarily on the radius of the tether, R. The floor length varies with distance from dumbbell center of rotation. As the rotation radius becomes large, the floor approaches a flat surface. Viewed from this side view in Figure 5-6, with an increasing radius, the floor would appear like a circle arc, a curved line, and a nearly flat line.

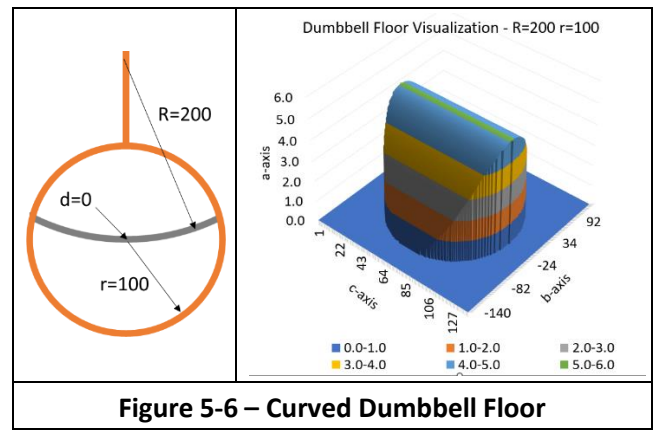

**Figure 5-6 – Curved Dumbbell Floor**

### 5.3.3 Multiple Curved Floors

Figure 5-7 shows an example of the half-volume flooring in the dumbbell module. The curved floors are dark grey lines. The top floor is located at the center of the spherical module of the dumbbell. The spherical modules rotate about the center of the dumbbell station. The floors are circular (or elliptical) within the dumbbell node when viewed from above along the tether a-axis. The floors are circular around the dumbbell rotation c-axis. The floors are flat when viewed from the third b-axis. The cylindrical floors would eliminate most of the flat floor slope and gravity problems. Instead of a curved floor, trusses and panels could follow the circle as a polygon and provide similar results.

The top floor is positioned near the center of the dumbbell node. This provides openness for the station population. The top floor would typically be used for open space, public use, and recreation. The floor lines illustrated in Figure 5-7 are mirrored in the other dumbbell module. The floor area changes with the depth in the cylinder. Floors reside from the center of the spherical module to the outer perimeter. There will be approximately $f_n = r / f_d$ floors in each node, where r is the radius of the dumbbell node and $f_d$ is the distance between floors. The areas of flat disks and curved disks are close enough for a first analysis and a rough engineering estimate.

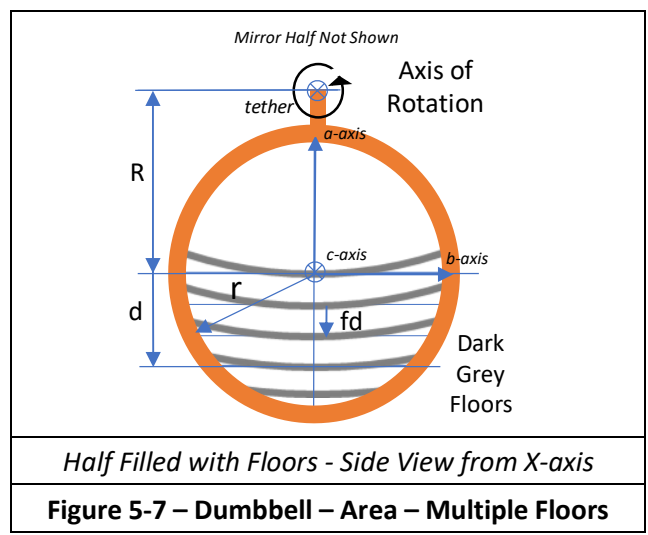

*Half Filled with Floors - Side View from X-axis*

**Figure 5-7 – Dumbbell – Area – Multiple Floors**

## 5.4 Stability and MOI Equations

A goal of this paper is to evaluate the rotational stability of all our space station geometries. To provide rotational stability, the desired axis of rotation should have an angular moment of inertia (MOI) at least 1.2 times greater than any other axis [Brown 2002]. Each of the station geometries will produce different MOI equations. Table 5-2 includes the MOI equations for various geometry shapes expected in the space station geometries. The MOI of the station is the sum of the MOIs for each of the component geometries in the station. The table includes equations for solid, thin-shell, and thick-shell geometries. The axes in Figure 2-1 define the x-, y-, and z-axes in Table 2-3. These equations come from multiple sources such as [Myers 1962] [Young and Budynas 2001] [Diaz, Herrera, and Martinez 2005] and [Globus et al. 2007]. We could not find sources for some elliptical geometry equations, so we derived our own using the difference between solid concentric ellipsoids. Table 5-2 identifies self-derived equations with asterisks.

The parallel axis theorem is also used to compute the MOIs with solids of revolution. This is calculated using a body of mass m with an MOI $I_m$ computed from rotating about an axis z passing through the body's center of mass. When the body rotates about a parallel axis z′, which is displaced by a distance d, the new MOI adds the body mass times the distance squared to the original MOI. As an equation, this is

| Table 5-2 – Moment of Inertia (MOI) Equations for Various Geometries ||||
|---|---|---|---|
| **Geometry** | **MOI for Solid Geometry** | **MOI for Thin Shell Geometry** | **MOI for Uniform Thick Shell Geometry** |
| **Cylinder**<br>Height: h/z<br>Radius: r/xy | $I_z = \frac{1}{2} m r^2$<br>$I_x = I_y = \frac{1}{12} m(3r^2 + h^2)$ | $I_z = m r^2$<br>$I_x = I_y = \frac{1}{12} m(6r^2 + h^2)$ | $I_z = \frac{1}{2} m(r_o^2 + r_i^2)$<br>$I_x = I_y = \frac{1}{12} m(3(r_o^2 + r_i^2) + h^2)$ |
| **Ellipsoid** *<br>Axes: a/x, b/y, c/z | $I_x = \frac{1}{5} m(b^2 + c^2)$<br>$I_y = \frac{1}{5} m(a^2 + c^2)$<br>$I_z = \frac{1}{5} m(a^2 + b^2)$ | $I_x = \frac{1}{3} m(b^2 + c^2)$<br>$I_y = \frac{1}{3} m(a^2 + c^2)$<br>$I_z = \frac{1}{3} m(a^2 + b^2)$ | $I_x = \frac{1}{5} m \frac{(a_o b_o c_o)(b_o^2 + c_o^2) - (a_i b_i c_i)(b_i^2 + c_i^2)}{a_o b_o c_o - a_i b_i c_i}$<br>$I_y = \frac{1}{5} m \frac{(a_o b_o c_o)(a_o^2 + c_o^2) - (a_i b_i c_i)(a_i^2 + c_i^2)}{a_o b_o c_o - a_i b_i c_i}$<br>$I_z = \frac{1}{5} m \frac{(a_o b_o c_o)(a_o^2 + b_o^2) - (a_i b_i c_i)(a_i^2 + b_i^2)}{a_o b_o c_o - a_i b_i c_i}$ |
| **Oblate Ellipsoid** *<br>Polar Axis: c/z<br>Other Axes: a/x=b/y | $I_z = \frac{2}{5} m a^2$<br>$I_x = I_y = \frac{1}{5} m(a^2 + c^2)$ | $I_z = \frac{2}{3} m a^2$<br>$I_x = I_y = \frac{1}{3} m (a^2 + c^2)$ | $I_z = \frac{2}{5} m \frac{c_o a_o^4 - c_i a_i^4}{c_o a_o^2 - c_i a_i^2}$<br>$I_x = I_y = \frac{1}{5} \frac{(c_o a_o^2)(c_o^2 + a_o^2) - (c_i a_i^2)(c_i^2 + a_i^2)}{c_o a_o^2 - c_i a_i^2}$ |
| **Sphere**<br>Radius: r/xyz | $I = \frac{2}{5} m r^2$ | $I = \frac{2}{3} m r^2$ | $I = \frac{2}{5} m((r_o^5 - r_i^5)/(r_o^3 - r_i^3))$ |
| **Torus**<br>Major Axis: R/xy<br>Minor Axis: r/xy | $I_z = \frac{1}{4} m(4R^2 + 3r^2)$<br>$I_x = I_y = \frac{1}{8} m(4R^2 + 5r^2)$ | $I_z = \frac{1}{2} m (2R^2 + 3r^2)$<br>$I_x = I_y = \frac{1}{4} m(2R^2 + 5r^2)$ | $I_z = m\left(R^2 + \frac{3}{4}(r_o^2 + r_i^2)\right)$<br>$I_x = I_y = \frac{1}{8} m\left(4R^2 + 5(r_o^2 + r_i^2)\right)$ |
| **Elliptical Torus** *<br>Major Axis: R/xy<br>Minor Axes: a/xy, c/z | $I_z = \frac{1}{4} m (4R^2 + 3a^2)$<br>$I_x = I_y = \frac{1}{8} m(4R^2 + 3a^2 + 2c^2)$ | $I_z = \frac{1}{2} m (2R^2 + 3a^2)$<br>$I_x = I_y = \frac{1}{4} m(2R^2 + 3a^2 + 2c^2)$ | $I_z = m\left(R^2 + \frac{3}{4}\left(\frac{a_o^3 c_o - a_i^3 c_i}{a_o c_o - a_i c_i}\right)\right)$<br>$I_x = I_y = \frac{1}{8} m\left(4R^2 + 3\frac{a_o^3 c_o - a_i^3 c_i}{a_o c_o - a_i c_i} + 2\frac{c_o^3 a_o - c_i^3 a_i}{a_o c_o - a_i c_i}\right)$ |
| **Rod**<br>Length: L/z<br>Radius: r/xy | $I_z = \frac{1}{2} m r^2$<br>$I_x = I_y = \frac{1}{12} mL^2$ | $I_z = m r^2$<br>$I_x = I_y = \frac{1}{12} mL^2$ | $I_z = \frac{1}{2} m(r_o^2 + r_i^2)$<br>$I_x = I_y = \frac{1}{12} mL^2$ |
| **Disk**<br>Height: h/z<br>Radius: r/xy | $I_z = \frac{1}{2} m r^2$<br>$I_x = I_y = \frac{1}{12} m(3r^2 + h^2)$ | $I_z = m r^2$<br>$I_x = I_y = \frac{1}{2} m r^2$ | $I_z = \frac{1}{2} m(r_o^2 + r_i^2)$<br>$I_x = I_y = \frac{1}{12} m(3(r_o^2 + r_i^2) + h^2)$ |
| **Elliptical Disk** *<br>Height: h/z<br>Radius: a/x, b/y | $I_z = \frac{1}{4} m(a^2 + b^2)$<br>$I_y = \frac{1}{12} m(3a^2 + h^2)$<br>$I_x = \frac{1}{12} m(3b^2 + h^2)$ | $I_z = \frac{1}{2} m(a^2 + b^2)$<br>$I_y = \frac{1}{12} m(6a^2 + h^2)$<br>$I_x = \frac{1}{12} m(6b^2 + h^2)$ | $I_z = \frac{1}{4} m(a_o^2 + a_i^2 + b_o^2 + b_i^2)$<br>$I_y = \frac{1}{12} m(3(a_o^2 + a_i^2) + h^2)$<br>$I_x = \frac{1}{12} m(3(b_o^2 + b_i^2) + h^2)$ |
| **Cuboid**<br>Axes: a/x, b/y, c/z | $I_z = \frac{1}{12} m(a^2 + b^2)$<br>$I_y = \frac{1}{12} m(a^2 + c^2)$<br>$I_x = \frac{1}{12} m(b^2 + c^2)$ | $I_z = \frac{1}{12} m(a^2 + b^2)$<br>$I_y = \frac{1}{12} m(a^2 + c^2)$<br>$I_x = \frac{1}{12} m(b^2 + c^2)$ | $I_z = \frac{1}{12} m(a^2 + b^2)(1 - S^5)$<br>$I_y = \frac{1}{12} m(a^2 + c^2)(1 - S^5)$<br>$I_x = \frac{1}{12} m(b^2 + c^2)(1 - S^5)$ |

*Geometries rotate about the z-axis; Inner dimensions are labeled with subscript "i". Outer dimensions are labeled with subscript "o". Outer dimension – Inner dimension = "t" (thickness); Scale sides using InnerDimension /OuterDimension = "S". Height of object = "h"*
*Asterisks identify geometries with self-derived inertia equations.*

$I_{mnew} = I_m + m\, d^2$. Details and derivations of this theorem are found in [Baker and Haynes 2024].

It is usually possible to create a closed-form equation for the station stability for a single-floor geometry with the thin shell (or thick shell) equations. For multiple-floor designs with varying densities, one can often replace the mass m in the equations and use a density $\rho$ (using the relationship mass equals density time volume). This often simplifies the effort to analytically combine the multiple inertias of the station components.

The multiple MOI equations can be used to compute the stability of the geometries with various radius and floor designs. Unlike the single-floor designs, creating closed-form equations for stability became impossible (at least tedious). In those cases, the geometries of some components were divided into many small pieces with well-defined masses and MOIs. The masses and moments of inertia were summed for all the pieces. Most often thin disks rotating about the z-axis were used for this effort. This would be a numerical Riemann sum of small parts of the component. The Newton Raphson numeric algorithm method (goal seek in Excel) was used to solve the stability equations with the many components and produce a rotationally stable station. The disk summation results matched the self-derived equation results in Table 5-2 (better than 99.9%).

# 6 Single Floor Rotational Stability

The station rotates to produce an Earthlike gravity for the residents. This is important for human long-term health. Perturbations will cause rotating systems in space to eventually rotate about the axis with the greatest angular moment of inertia [Globus et al. 2007]. There is the risk that the changing axis of rotation would cause the station to catastrophically fail and tumble end-over-end. Globus and his co-authors designed a cylinder station that would not have this risk [Globus et al. 2007]. They designed the axis of rotation to be 1.2 times greater than the other rotation axes [Brown 2002]. They analytically derived the geometry ratio of cylinder length-to-radius to produce this stable station.

This section extends the Global cylinder analysis approach [Globus et al. 2007]. It uses the refined set of station geometries; see Figure 2-1. This section analyzes the stations with thin and thick shells. Like the Globus analysis, this section's analysis only considers the outer shell of the geometries.

## 6.1 Assumptions

The station is assumed to have a habitable gravity range. On the cylinder geometry, the single floor is on the outer shell and has a gravity of 1.0g. The other geometries are curved on all three axes and the single floor ranges from 0.95g to 1.05g. This limits the usable distance up the curved outer shell. As an example, to support that gravity range the habitable region height of the ellipsoid has a height of the rotating radius over 10.5; see *§3.1 Gravity Limits*.

The goal of this analysis is to determine the geometry dimensions to provide passive rotational stability. Again, the desired axis of rotation should be 1.2 times greater than the other rotation axes [Brown 2002]. To compute the stable designs, the Moments of Inertia (MOI) equations from Table 5-2 are used for the analytic analysis. For consistency, the axis of rotation is labeled the Z-axis. The moment of inertia about that preferred axis of rotation is labeled Iz. For the cylinder, sphere, and ellipsoid station geometries, the MOIs of the other two axes are equal. The station is designed so the other two axes have smaller MOIs than the z-axis. In the upcoming dumbbell subsection, the MOI of the rotation axis Iz is the same as Iy. Even though Ix and Iy are not equal, this matching Iz and Iy MOIs introduces a passive stability problem in the dumbbell subsection.

## 6.2 Stability of Cylinder

Al Globus and his team evaluated the stability of the cylinder space station geometry [Globus et al. 2007]. They identified stability issues with long, narrow, rotating cylinders. This section presents the Globus approach and then extends that approach to a thick shell cylinder. Given his results, a stable cylinder is short and squatty (hatbox) and has a rotation radius (R) larger than the height (h); see Figure 6-1. A line drawing of the cylinder is in Figure 2-1. The moment of inertia (MOI) for the cylinder uses the mass (M), radius (R), and length (L).

The inertias for a thin-shelled cylinder using the axes and dimensions in Figure 6-1 are:

$$I_z = M_{shell}\, R^2$$
$$I_x = I_y = M_{shell}\left(\tfrac{1}{2}R^2 + \tfrac{1}{12}L^2\right)$$

The endcaps are offset by L/2, as shown in Figure 6-1. Their inertias are increased using the parallel axis theorem:

$$I_z = \tfrac{1}{2} M_{cap}\, R^2$$
$$I_x = I_y = \tfrac{1}{4} M_{cap}\, R^2 + \tfrac{1}{4} M_{cap}\, L^2$$

Globus uses thin shell inertia equations and assumes the disks and shell have the same thickness and density. For the thin shell cylinder, he combines the endcap and shell to find these MOI equations:

$$I_z = M_{shell}\, R^2 \left(1 + \tfrac{R}{2L}\right)$$
$$I_x = I_y = M_{shell}\left[\tfrac{R^2}{2} + \tfrac{L^2}{4} + \tfrac{R^3}{4L} + \tfrac{RL}{4}\right]$$

Using the stability rule of Iz >= 1.2 Ix, the single floor thin shell cylinder stability is met when:

$$R^2 + \tfrac{R^3}{2L} \geq 1.2 \left(\tfrac{R^2}{2} + \tfrac{L^2}{12} + \tfrac{R^3}{4L} + \tfrac{RL}{4}\right)$$
$$4R^2 + \tfrac{2R^3}{L} - L^2 - 3RL \geq 0$$

The Newton-Raphson method determines that the system would be stable when L is approximately 1.3R. This corroborates the result in [Globus et al. 2007].

The thin shell analysis is extended to evaluate a thick shell cylinder. This uses the axes and dimensions for the cylinder in Figure 6-1. The radius R is subscripted as $R_o$ for the outer

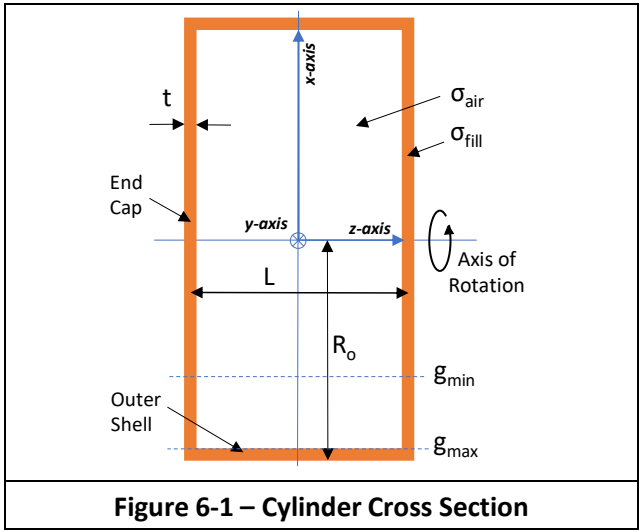

**Figure 6-1 – Cylinder Cross Section**

radius and $R_i$ for the inner radius. In this thick shell case, the inertias for the cylinder are:

$$I_z = \tfrac{1}{2} M_{shell}\,(R_o^2 + R_i^2)$$

$$I_x = I_y = M_{shell}\left(\tfrac{1}{4}(R_o^2 + R_i^2) + \tfrac{1}{12}L^2\right)$$

The inertias for the endcaps, as shown in Figure 6-1, are:

$$I_z = \tfrac{1}{2} M_{cap}\,R_o^2$$

$$I_x = I_y = \tfrac{1}{4} M_{cap}\,R_o^2 + \tfrac{1}{4} M_{cap}\,L^2$$

Using the average distance $(L_o+L_i)/2$ instead of L in these thick shell inertia equations would be more accurate. When $R_o=R_i=R$, the equations match the thin shell moment of inertia equations for the cylinder and disks. Combining the cylinder shell and the two disk endcaps:

$$I_z = \tfrac{1}{2} M_{shell}(R_o^2 + R_i^2) + 2\left[\tfrac{1}{2} M_{cap} R_o^2\right]$$

$$I_x = I_y = M_{shell}\left(\tfrac{1}{4}(R_o^2 + R_i^2) + \tfrac{1}{12}L^2\right) + 2\left[\tfrac{1}{4} M_{cap} R_o^2 + M_{cap}\left(\tfrac{L}{2}\right)^2\right]$$

Assuming the endcaps and shell have the same thickness, volumes and equal densities are used to find the relationship between their masses:

$$M_{cap} = M_{shell}\,\frac{R_o^2}{L\,(R_o + R_i)}$$

Combining these equations, the moments of inertias for the thick shell cylinder station become:

$$I_z = \tfrac{1}{2} M_{shell}\left(R_o^2 + R_i^2 + 2\,\frac{R_o^4}{L\,(R_o + R_i)}\right)$$

$$I_x = I_y = \tfrac{1}{12} M_{shell}\left[(3(R_o^2 + R_i^2) + L^2) + 6\,\frac{R_o^2}{L\,(R_o + R_i)}(R_o^2 + L^2)\right]$$

Evaluating as $R_o$ and $R_i$ approach R, the thick shell equations match the original thin shell MOI equations. The thick shell stability uses $I_z \geq 1.2\,I_x$:

$$\tfrac{1}{2}(R_o^2 + R_i^2) + \frac{R_o^4}{L\,(R_o + R_i)} \geq$$

$$1.2\left[\tfrac{1}{12}(3(R_o^2 + R_i^2) + L^2) + \tfrac{1}{2}\frac{R_o^2}{L\,(R_o + R_i)}(R_o^2 + L^2)\right]$$

$$2\,(R_o^2 + R_i^2) + 4\,\frac{R_o^4}{L\,(R_o + R_i)} - L^2 - 6\,\frac{L\,R_o^2}{(R_o^2 + R_i^2)} \geq 0$$

With a thin shell and $R_o$ and $R_i$ approaching R, the thick shell stability equation reduces to:

$$4R^2 + \frac{2R^3}{L} - L^2 - 3LR \geq 0$$

This matches the previous thin shell stability equation. Globus found for cylinders with thin shells and endcaps that $I_z \geq 1.2\,I_x$ when $L \leq 1.3R$ [Globus et al. 2007]. For thick shells and endcaps, $I_z \geq 1.2\,I_x$ consistently when $L \leq 1.29R$ over a wide range of radii and wall thicknesses. The length-to-radius ratio increased slightly with very thick shells. The graphs in Figure 6-2 show this geometry increase. The graphs show the length-to-radius ratio (L/R) along the y-axis. All cylinders are designed to be rotationally stable with $I_z/I_x \geq 1.2$. The charts show differences between the thick and thin shells. The y-axis only ranges from 1.2 to 1.45. This geometry ratio is used to control the stability of the cylinder. Both charts in Figure 6-2 show an L/R of about 1.29 for most of their data.

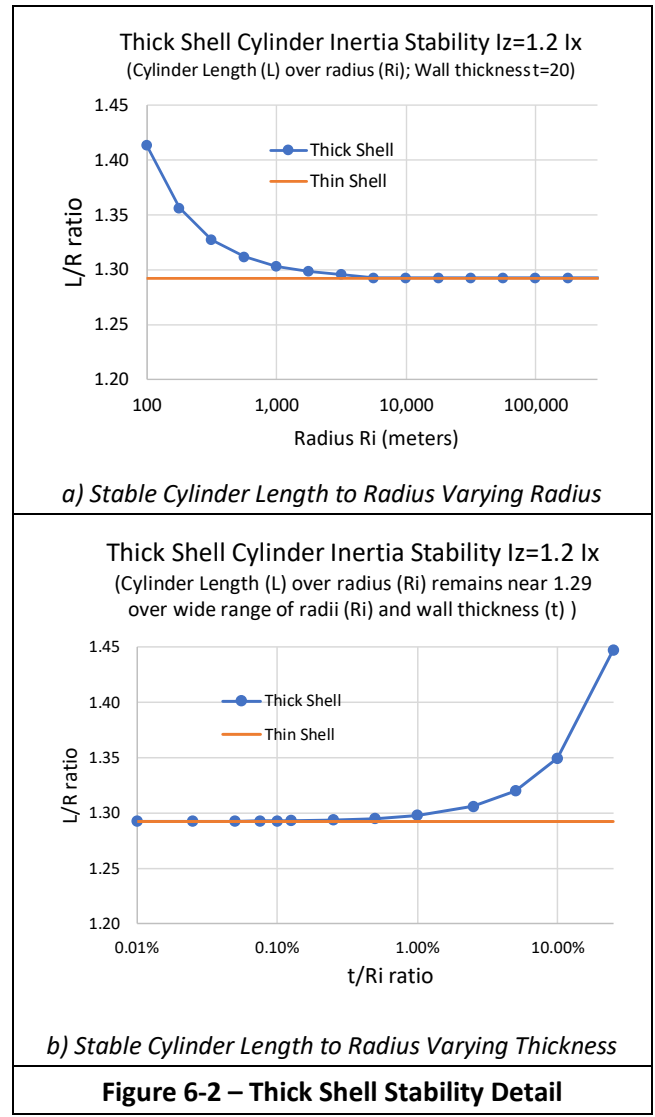

*a) Stable Cylinder Length to Radius Varying Radius*

*b) Stable Cylinder Length to Radius Varying Thickness*

**Figure 6-2 – Thick Shell Stability Detail**

The x-axis in Figure 6-2a shows the cylinder radius ranging from 100 to 300,000 meters. The shell thickness is a constant 20 meters for the range of cylinder radii. The shell becomes more proportionately thick for the small radii. The geometry ratio L/R increases from 1.29 to 1.42 as the radii become small in Figure 6-2a.

The x-axis in Figure 6-2b shows a normalized thickness scaled with the radius. Shells with thicknesses up to 1% of the radius require a length-to-radius ratio (L/R) of 1.29, like the thin shell geometry ratio. Shells with thicknesses greater than 1% of the radius support longer-length cylinder sides and remain rotationally stable. A shell with a thickness of 10% of the radius permits a ratio of L up to 1.35R.

## 6.3   Stability of Sphere and Ellipsoid

This section applies the cylinder stability analysis to an ellipsoid-shaped rotating space station. As noted in the Kalpana paper references, a spherical space station would not be rotationally stable [Globus et al. 2007]. The moments of inertia for an ellipsoid instead of the sphere are used to provide passive control. This single-floor section considers just the ellipsoid shell. The two rotation axes are kept the same length to provide double symmetry. This supports stable spinning and the production of centripetal gravity.

As an initial investigation, assume the ellipsoid wall thickness is much less than the radius. This can be modeled as a thin shell. An oblate ellipsoid has two major axes that are the rotation radius length (a=b) and a third minor axis with a length (c) that is shorter than the other two; see Figure 6-3. A line drawing of the ellipsoid is in Figure 2-1. The moments of inertia for the oblate ellipsoid use the mass (M) and the radii (a, b, and c). The ellipsoid rotates about the minor axis. The MOIs of the ellipsoid around its axes are:

$$I_z = \frac{2}{3} Mshell \, a^2$$

$$I_x = I_y = \frac{1}{3} Mshell \, (a^2 + c^2)$$

To be stable, Iz ≥1.2 Ix, and this results in:

$$\frac{2}{3} Mshell \, a^2 \geq 1.2 \left( \frac{1}{3} Mshell \, (a^2 + c^2) \right)$$

$$a^2 \geq 1.5 \, c^2 \text{ or } a \geq 1.225 \, c \text{ or } c \leq 0.8165 \, a$$

The length of the minor axis c must be less than 0.8165 times the major axis (a or b) length. This shows that an ellipsoid can be stable. This also mathematically validates that a thin shell sphere would not be stable.

Table 5-2 contains the MOI equations for triaxial elliptical thick shells. The shell thickness is assumed to be a uniform thickness. Where essential and for validation, these inertia values were computed numerically with a Riemann sum of small parts of the uniform thickness shell. These MOI equations produce analytic results that are nearly identical (better than 99%) to the numeric summation results.

Figure 6-3 contains the dimensions and axes for our thick shell analysis. The polar axis c is shorter than the other radial axes a and b in that diagram. The MOIs of the oblate ellipsoid

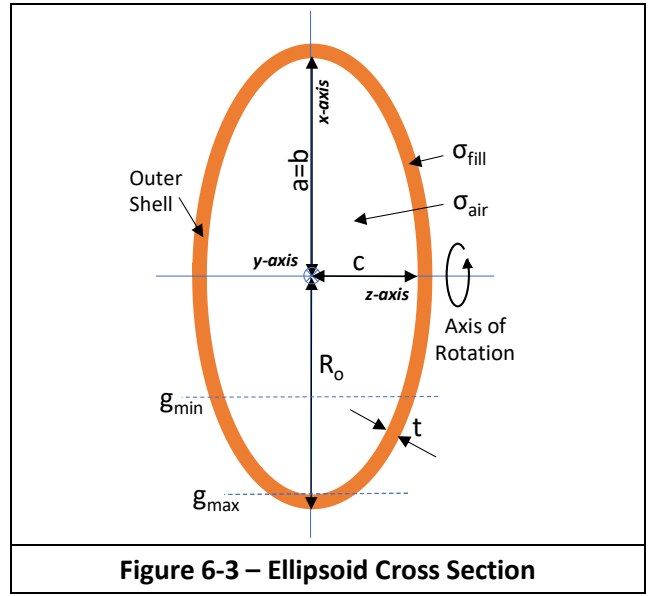

**Figure 6-3 – Ellipsoid Cross Section**

around the rotation axis (z-axis) are used. The radii of the other two axes and their inertias are equal (a=b). This results in:

$$I_z = \frac{2}{5} m_{shell} \frac{c_o a_o^4 - c_i a_i^4}{c_o a_o^2 - c_i a_i^2}$$

$$I_x = I_y = \frac{1}{5} m_{shell} \frac{(c_o a_o^2)(c_o^2 + a_o^2) - (c_i a_i^2)(c_i^2 + a_i^2)}{c_o a_o^2 - c_i a_i^2}$$

Using these inertia equations for the ellipsoid to be rotationally stable, Iz ≥1.2 Ix, which results in:

$$\frac{2}{5} \frac{c_o a_o^4 - c_i a_i^4}{c_o a_o^2 - c_i a_i^2} \geq 1.2 \left[ \frac{1}{5} \frac{(c_o a_o^2)(c_o^2 + a_o^2) - (c_i a_i^2)(c_i^2 + a_i^2)}{c_o a_o^2 - c_i a_i^2} \right]$$

$$0.4 \left[ c_o a_o^4 - c_i a_i^4 \right] \geq 0.24 \left[ (c_o a_o^2)(c_o^2 + a_o^2) - (c_i a_i^2)(c_i^2 + a_i^2) \right]$$

$$0.4 \left[ c_o a_o^4 - c_i a_i^4 \right] \geq 0.24 \left[ (c_o^3 a_o^2 + c_o a_o^4) - (c_i^3 a_i^2 + c_i a_i^4) \right]$$

$$\left[ c_o a_o^4 - c_i a_i^4 \right] \geq 0.6 \left[ (c_o^3 a_o^2 - c_i^3 a_i^2) \right] + 0.6 \left[ (c_o a_o^4 - c_i a_i^4) \right]$$

$$0.4 \left[ c_o a_o^4 - c_i a_i^4 \right] \geq 0.6 \left[ (c_o^3 a_o^2 - c_i^3 a_i^2) \right]$$

$$c_o a_o^4 - c_i a_i^4 \geq 1.5 \left[ a_o^2 c_o^3 - a_i^2 c_i^3 \right]$$

Our evaluation used a broad range of thicknesses and the Newton Raphson numeric algorithm method (Excel goal seek) and found that:

$$a_o \geq 1.29 \, c_o \text{ or } c_o \leq 0.775 a_o$$

And as a reminder, for a thin shell ellipsoid:

$$a \geq 1.225 \, c \text{ or } c \leq 0.8165 \, a$$

The length of the polar axis c must be less than 0.8165 times the radial axis (a or b) length. This shows that the stability relationships of the thin shell and thick shell ellipsoids are similar. This also implies that a thick shell sphere would not be rotationally stable.

Analytically the solid ellipsoid formula matches the thick shell results using inner dimensions of zero. Using:

$$I_{ashell} = \frac{1}{5} m_{shell} \frac{(c_o a_o^2)(c_o^2 + a_o^2) - (c_i a_i^2)(c_i^2 + a_i^2)}{c_o a_o^2 - c_i a_i^2}$$

With $a_i$, $b_i$, and $c_i$ equal to zero:

$$I_{ashell} = \frac{1}{5} m_{shell}(a_o^2 + c_o^2) = I_{solidshell}$$

A similar comparison to the thin shell equations is attempted by having the outer radii dimensions approach the inner radii dimensions:

$$I_{athin} = \frac{1}{5} m_{shell} \frac{(c_o a_o^2)(c_o^2 + a_o^2) - (c_i a_i^2)(c_i^2 + a_i^2)}{c_o a_o^2 - c_i a_i^2}$$

This is zero over zero and this analysis likely needs to set the inner dimensions equal to the outer dimensions less the thickness as the thickness approaches zero. We attempted this approach analytically and computationally. Those approaches did not simplify to the thin shell oblate ellipsoid inertia equation results. This is one of the cases where such complexities are avoided by using numeric summations of small disks about the center of rotation to compute the station masses and MOIs. The stability was evaluated numerically using this approach. The evaluation used the inertia equations in this subsection and used goal-seeking to obtain the stability ratio. This supported comparing the scaled and uniform shell thickness and their effect on the cross-section ratio. *Section 5.2 Ellipsoid Analysis* introduced different ellipsoid models. The chart in Figure 6-4 shows the difference between a thick shell that is scaled with the shell axes dimensions and a thick shell that is uniform over the entire shell. This ellipsoid difference was covered analytically in *§5.1 Station Component Equations.*

Figure 6-4 contains data from an ellipsoid station with a polar radius ($c_i$) equal to 2,000 meters. The x-axis of the chart shows the shell thickness of an ellipsoid. The thickness ranges from 1 meter to 1,000 meters. The thickness affects the stability and the dimension of the longer radial axis ($a_i$) is adjusted to maintain the stability ($I_z$=1.2$I_x$). The y-axis shows the ratio of the two axes (a/c). The evaluation of multiple-size stations produced nearly identical results in this chart. For reference, a black dashed line on the chart shows the thin shell stability result of 1.225. The scaled shell ratio matches the thin shell result. Until the shell thickness on the uniform thickness is more than 5% of the station radius, there is little change in its ratio. The radial length can be longer with a thicker shell. Thicker shells provide more material on the outer rim area of the ellipsoid and provide more inertial stability. The inertias of many small disks representing the uniform and the scaled thick ellipsoid shell were summed to compute the thick shell stability. The summed inertia of the disks matched the inertia equation results and helped validate the ellipsoid shell inertia equations.

## 6.4 Stability of Elliptical Torus

This study has considered tori with circular and elliptic cross-sections. The elliptic cross-section provides more habitable space than the circular cross-section. Figure 6-5 shows the elliptic cross-section. This subsection again applies the stability analysis to this station. A stable station is designed so the desired axis of rotation has an angular MOI at least 1.2 times greater than any other axis [Brown 2002].

### 6.4.1 Thin Shell

The initial investigation assumes the torus wall thickness is much less than the radius and can be modeled as a thin shell. A torus has a major axis R, and the elliptical cross-section has lengths (a) and (c); see Figure 6-5. A line drawing of the torus is in Figure 2-1. The moments of inertia for the torus use the mass (M) and the radii (R, a, and c). The elliptical torus rotates about the z-axis.

Using the moment of inertia equations from Table 5-2, the MOIs of the torus are:

$$I_z = m_{shell}(2R^2 + 3a^2)/2$$
$$I_x = I_y = m_{shell}(2R^2 + 3a^2 + 2c^2)/4$$

These MOIs are used to compute the stability. To be stable, $I_z \geq 1.2\, I_x$ and results in:

$$\frac{1}{2} m_{shell}(2R^2 + 3a^2) \geq 1.2 \left(\frac{1}{4} m_{shell}(2R^2 + 3a^2 + 2c^2)\right)$$

$$2R^2 + 3a^2 \geq 0.6\,(2R^2 + 3a^2 + 2c^2)$$

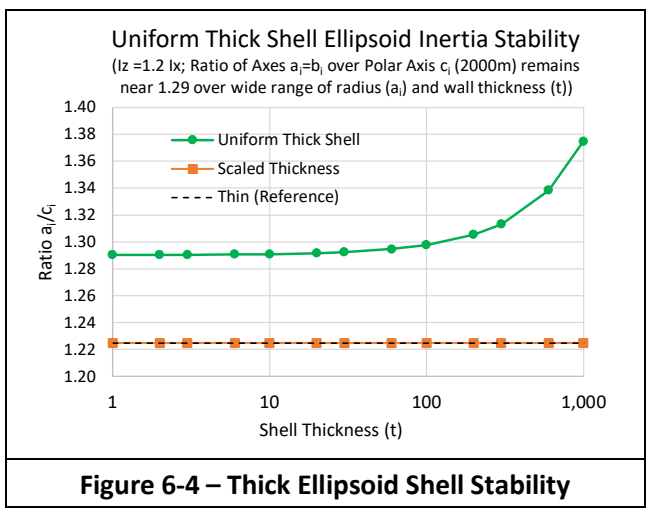

Figure 6-4 – Thick Ellipsoid Shell Stability

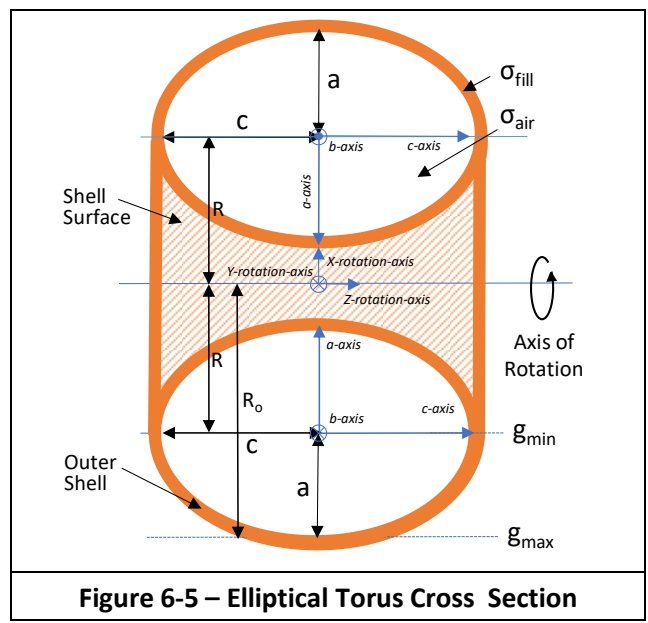

Figure 6-5 – Elliptical Torus Cross Section

$$8R^2 \geq 12c^2 - 12a^2$$
$$R^2 \geq \frac{3}{2}(c^2 - a^2)$$

The design of the elliptical torus station often uses c set to 2a or 3a. Using those assignments:

| With c=2a | With c=3a |
|---|---|
| $R^2 \geq \frac{3}{2}(4a^2 - a^2)$ | $R^2 \geq \frac{3}{2}(9a^2 - a^2)$ |
| $R^2 \geq 4.5\, a^2$ | $R^2 \geq 12 a^2$ |
| $R \geq 2.12\, a$ | $R \geq 3.46\, a$ |

Analytically solve for Iz/Ix:

$$Iz/Ix = \tfrac{1}{2} m_{shell}(2R^2 + 3a^2) / \tfrac{1}{4} m_{shell}(2R^2 + 3a^2 + 2c^2)$$
$$Iz/Ix = 2\,(2R^2 + 3a^2)/(2R^2 + 3a^2 + 2c^2)$$
$$Iz/Ix = 2\left(1 - \frac{2c^2}{2R^2 + 3a^2 + 2c^2}\right)$$

When R=6.33a or 9.5a and c=2a or c=3a the stability Iz/Ix is greater than 1.2. As the c becomes larger, the stability degrades. When c=7.8a, the stability reaches 1.2 with R=9.5a. The stability is less than 1.2 with greater values of c. With R=9.5a and c=2a, the stability Iz/Ix equals 1.916. With R=6.33a and c=2a, the stability Iz/Ix equals 1.824. In general, the thin shell is stable for our planned torus designs.

### 6.4.2 Thick Shell

Table 5-2 shows the moments of inertia for a thick shell elliptical torus along the minor rotation axis (Iz) and perpendicular major axes (Ix and Iy) are:

$$I_x = I_y = \frac{1}{8} m_{shell}\left[4R^2 + 3\frac{a_o^3 c_o - a_i^3 c_i}{a_o c_o - a_i c_i} + 2\frac{c_o^3 a_o - c_i^3 a_i}{a_o c_o - a_i c_i}\right]$$

$$I_z = m_{shell}\left(R^2 + \frac{3}{4}\left(\frac{a_o^3 c_o - a_i^3 c_i}{a_o c_o - a_i c_i}\right)\right)$$

The torus's shell is defined with concentric ellipses, with $c_i$ scaled as $3a_i$. Adding the thickness t to the inner dimensions obtains $a_o = a_i + t$ and $c_o = c_i + t$. The results of the thick shell ellipsoid equations were validated against the results from summing the masses and MOIs of many small disks representing the uniform-thickness shell.

The 1.20 metric for rotational stability rule mandates $I_z \geq 1.2\, I_x$. Explicitly evaluating the $I_y$ MOI is unnecessary because it is equal to $I_x$. Given this stability rule, the thick shell elliptical torus would be stable when:

$$m_{shell}\left(R^2 + \frac{3}{4}\left(\frac{a_o^3 c_o - a_i^3 c_i}{a_o c_o - a_i c_i}\right)\right)$$
$$\geq 1.2\left(\frac{1}{8} m_{shell}\left[4R^2 + 3\frac{a_o^3 c_o - a_i^3 c_i}{a_o c_o - a_i c_i} + 2\frac{c_o^3 a_o - c_i^3 a_i}{a_o c_o - a_i c_i}\right]\right)$$

$$R^2 - \frac{12}{20} R^2 \geq \frac{3}{20}\left[3\frac{a_o^3 c_o - a_i^3 c_i}{a_o c_o - a_i c_i} + 2\frac{c_o^3 a_o - c_i^3 a_i}{a_o c_o - a_i c_i}\right] - \frac{3}{4}\left(\frac{a_o^3 c_o - a_i^3 c_i}{a_o c_o - a_i c_i}\right)$$

$$8R^2 \geq 6\frac{c_o^3 a_o - c_i^3 a_i}{a_o c_o - a_i c_i} - 6\left(\frac{a_o^3 c_o - a_i^3 c_i}{a_o c_o - a_i c_i}\right)$$

$$R^2 \geq \frac{3}{4}\left(\frac{c_o^3 a_o - c_i^3 a_i}{a_o c_o - a_i c_i} - \frac{a_o^3 c_o - a_i^3 c_i}{a_o c_o - a_i c_i}\right)$$

This study's elliptical torus designs often set $c_i$ to be $2a_i$ or $3a_i$. A scaled, thick shell elliptical torus uses $c_i$ equal to $3a_i$ and $c_o$ to be $3a_o$. This elliptical torus would be stable when:

$$R^2 \geq \frac{3}{4}\left(\frac{27 a_o^3 a_o - 27 a_i^3 a_i}{a_o 3 a_o - a_i 3 a_i} - \frac{a_o^3 3 a_o - a_i^3 3 a_i}{a_o 3 a_o - a_i 3 a_i}\right)$$

$$R^2 \geq \frac{3}{4}\left(9\frac{a_o^4 - a_i^4}{a_o^2 - a_i^2} - \frac{a_o^4 - a_i^4}{a_o^2 - a_i^2}\right)$$

$$R^2 \geq 6(a_o^2 + a_i^2)$$

Using a thin shell where $a_i$ is approximately equal to $a_o$, the torus would be stable when:

$$R^2 \geq 12\, a_i^2 \text{ and } R \geq 3.46\, a_i$$

Which matches our thin shell result with c=3a.

Setting $c_o$ equal to $2a_o$, the scaled thick shell elliptical torus would then be stable when:

$$R^2 \geq \frac{3}{4}\left(\frac{c_o^3 a_o - c_i^3 a_i}{a_o c_o - a_i c_i} - \frac{a_o^3 c_o - a_i^3 c_i}{a_o c_o - a_i c_i}\right)$$

$$R^2 \geq \frac{3}{4}\left(\frac{8 a_o^3 a_o - 8 a_i^3 a_i}{a_o 2 a_o - a_i 2 a_i} - \frac{a_o^3 2 a_o - a_i^3 2 a_i}{a_o 2 a_o - a_i 2 a_i}\right)$$

$$R^2 \geq \frac{9}{4}\left(\frac{a_o^4 - a_i^4}{a_o^2 - a_i^2}\right)$$

$$R^2 \geq \frac{9}{4}(a_o^2 + a_i^2)$$

Evaluating a thin shell where $a_i$ is approximately equal to $a_o$, the torus would be stable when:

$$R^2 \geq 4.5\, a_i^2 \text{ and } R \geq 2.12\, a_i$$

Which matches the thin shell result for c=2a. These matches provide us with confidence in our analysis and equations.

Figure 6-6 includes more data on the elliptical torus stability and the torus geometry. The graph shows the stability (Iz/Ix) on the left axis and ranges from 0 to 2.5. The x-axis shows the ratio of the major radius over the minor radius. This ratio defines the gravity range in our stations; see *§3.1 Gravity Limits*. The chart includes the value of Iz/Ix=1.2 as a minimum stability value for reference. The graph includes stability values for the thin, scaled thick, and uniform thick shell models. It also includes the stability ratio using the

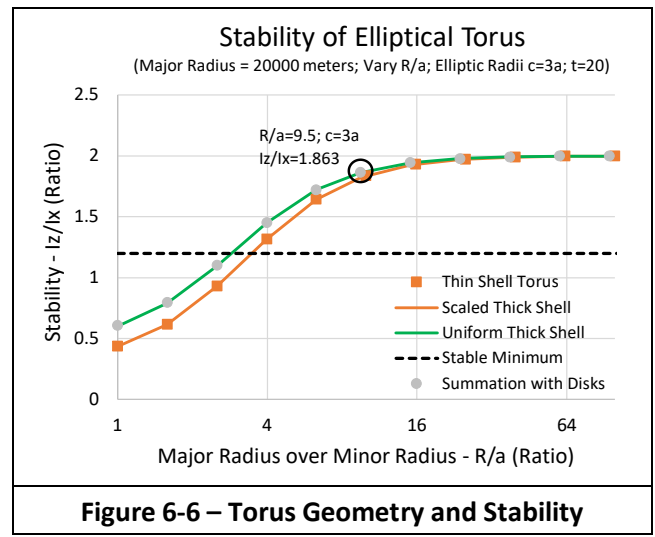

**Figure 6-6 – Torus Geometry and Stability**

summation of many small disks. All three models become stable as the major-to-minor radius ratio becomes greater than 4. The summation of the small disks matches the uniform thick shell model results.

### 6.4.3 Stability Equation

Table 5-2 provides the moments of inertia for a thick shell elliptical torus along the minor rotation axis ($I_z$) and perpendicular major axes ($I_x$ and $I_y$) are:

$$I_x = I_y = \frac{1}{8} m_{shell} \left[ 4R^2 + 3 \frac{a_o^3 c_o - a_i^3 c_i}{a_o c_o - a_i c_i} + 2 \frac{c_o^3 a_o - c_i^3 a_i}{a_o c_o - a_i c_i} \right]$$

$$I_z = m_{shell} \left( R^2 + \frac{3}{4} \left( \frac{a_o^3 c_o - a_i^3 c_i}{a_o c_o - a_i c_i} \right) \right)$$

The stability ratio for the thick shell elliptical torus would be:

$$\frac{I_z}{I_x} = \frac{m_{shell} \left( R^2 + \frac{3}{4} \left( \frac{a_o^3 c_o - a_i^3 c_i}{a_o c_o - a_i c_i} \right) \right)}{\frac{1}{8} m_{shell} \left[ 4R^2 + 3 \frac{a_o^3 c_o - a_i^3 c_i}{a_o c_o - a_i c_i} + 2 \frac{c_o^3 a_o - c_i^3 a_i}{a_o c_o - a_i c_i} \right]}$$

$$\frac{I_z}{I_x} = \frac{8R^2 + 6 \left( \frac{a_o^3 c_o - a_i^3 c_i}{a_o c_o - a_i c_i} \right)}{4R^2 + 3 \frac{a_o^3 c_o - a_i^3 c_i}{a_o c_o - a_i c_i} + 2 \frac{c_o^3 a_o - c_i^3 a_i}{a_o c_o - a_i c_i}}$$

$$\frac{I_z}{I_x} = 2 - \frac{4 \left( \frac{c_o^3 a_o - c_i^3 a_i}{a_o c_o - a_i c_i} \right)}{4R^2 + 3 \frac{a_o^3 c_o - a_i^3 c_i}{a_o c_o - a_i c_i} + 2 \frac{c_o^3 a_o - c_i^3 a_i}{a_o c_o - a_i c_i}}$$

This equation is used to evaluate the stability of an elliptical torus for several specific examples. First, the analysis sets the major radius R equal to 9.5 times the minor radius a. The perpendicular minor radius c is 2 times the minor radius a. This uses a uniform thickness of t added to the minor inner radii to set the minor outer radii. Using Wolfram Alpha to find the $I_z/I_x$ ratio produces a Taylor series of:

$$\frac{I_z}{I_x} = 1.93007 - \frac{0.0370189 t}{a} - \frac{0.00898252 \, t^2}{a^2} - \frac{0.00108674 \, t^3}{a^3} - O(t^4)$$

Next, the analysis sets the major radius R equal to 6.33 times the minor radius a. The perpendicular minor radius c is 2 times the minor radius a. Again, a uniform thickness of t is added to the minor inner radii to set the minor outer radii. The $I_z/I_x$ ratio produces a Taylor series of:

$$\frac{I_z}{I_x} = 1.85235 - \frac{0.0720419 t}{a} - \frac{0.0140877 \, t^2}{a^2} - \frac{0.00364119 \, t^3}{a^3} - O(t^4)$$

As a third example, the analysis sets the major radius R equal to 9.5 times the minor radius a. The perpendicular minor radius c is 3 times the minor radius a. Again, a uniform thickness of t is added to the minor inner radii to set the minor outer radii. The $I_z/I_x$ ratio produces a Taylor series of:

$$\frac{I_z}{I_x} = 1.86346 - \frac{0.0505465 t}{a} - \frac{0.00701979 \, t^2}{a^2} - \frac{0.00119831 \, t^3}{a^3} - O(t^4)$$

These three example values match the stability values from the summation of the disks comprising the uniform shell. Figure 6-6 includes the third Taylor series $I_z/I_x$ stability value of 1.863 at R/a equal to 9.5. Again, these matches provide confidence in the analysis and equations.

### 6.4.4 Thickness

Evaluating the stability for different thicknesses, t, provides an additional comparison. Consider the stability equation using the thickness when $c_o=3a_o$. Using the previous thick shell stability equation result:

$$R^2 > 6(a_o^2 + a_i^2)$$

Setting $a_o=a_i+t$:

$$R^2 > 6((a_i + t)^2 + a_i^2)$$
$$R^2 > 6((a_i^2 + 2a_i t + t^2) + a_i^2)$$
$$R^2 > 12a_i^2 + 12a_i t + 6t^2$$

This shows the relationship of the major radius R as a function of the minor axis a. For two example thicknesses, there is only a small difference in the radii relationship:

| For $t=a_i/10$ | For $t=a_i/100$ |
|---|---|
| $R^2 > 12a_i^2 + 12a_i \frac{a_i}{10} + 6 \frac{a_i^2}{100}$ | $R^2 > 12a_i^2 + 12a_i \frac{a_i}{100} + 6 \frac{a_i^2}{10000}$ |
| $R^2 > 12a_i^2 + 1.2 \, a_i^2 + 0.06 a_i^2$ | $R^2 > 12a_i^2 + 0.12 \, a_i^2 + 0.0006 a_i^2$ |
| $R^2 > 13.26 \, a_i^2$ | $R^2 > 12.1206 \, a_i^2$ |
| $R > 3.64 a_i$ | $R > 3.48 \, a_i$ |

Figure 6-7 shows the effect of varying the shell thickness on the stability of the elliptical torus. The graph compares scaled and uniform shell thickness and their effect on the cross-section ratio. The scaled ellipsoid has more material close to the rotation axis; see Figure 5-2. The smaller MOI is appropriate because the MOI is proportional to mass times the distance squared. For reference, the chart includes the thin shell cross-section ratio to provide the $I_z/I_x=1.2$ stability ratio. The analysis of the scaled and uniform models uses the same equations from Table 5-2. The inner and outer axes are set appropriately to provide the scaled and uniform shell thickness.

### 6.4.5 Stability Overview

Table 6-1 provides an overview of the elliptical torus stability. This table includes results for three different shell models. This compares the stability of stations using the thin shell, the scaled thick shell, and the uniform thick shell models. The table has a column for the shell model, the ratio of the elliptical cross-section axes, and the resulting stability ($I_z/I_x$). A major radius of 2000 meters is used for these data.

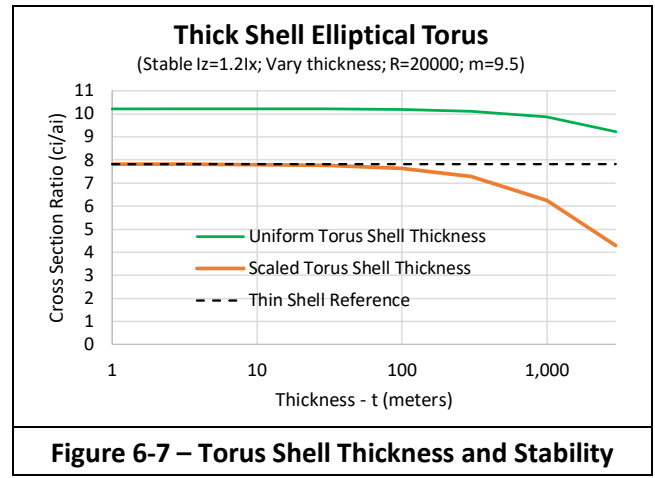

**Figure 6-7 – Torus Shell Thickness and Stability**

| Table 6-1 – Stability for Different Shell Models | | |
|---|---|---|
| Shell Model | Elliptical Axes Ratio (ci/ai) | Stability (Iz/Ix=1.2) |
| Thin | 7.82 | 1.20 |
| Scaled | 7.82 | 1.15 |
| Scaled | 7.46 | 1.20 |
| Uniform | 7.82 | 1.45 |
| Uniform | 10.15 | 1.20 |

The gravity ratio is $a_i$=R/9.5. The thin shell model is stable (Iz/Ix=1.2) when the elliptical cross-section ratio (ci/ai) is 7.82 or less. Using the same thin cross-section ratio (7.82) with the scaled thickness shell model, the stability reduces to 1.15. The scaled thickness model needs the elliptical axes ratio to be 7.46 or less to be stable. The table data also shows the uniform thickness shell model is stable when the elliptical cross-section ratio (ci/ai) is 10.15 or less. The stability improves to 1.45 using the same thin cross-section ratio (7.82) with the uniform thickness shell model. This table shows that the uniform thickness elliptical torus becomes more rotationally stable with the thick shell. This is desirable because the station's elliptical cross-section dimensions can be increased to support larger populations. The scaled thickness elliptical torus model requires smaller elliptical axes than the uniform thickness model to retain the same rotational stability.

The torus designs are rotationally stable for the desired dimensions of elliptic cross-sections. Figure 6-8 shows two regions of stability and instability with the uniform thick shell elliptical torus. The y-axis along the right side of the chart represents the ratio of the major radius R over the minor radius a. The y-axis ratio ranges from 1 to 40. This R/a ratio typically defines the gravity range on the elliptical station. The x-axis across the bottom of the chart represents the c/a ratio of the elliptical radii. The minor radius a is coincident with the longer major axis R. The minor radius c is perpendicular to the major axis R. The x-axis ranges from 1 to 70.

The data in Figure 6-8 was produced with the major radius R equal to 2000 meters. As a test, the radius was changed to values between 1000 and 20000 meters, and the chart was visually identical. Changing the large major radii appear to have minimal effect on the station stability. These tests were done with a 20-meter shell thickness. The stability was tested using $I_z \geq 1.2\, I_x$ as derived earlier in this subsection:

$$R^2 > \frac{3}{4}\left(\frac{c_o^3 a_o - c_i^3 a_i}{a_o c_o - a_i c_i} - \frac{a_o^3 c_o - a_i^3 c_i}{a_o c_o - a_i c_i}\right)$$

The analysis used a uniform shell thickness. The outer a-axis and outer c-axis dimensions are the inner dimensions plus the thickness. The chart has four white lines and intersections. The two vertical lines are at c=2a and c=3a. Both are typical design ratios used for the elliptical cross-section of the torus designs. The two horizontal lines are at R=9.5a and R=6.33a. Both are typical ratios in our designs to define a habitable gravity range over our outer torus. Those four intersections on the chart represent four of our common station designs. Those four designs are well within the stable region.

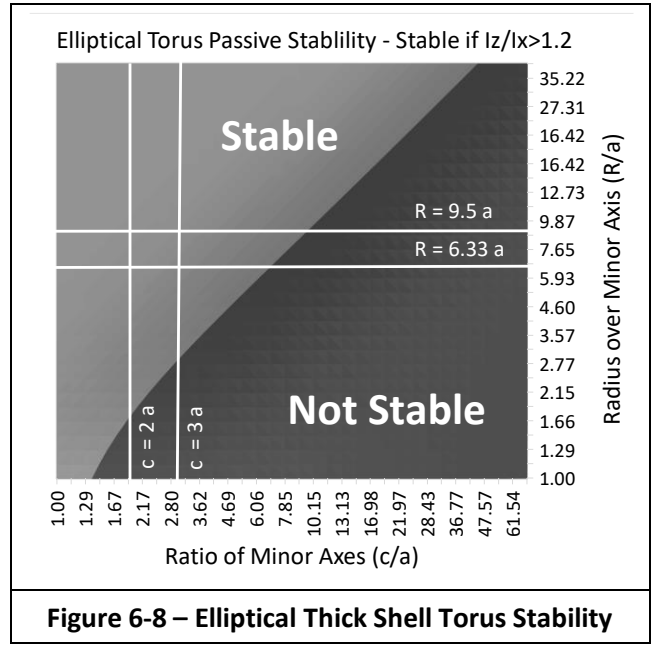

Figure 6-8 – Elliptical Thick Shell Torus Stability

## 6.5 Stability of Dumbbell

This study has analyzed dumbbell geometries with spherical and ellipsoid nodes. Figure 6-9 shows the cross-section of a dumbbell space station with spherical nodes. The stability analysis is again applied to this geometry. To be stable, the station design should have the desired axis of rotation with an angular moment of inertia at least 1.2 times greater than any other axis [Brown 2002]. The z-axis is defined as the preferred axis of rotation, and its MOI is Iz. With cylinder, sphere, and ellipsoid station geometries, the other two MOIs (Ix and Iy) are equal because of symmetry. Those geometries could be stable, with the other two axes having smaller MOIs than the z-axis. The Ix and Iy for the dumbbell are not equal and appear to introduce an unsolvable stability problem.

### 6.5.1 Are dumbbell stations rotationally stable?

There are many posts and papers on dumbbell or bolo geometry space stations. Such stations have been advocated since the 1920s. It is hard to believe they are not rotationally stable. To address this question, the stability rules presented at the beginning of this section are used.

Figure 6-10 previews the rotational stability of a dumbbell with a diagram and equations for a simple dumbbell model. This representation shows two spheres connected with a very low-mass rod. The MOI about the z and y axes are equal for this model. The MOI about the x-axis is much smaller than the moment of inertia about the z-axis.

In the diagram, the station is rotating about the z-axis. According to the Fitzpatrick rule, the station should not be stable because the principal axis moment Iz is not distinct from the moments of the other two axes (Iz is equal to Iy). According to the stability rule, the station should not be stable because Iz is not greater than 1.2 times Iy (again, Iz is equal to Iy).

An advanced dynamics class, the University of Manchester also offers analysis and guidelines for rotational stability.

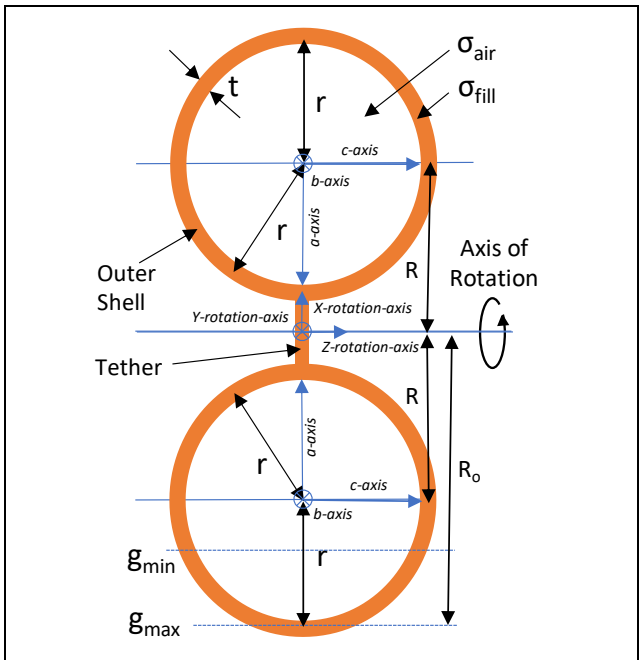

**Figure 6-9 – Dumbbell Cross Section**

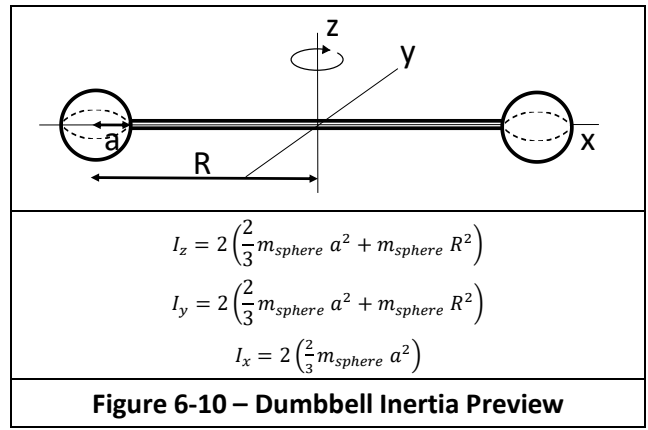

$$I_z = 2\left(\frac{2}{3}m_{sphere}\,a^2 + m_{sphere}\,R^2\right)$$

$$I_y = 2\left(\frac{2}{3}m_{sphere}\,a^2 + m_{sphere}\,R^2\right)$$

$$I_x = 2\left(\frac{2}{3}m_{sphere}\,a^2\right)$$

**Figure 6-10 – Dumbbell Inertia Preview**

Mike Birse writes "*For rotations about the other two principal axes, the frequency of small perturbations is zero. This implies that these rotations are marginally (un)stable: small perturbations do grow with time, but only linearly.*" [Birse 2000]. He also offers analysis on precession about the axis of symmetry [Birse 2000p]. Birse's analysis would support the result that the dumbbell space station is not rotationally stable but would only have small perturbations.

To validate the applicability of the [Fitzpatrick 2011] analysis guidelines to the dumbbell, we communicated with Dr. Fitzpatrick. In his response, he wrote: *"The dumbbell space station would appear to be rotationally unstable under the standard criterion because the principal moments of inertia about the y and z axes are the same, which means that there is no reason why the station should prefer to rotate about any particular axis perpendicular to the x axis. However, the instability with two equal moments of inertia is algebraic rather than exponential. Also, even if the rotation is unstable that only means that the rotation will settle into a limit cycle in which the rotational axis wobbles slightly."* [Fitzpatrick 2023].

These guidelines, analysis, and communication all lead to the conclusion that the dumbbell station is not rotationally stable. Its instability grows only linearly and would introduce a wobble into the station. This section continues to analyze the single dumbbell station, knowing this wobble could exist and cause issues for the residents. A later section introduces a simple extension to the dumbbell station to eliminate this instability and wobble.

### 6.5.2 Two Spherical Thin Shell Nodes

As an initial investigation the dumbbell is modeled with thin shell nodes. The shell thickness is much less than the radius of the station. A dumbbell has a major axis R, and spherical nodes have a minor radius of a. A line drawing of the spherical node dumbbell is in Figure 6-10. Prolate ellipsoid nodes have radial minor radius (a) and perpendicular radius (c). This first analysis ignores the tether mass and moment of inertia. The MOI for the dumbbell uses the mass (M) and the radii (R, a, and c). The dumbbell rotates about the z-axis. The MOIs about the z-axis and the y-axis are equal in our rotating framework analysis [Temples 2015]. Table 2 3 provides the moment of inertia equations.

This analysis uses the same approach used with the previous geometries. The station stability uses the MOI equations and the stability ratio $I_z \geq 1.2\, I_x$. The analysis finds:

$$\frac{2}{3}m_{sphere}\,a^2 + m_{sphere}\,R^2 \geq 1.2\left(\frac{2}{3}m_{sphere}\,a^2\right)$$

$$\frac{2}{3}a^2 + R^2 \geq \frac{2.4}{3}a^2$$

$$R^2 \geq \frac{0.4}{3}a^2$$

The rotating station with two thin shell spheres would be stable when $R^2 \geq \frac{0.4}{3}a^2$ or $R \geq 0.365\,a$. The radius R is typically much greater than the minor radius a. This would nearly always be true. Unfortunately, Iz and Iy are equal for our simple model of the dumbbell system and would not meet the Brown stability criteria for Iz>1.2Iy. This system also violates the Fitzpatrick's stability rule. This paper accepts that the dumbbell system would be unstable and assumes it would have a rotational wobble.

### 6.5.3 Two Spherical Thick Shell Nodes

More complex dumbbell models were analyzed with the hope of finding a rotating system with passive stability across all three axes. Including the tether does not help the stability because the Iy and Iz remain equal.

Evaluating the thick shell spherical nodes found the inertias for two thick shell spherical nodes would be:

$$I_z = 2\left(\frac{2}{5}m_{sphere}\left((a_o^5 - a_i^5)/(a_o^3 - a_i^3)\right) + m_{sphere}\,R^2\right)$$

$$I_y = 2\left(\frac{2}{5}m_{sphere}\left((a_o^5 - a_i^5)/(a_o^3 - a_i^3)\right) + m_{sphere}\,R^2\right)$$

$$I_x = 2\left(\frac{2}{5}m_{sphere}\left((a_o^5 - a_i^5)/(a_o^3 - a_i^3)\right)\right)$$

Our analysis found that Iz>=1.2Ix. Unfortunately, with Iz equal to Iy, the thick shell spherical node system would still violate the Iz=1.2Iy stability rule and not be passively stable.

### 6.5.4 Other Thick Shell Ellipsoid Nodes

Analysis was performed using other geometry configurations. This included thin and thick shell ellipsoid nodes using various thicknesses and axes lengths. This also included adding tethers and a cylindrical shuttle bay. None helped to solve the passive stability problem in the dumbbell station. In all these cases, the Iz < 1.2 Iy and the dumbbell system would not be rotationally stable. These simple dumbbell models corroborate the initial analysis showing that the dumbbell would not be rotationally stable. Iz is greater than Ix but equal or nearly equal to Iy.

Considering the ellipsoid nodes in the dumbbell station, it was initially thought that the asymmetric shape of the nodes could lead to a stable rotation. Using the MOIs for the ellipsoid nodes:

$$I_x = I_y = \frac{1}{5} m \frac{(c_o a_o^3)(c_o^2 + a_o^2) - (c_i a_i^2)(c_i^2 + a_i^2)}{c_o a_o^2 - c_i a_i^2}$$

$$I_z = \frac{2}{5} m \frac{c_o a_o^4 - c_i a_i^4}{c_o a_o^2 - c_i a_i^2}$$

With equal minor axes, a=c, these ellipsoid node MOI equations match the spherical-nodes MOIs as expected. Using the ellipsoid nodes in the stability ratio equation.

$$I_z = 2 \left( \frac{2}{5} m \frac{c_o a_o^4 - c_i a_i^4}{c_o a_o^2 - c_i a_i^2} + m_{node} R^2 \right)$$

$$I_y = 2 \left( \frac{1}{5} m \frac{(c_o a_o^2)(c_o^2 + a_o^2) - (c_i a_i^2)(c_i^2 + a_i^2)}{c_o a_o^2 - c_i a_i^2} + m_{node} R^2 \right)$$

$$I_x = 2 \left( \frac{1}{5} m \frac{(c_o a_o^2)(c_o^2 + a_o^2) - (c_i a_i^2)(c_i^2 + a_i^2)}{c_o a_o^2 - c_i a_i^2} \right)$$

As with the spherical nodes, the inertia about the z-axis Iz is generally greater than the inertia Ix. Unlike the spherical nodes where the Iz was equal to Iy, the MOIs for the ellipsoid nodes are different. Applying those MOIs to the stability ratio $I_z \geq 1.2 I_y$, analysis finds:

$$\frac{4}{5} \frac{c_o a_o^4 - c_i a_i^4}{c_o a_o^2 - c_i a_i^2} + 2 R^2 >$$

$$1.2 \left[ \left( \frac{2}{5} \frac{(c_o a_o^2)(c_o^2 + a_o^2) - (c_i a_i^2)(c_i^2 + a_i^2)}{c_o a_o^2 - c_i a_i^2} + 2 R^2 \right) \right]$$

$$\frac{4}{5} \frac{c_o a_o^4 - c_i a_i^4}{c_o a_o^2 - c_i a_i^2} + 2 R^2 > \frac{2.4}{5} \frac{(c_o a_o^2)(c_o^2 + a_o^2) - (c_i a_i^2)(c_i^2 + a_i^2)}{c_o a_o^2 - c_i a_i^2} + 2.4 R^2$$

$$2 \frac{c_o a_o^4 - c_i a_i^4}{c_o a_o^2 - c_i a_i^2} - 1.2 \frac{(c_o a_o^4 - c_i a_i^4)}{c_o a_o^2 - c_i a_i^2} + 1.2 \frac{(a_i^2 c_i^3 - a_o^2 c_o^3)}{c_o a_o^2 - c_i a_i^2} > R^2$$

$$0.8 \frac{c_o a_o^4 - c_i a_i^4}{c_o a_o^2 - c_i a_i^2} + 1.2 \frac{(a_i^2 c_i^3 - a_o^2 c_o^3)}{c_o a_o^2 - c_i a_i^2} > R^2$$

Evaluating the equation using c=2a:

$$0.8 \frac{2 a_o a_o^4 - 2 a_i a_i^4}{2 a_o a_o^2 - 2 a_i a_i^2} + 1.2 \frac{a_i^2 8 a_i^3 - a_o^2 8 a_o^3}{2 a_o a_o^2 - 2 a_i a_i^2} > R^2$$

$$0.8 \frac{a_o^5 - a_i^5}{a_o^3 - a_i^3} - 4.8 \frac{a_o^5 - a_i^5}{a_o^3 - a_i^3} > R^2$$

$$-4 \left( \frac{a_o^5 - a_i^5}{a_o^3 - a_i^3} \right) > R^2$$

This is never true for c=2a. Evaluating other ratios and considering previous results, it appears that dumbbells using ellipsoid nodes are also not rotationally stable.

### 6.6 Single Floor Stability Summary

Table 6-2 contains a summary of our current stability equations for the four station geometry types. These equations represent moments of inertia for thin-shell hollow geometries. This table was adapted from [Jensen 2023] with minor description changes. Our analysis in this section includes thick-shell geometries. The thick shell results are typically close to the thin shell results when the thickness is much less than the rotation radius. Our results match and extend the cylinder results from [Globus et al. 2007]. Ellipsoid geometry stations can be rotationally stable; however, spherical stations are not. Torus designs would be rotationally stable for the desired station dimensions. Dumbbell stations would not be rotationally stable. They would have a wobble that might impact the station residents.

## 7 Multiple Component Rotational Stability

The previous section considered single-floor stations with thin and thick shells. The analysis considered only the outer shell and ignored other components such as spokes, shuttle bays, multiple floors, and air. The analysis in this section includes the mass and MOIs of all the station components. The first subsection introduces common analysis details. This includes the analysis of each of the station components. The following subsections analyze the stability of our geometries with all components. This generates the station components, mass and inertia equations, top floor limits, mass results, and balance results for each of the geometries.

| Table 6-2 – Geometries and Rotational Stability for Thin Shell Geometries | | | |
|---|---|---|---|
| **Geometry** | **Key Stability Factor** | **Rotational Stability** | **Notes** |
| Cylinder | L < 1.3 R | Hatbox cylinders can be stable | Flat endcaps |
| Ellipsoids | c < 0.8165 a | Oblate ellipsoids can be stable | Sphere stations are not stable |
| Elliptical Torus | $R^2 > 1.5 (c^2-a^2)$ | Elliptical tori are stable where R>=6.33a and c=2a or c=3a | Torus only – inner docking station and spokes not included |
| Dumbbell | Iz/Ix=1.2 when $R^2 \geq \frac{1}{3}(1.2 c^2 - 0.8 a^2)$<br>Iz/Iy is less than 1.2 for all R | Instability from equal MOIs on rotation and radial axes grows algebraically, and the rotational axis will wobble. | A wobble will remain an open issue for the single dumbbell system. |
| *Credit: Adapted from Table 3-3 [Jensen 2023] [CC BY-SA 4.0] and by extending cylinder concepts from [Globus et al. 2007] [Facts]* | | | |

## 7.1 Common Analysis Details

This subsection covers much of the common analysis of all four geometries: cylinders, ellipsoids, toruses, and dumbbells. This includes the analysis approach for components, using material densities, using variable air density, defining the top floor of the multiple floors in the stations, and previewing the component mass and inertia allocations.

The variables in this section include m for mass, V for volume, ρ for density, r for cylinder radius, I for the rotational moment of inertia, and L for the cylinder length. Variable subscripts provide an orientation, indicate inner and outer dimensions, or provide a description.

### 7.1.1 Multiple Components Analysis Approach

This section extends the stability analysis to include more of the station components. The major components in the station have different densities, geometries, and dimensions. These components include the outer shell, multiple floors, the main (top) floor, the shuttle bay, spokes, and air.

Our single-floor stability analysis assumed homogeneous outer shell densities to compute the station's moments of inertia (MOI). Table 5-2 provides the Moments of Inertia (MOI) equations used as an engineering estimate for analytic analysis. The analysis produced a set of closed-form equations to evaluate the stability. The analysis ignored other station components. This single-floor approach provides a quick approximation for initial stability analysis.

We first tried to use a similar numeric analysis to compute the station inertia values for the multiple-component designs. Closed-form equations for the mass and inertia are available for many subcomponents. Combining those closed-form equations became unyielding for some of the more complex components. Another approach was used to compute the rotational moments of inertia. Complex components were decomposed into many small pieces. The analysis summed the rotational MOIs and masses of all those individual pieces. This can be considered a Riemann sum of the MOIs and masses of those pieces. Where possible, these results were compared to analytic equation results for portions of those more complex components to validate our results.

### 7.1.2 Density and Moment of Inertia Analysis

The single-floor analysis could usually derive closed-form equations for the inertia using the mass of the shells. Different material densities are used with the multiple component analysis. This analysis uses the foundation of equations and densities as introduced in *§5.1 Station Component EquationsThe stability analysis of stations with multiple components requires mass and MOIs for all the components. The station's mass and MOIs are the sum of the masses and MOIs of its components. Similarly, the components' mass and MOIs are the sum of the masses and MOIs of their* constituent parts. The analysis uses different densities for the various components. Table 5-1 provides the different densities. The mass variables in the inertia equations of Table 5-2 can be replaced with the volumes and densities of the components and their constituent parts.

### 7.1.3 Component Mass and Inertia Preview

There is similarity in the components across the four different station geometries. The following subsections briefly preview the mass and moment of inertia equations for the components in the four geometries. The analysis of some components is essentially identical across all the station geometries. The analysis of those nearly identical components is detailed in the following subsections. Station components such as the shell, the multiple floors, the air, and the dividers require geometry (and station) specific analysis. They are briefly introduced and then detailed separately.

#### 7.1.3.1 Outer Shell

The outer shell is different for each of the four geometries. The cylinder geometry station uses a hollow cylinder as the outer shell and solid cylinders as the thick endcaps. The ellipsoid station uses the difference between concentric ellipsoids to compute its mass and MOIs. The elliptical torus is modeled with multiple thin disks that are summed to calculate its mass and MOIs. Dumbbells also require the summing of many thin disks to compute the mass and MOIs of their shells. Each station geometry section provides specifics on the mass and MOIs of their outer shells.

The shell thickness is typically 20 meters in our analysis. The outer shell contains a thick layer of regolith to provide shielding from radiation and debris and provides protection to our station's residents. The shell has a truss framework to provide most of the structural integrity. Ten-meter walls would be sufficient for radiation protection, but we prefer greater thickness to provide additional collision safety.

#### 7.1.3.2 Spokes

The spokes of the station are modeled as thick shell cylinders. Our research has modeled these cylinders using a single density and radius, as well as an outer-filled thick shell cylinder with an inner hollow structure. The equation $m_{spoke} = \rho_{spoke}\, pi\, L\, (r_o^2 - r_i^2)$ is used to compute each of these thick shell cylinders. The inner and outer cylinders are computed separately with different densities to provide more generality with changing radius size. The length of the spoke and the cylinder is bound by the station shell. Table 5-2 shows the MOIs of a cylinder along the x-axis length is $I_x = \frac{1}{2}m(r_o^2 + r_i^2)$ for the x-axis and $I_z = I_y = \frac{1}{12}m(3(r_o^2 + r_i^2) + L^2)$ for the y- and z-axis.

All the geometries use multiple sets of these spokes to provide structural strength. The shells themselves provide structural integrity except on the dumbbell stations. These spokes are placed at equal angular distances except on the dumbbell stations. Figure 7-1 shows 4 sets of spokes at 45-degree spacing. We derived the following MOI equations for the spokes angled about the z-axis:

$$I_z = \frac{1}{12}m(3r^2 + L^2) \text{ for the z-axis}$$

$$I_{x\alpha} = \frac{1}{12}m(3r^2 \cos^2\alpha + 3r^2 + L^2 \sin^2\alpha) \text{ for the x-axis}$$

$$I_{y\alpha} = \frac{1}{12}m(3r^2 \sin^2\alpha + 3r^2 + L^2 \cos^2\alpha) \text{ for the y-axis}$$

The variable α represents the angle about the z-axis. The $I_z$ is not dependent on the angle. The x-axis and y-axis MOIs vary with the angle but are equivalent for spokes offset by 90 degrees. This provides an angle identity rule:

$$I_{y\alpha} = I_{x\alpha+90}$$

Summing the mass and MOI values for all the spokes computes the total spoke mass and MOIs. Summing the 4 spokes at 0, 45, 90, and 135 degrees and using the previous equations and angle identity rule produces:

$$I_{z4spokes} = 4\left[\frac{1}{12}m\left(3r^2 + L^2\right)\right]$$

$$I_{x4spokes} = I_{y4spokes} = 4I_{x45} = \frac{1}{12}m(18r^2 + 2L^2)$$

Side spokes are offset from the x-axis and the y-axis by a distance, d, as shown in the side view of Figure 7-1. The parallel axis theorem adds the $m_{spoke}$ times d² to the side spokes Ix and Iy.

In large stations, multiple sets of spokes are used to provide more strength for the expected stresses from centripetal gravity. These mass and inertias scale with the sets of spokes.

### 7.1.3.3 Shuttle Bay

A shuttle bay is included along the rotation axis in each station geometries. Figure 7-2 illustrates interior concepts in a shuttle bay. This level of detail supported our density analysis. The illustration removes the endcap to show the thick exterior shell, a series of concentric cylinder floors, and arrival stations (yellow thick-walled cubes). The floors are 5 meters apart and provide work areas for arrivals and departures. The thick exterior shell provides radiation and collision protection. The arrival stations provide space for shuttles, jet bridges, and service equipment.

The shuttle bay is modeled as a cylinder located at the center of rotation along the z-axis direction; see Figure 7-3. The shuttle bay is modeled in larger stations with two cylinders and a connecting spoke. The two cylinders are positioned at the outer and opposite edges of the station, and the spoke connects them; see Figure 7-3. The shuttle bay geometry and analysis are nearly identical for all the station geometries. The shuttle bay cylinder is comprised of two concentric cylinders. The outer cylinder is a filled shell; the inner cylinder is more like a spoke structure with interior walls. Originally, we considered extending a single shuttle bay cylinder across the entire station length. With large stations, the shuttle bay would be very long, use considerable material, and would not be useful space because of low gravity.

Each shuttle bay is typically limited to a maximum diameter of 360 meters and a maximum length of 200 meters. This size could support multiple space shuttle-sized crafts. The length and diameter are less for very small stations. Large stations use the connecting spoke to extend along the z-axis between the two bays. This provides a structural connection between the two bay cylinders and uses less mass than extending the shuttle bays. The spoke would eventually contain ventilation and illumination in the cylinder and ellipsoid stations.

Our shuttle bay typically uses an exterior 20-meter-thick outer shell and endcaps. Its interior has a density that is more like a spoke structure with additional walls. Each of the shuttle bay cylinders would have a mass of $m_{bay} = \rho_{bay}\, pi\, L_{bay}\, (r_o^2 - r_i^2)$ where $r_o$ is the outer radius, and $r_i$ is the inner radius. The density of the shuttle bay varies when interior to the station outer shell (cylinder and ellipsoid) and exterior to the station outer shell (dumbbell and torus). A density of 337.4 kilograms per cubic meter is used for the interior cylinder. The shuttle bay's exterior cylinder would have a density of 1721 kilograms per cubic meter (kg/m3) when it is exterior to the outer shell of the station. This provides more protection from radiation and collisions.

The connecting spoke has a filled outer cylinder that is 5-meters thick, and the interior cylinder is assumed to have a floor-like structure separated by 5 meters. The connecting spoke is modeled as a cylinder. It has the same radius as other

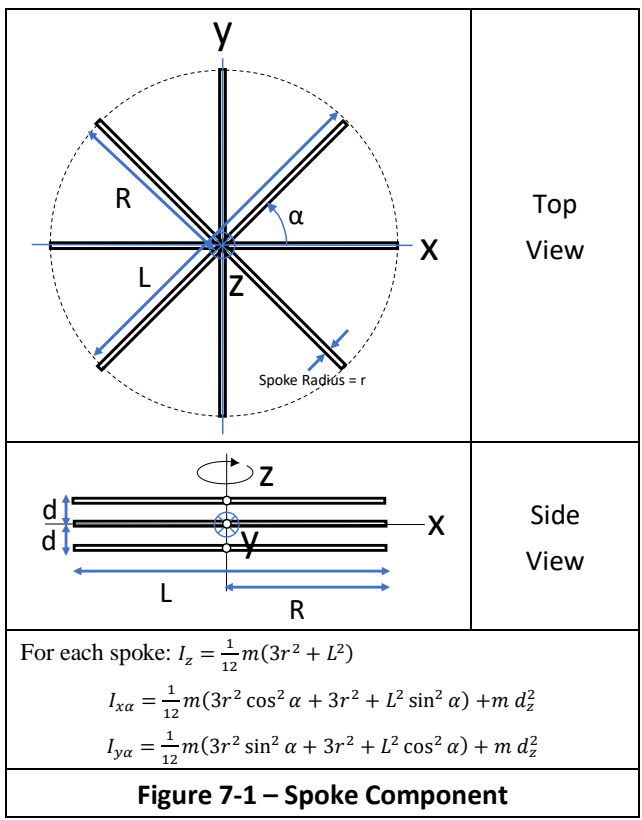

**Figure 7-1 – Spoke Component**

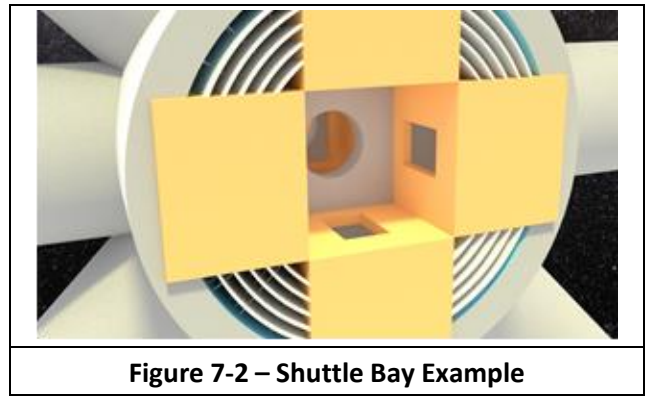

**Figure 7-2 – Shuttle Bay Example**

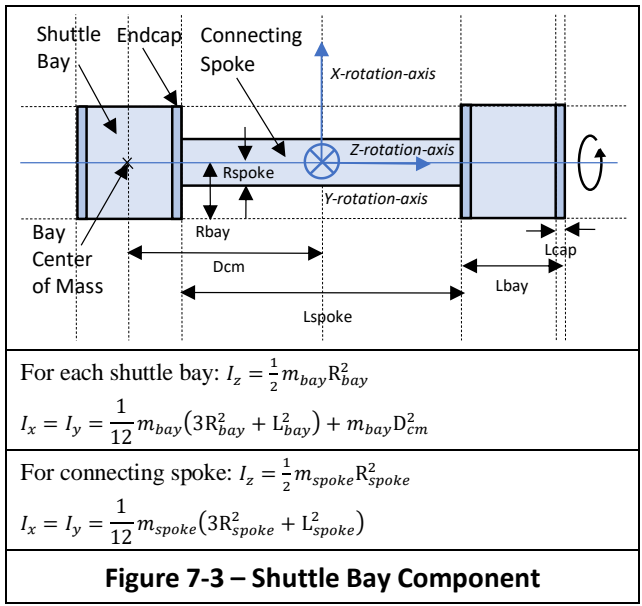

For each shuttle bay: $I_z = \frac{1}{2} m_{bay} R_{bay}^2$

$I_x = I_y = \frac{1}{12} m_{bay}(3R_{bay}^2 + L_{bay}^2) + m_{bay} D_{cm}^2$

For connecting spoke: $I_z = \frac{1}{2} m_{spoke} R_{spoke}^2$

$I_x = I_y = \frac{1}{12} m_{spoke}(3R_{spoke}^2 + L_{spoke}^2)$

**Figure 7-3 – Shuttle Bay Component**

spokes in the station. The spokes do not have thick-filled shells for radiation and collision protection and have smaller densities than the shuttle bay. The exterior cylinder has a density of 337.4 kg/m3, and the interior would have a density of 32.2 kg/m3. The exterior endcaps of the shuttle bays have a density of 1721 kg/m3.

The MOIs of all the modeled cylinders use $I_z = \frac{1}{2} m(r_o^2 + r_i^2)$ for the z-axis and $I_x = I_y = \frac{1}{12} m(3(r_o^2 + r_i^2) + L^2)$ for the x- and y-axis. The Ix and Iy for the bays and endcaps also require using the parallel axis theorem dependent on the station geometry. Figure 7-3 shows the location of the shuttle bay center of mass for that computation.

#### 7.1.3.4 Main Floor

The main floor is considered separately from the multiple-floors component. The main floor is positioned to create a habitable gravity range; see *§3.1 Gravity Limits.* The multiple floors are placed under the main floor. A gravity range between 0.95g and 1.05g in a cylinder station would use a ratio of the station radius over the floor height of 10.5. Figure 7-4 shows this height as R/m, where m is the gravity scaling factor 10.5. The main floor is placed at this height and would have the minimum gravity of 0.95g.

The top floor is thicker than other floors in the station. The top floor provides open aesthetics for psychological well-being. In general, for modeling, the main floor has the same thickness and density as the shell and endcaps. The floor thickness is set to 1% of the rotation radius of the station shell. For large stations, the main floor thickness has a maximum of 20 meters. For small stations the main floor thickness has a minimum of 2 meters. The thick floor structure would also support 2 meters of topsoil. The structure and topsoil could be designed to form small hills and valleys. The length of the main floor is represented by the variable L and varies for the various geometries. Figure 7-4 illustrates a side view of a simple main floor in a rotating cylinder or ellipsoid station.

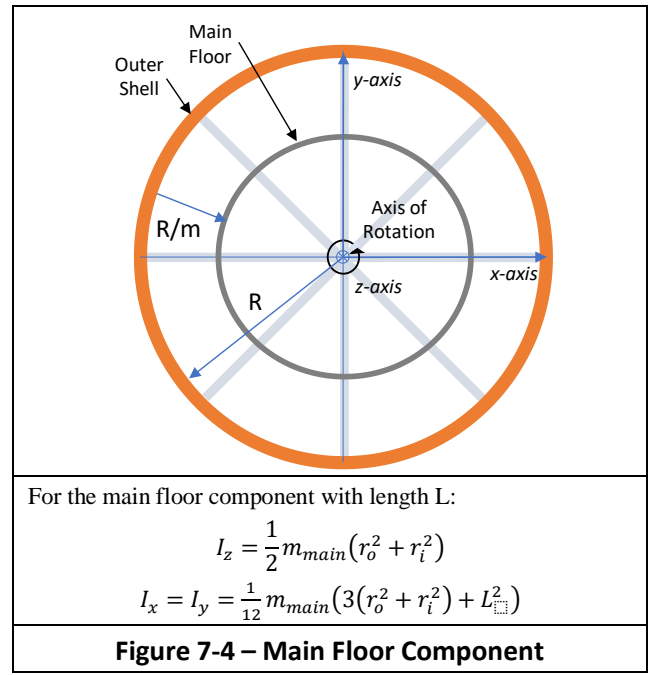

For the main floor component with length L:

$$I_z = \frac{1}{2} m_{main}(r_o^2 + r_i^2)$$

$$I_x = I_y = \frac{1}{12} m_{main}(3(r_o^2 + r_i^2) + L^2)$$

**Figure 7-4 – Main Floor Component**

The mass of the main floor (cylinder) is modeled as two hollow, thick-shell cylinders. One cylinder represents the topsoil, and the other represents the supporting structure. The mass of those cylinders would be computed with:

$$V_{main} = pi\, L\, (r_o^2 - r_i^2)$$
$$m_{main} = \rho_{fill} V_{main} = \rho_{fill}\, pi\, L\, ((r_i + t_m)^2 - r_i^2)$$

The variables $r_o$ and $r_i$ represent the outer and inner radius of the main floor. The two cylinders are summed to produce the total main floor mass. The thickness of the structure floor is $t_m$ and is equal to $r_o$-$r_i$. The topsoil uses the same equations, except the thickness is 2 meters. The length of the cylinders representing the floor is L. The main floor inertias use the thick shell cylinder MOIs from Table 2-3. The MOIs for the main floor would be:

$$I_x = I_y = \frac{1}{12} m_{main}(3(r_o^2 + r_i^2) + L^2)$$

$$I_z = \frac{1}{2} m_{main}((r_i + t_m)^2 - r_i^2)$$

Again, the main floor structure and the top soil inertias are summed to produce the total inertias for the main floor. In the dumbbell, the main floor is different from the other geometries. The dumbbell models the mass and MOIs of the main floor using a flat elliptical disk. This floor design model was previewed in *§5.3 Dumbbell Modeling Details*. Additional details on the different floor models for the dumbbell are covered in the later dumbbell geometry section.

#### 7.1.3.5 Multiple Floors

This component of the station typically has floors that are 5 meters apart and are bounded by the station shell. The multiple floors are modeled with a series of concentric cylinders. The concentric cylinders are closer together to model the multiple floors component more accurately. This is required because a curved shell bounds its sides. The floors and the modeled cylinders are between the main floor and the outer rim over a distance of R/m. The cylinders are analyzed like the main floor cylinders. Each cylinder would have a mass of $m_{floor5} = \rho_{floor5}\, pi\, L_n\, (r_o^2 - r_i^2)$ where ro is the outer radius of the cylinder, and ri is the inner radius. The density of the multiple floors is from Table 5-1. This floor density uses the mass and volumes of the constituent trusses, panels, columns, and open space between floors. The length of each cylinder is bound by the station shell or endcaps. The radii range from the main floor to the outer rim. The mass and MOIs are computed for the series of concentric cylinders. The air density is excluded from the multiple-floors analysis; and instead, it is used in the air analysis. The MOI equation about the z-axis for the cylinder would be $I_z = \frac{1}{2}m(r_o^2 + r_i^2)$ and about the x (or y) axis the MOI equation would be $I_x = I_y = \frac{1}{12}m(3(r_o^2 + r_i^2) + L_n^2)$. Figure 7-5 shows a simple drawing of the multiple floors in a cylinder or ellipsoid station. The length, $L_n$, would be a constant in the cylinder station and vary with height in the ellipsoid station. Each station geometry section provides specifics on the lengths, masses, and MOIs of their multiple floors. Those multiple cylinder values are summed to determine the total mass and MOIs of the multiple-floors component.

#### 7.1.3.6  Air

The outer rim of the station uses the sea level air density of 1.225 kg/m3 and the maximum gravity. The air density decreases with increasing height above the outer rim. The station geometries use the equations described in *§3.2 Air Pressure Limits*.

Figure 7-6 shows the decreasing air density of an ellipsoid and a torus. Both geometries have an outer rotation radius of 2000 meters. The air density in the ellipsoid continues to decrease to the center of rotation (radius R=0 and h=R). The air density in the torus decreases at the same rate but ends abruptly at the inner edge of the torus tube.

In larger stations, many floors have habitable gravity. Top floors in those very large stations would have unacceptably low air pressure. The *§3.5 Top Floor Limits* introduces the concept of multiple airtight floor layers to provide habitable air pressure on all the floors in those large stations. Figure 3-3 illustrated the air density being reset with each of the airtight layers. When the air density becomes the Denver limit (1.01 kg/m3), the airtight layer resets the air density to sea level (1.225 kg/m3). The height of the airtight layers varies with the geometry but is typically about 1600 meters high in the large rotating stations.

The air component is modeled and analyzed using concentric cylinders; see Figure 7-7. These cylinders of air extend from the station rotation center to the outer rim. The mass and MOIs of the concentric cylinders are calculated in the same fashion as the multiple floors. Each station geometry section provides specifics on the lengths, masses, and MOIs of their multiple air cylinders. The air cylinder MOI equation about the z-axis for the cylinder would be $I_z = \frac{1}{2}m(r_o^2 + r_i^2)$ and about the x (or y) axis, the MOI equation would be $I_x = I_y = \frac{1}{12}m(3(r_o^2 + r_i^2) + L^2)$. Those multiple cylinder values are summed to determine the total mass and MOIs of the air component.

#### 7.1.3.7  Divider

Only the torus design has dividers. The dividers separate the torus tube into airtight compartments as part of a fail-safe design. A divider is comprised of two vertical walls with a structure of floors between them. In case of a catastrophic impact and loss of atmosphere, only one section of the torus tube would be lost.

A divider in the torus is modeled as two filled elliptical disks with a structure of trusses between them. The disks are represented with different densities and thicknesses. The divider ellipse rotates about the z-axis at a distance of R; see Figure 7-8. The dividers enclose the spokes; as such, their thickness is about the diameter of the spokes.

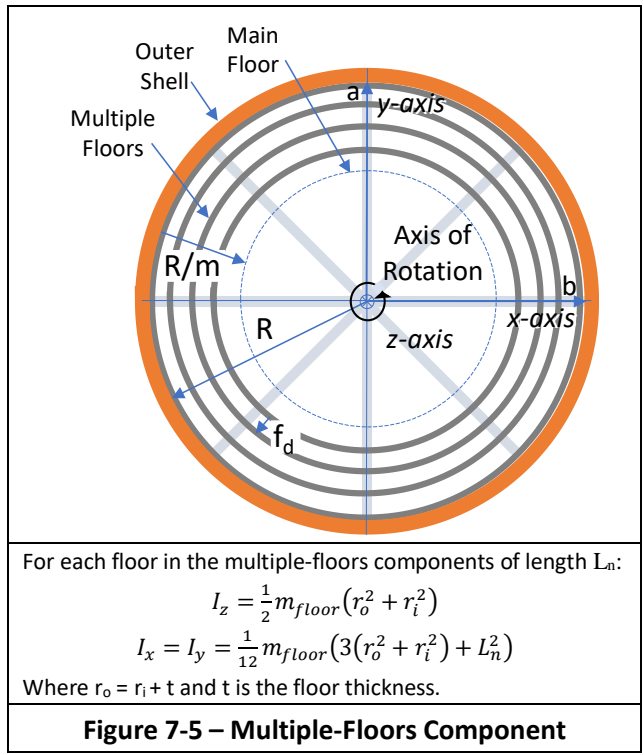

For each floor in the multiple-floors components of length $L_n$:
$$I_z = \tfrac{1}{2}m_{floor}(r_o^2 + r_i^2)$$
$$I_x = I_y = \tfrac{1}{12}m_{floor}(3(r_o^2 + r_i^2) + L_n^2)$$
Where $r_o = r_i + t$ and t is the floor thickness.

**Figure 7-5 – Multiple-Floors Component**

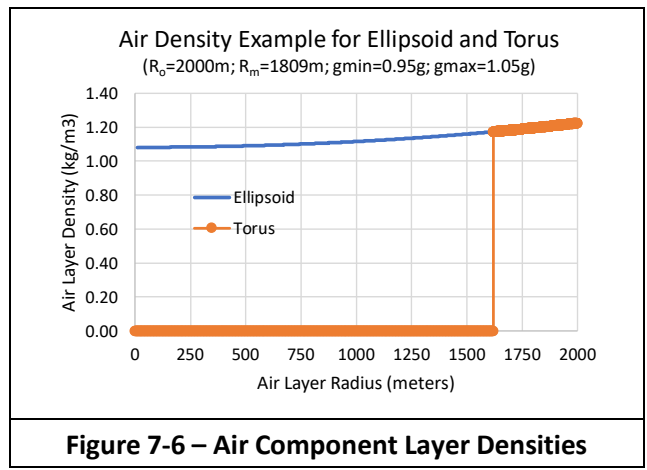

**Figure 7-6 – Air Component Layer Densities**

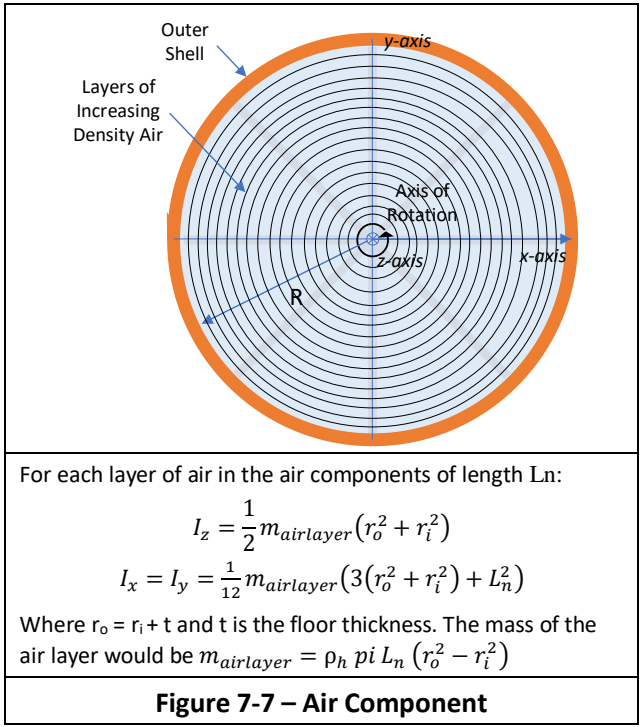

For each layer of air in the air components of length Ln:

$$I_z = \frac{1}{2} m_{airlayer}(r_o^2 + r_i^2)$$

$$I_x = I_y = \frac{1}{12} m_{airlayer}(3(r_o^2 + r_i^2) + L_n^2)$$

Where $r_o = r_i + t$ and t is the floor thickness. The mass of the air layer would be $m_{airlayer} = \rho_h\, pi\, L_n\, (r_o^2 - r_i^2)$

**Figure 7-7 – Air Component**

The following describes one divider design. The divider would have a center interior structure built from trusses and panels, providing floors with a spacing of 5 meters. This divider has two vertical walls filled with regolith and is 5 meters thick each. The thickness of the interior structure and the exterior walls sum to the diameter of the station spoke. This design also includes another 10 meters of floor structure on the exterior of the divider walls. Figure 2-2 shows this design with the many floors; the floors would provide a spectacular view and could be used for living quarters, shops, and offices.

Other thicknesses could be used, and we recommend that the divider strength be designed and analyzed more rigorously. The divider must support a catastrophic outer shell failure. The width of our design is likely to be conservative, and a more refined design would provide additional building material for other components.

The mass of the divider is computed using different densities for the three different types of disks. The mass of the interior divider structure would be the divider ellipse area ($\pi\, a\, c$) times the structure thickness times the multiple-floors density, $\rho_{floor5}$. The two filled disks would use a density of $\rho_{fill}$. The two outer floor structures also use the floor density, $\rho_{floor5}$, in our analysis. The 5 walls are summed to compute the divider mass, $m_{div}$. Conceptually, the mass $m_{div} = \rho\, t_{div}\, \pi\, a_i\, c_i$.

The MOI analysis of the divider begins by using the elliptical disk equations from Table 5-2. The MOI equations for the axes of the elliptical disk positioned at its center of gravity at the elliptical torus origin are:

$$I_z = \frac{1}{12} m_{div}(3a_i^2 + t_{div}^2),$$

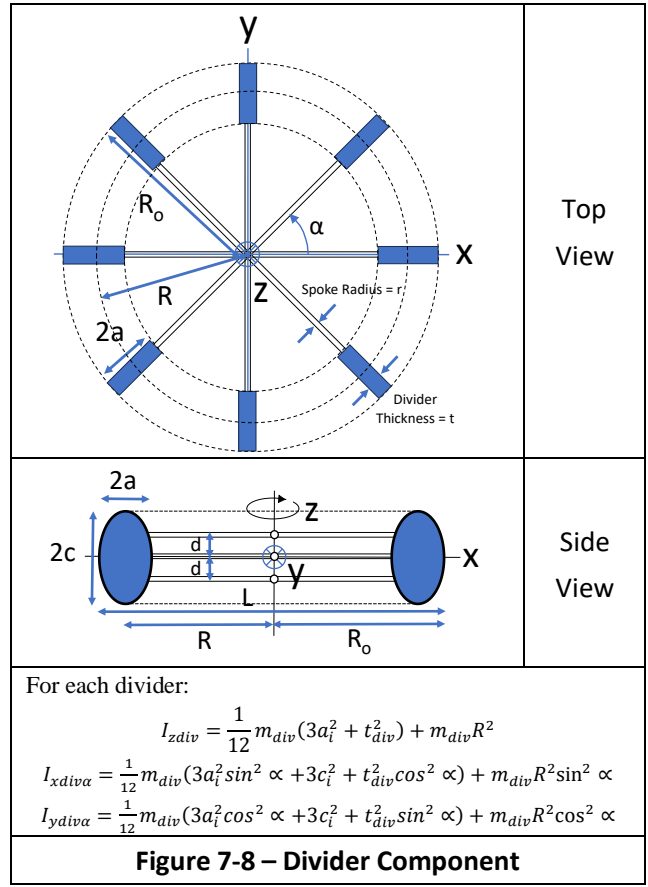

For each divider:

$$I_{zdiv} = \frac{1}{12} m_{div}(3a_i^2 + t_{div}^2) + m_{div}R^2$$

$$I_{xdiv\alpha} = \frac{1}{12} m_{div}(3a_i^2 \sin^2\alpha + 3c_i^2 + t_{div}^2 \cos^2\alpha) + m_{div}R^2\sin^2\alpha$$

$$I_{ydiv\alpha} = \frac{1}{12} m_{div}(3a_i^2 \cos^2\alpha + 3c_i^2 + t_{div}^2 \sin^2\alpha) + m_{div}R^2\cos^2\alpha$$

**Figure 7-8 – Divider Component**

$$I_y = \frac{1}{4} m_{div}(a_i^2 + c_i^2), \text{ and}$$

$$I_x = \frac{1}{12} m_{div}(3c_i^2 + t_{div}^2).$$

Because the 8 disks rotate about the z-axis at an equal distance of R, the parallel axis theorem is used to determine the total MOI for the system Iz:

$$I_{zdiv8} = 8(Iz + m_{div}R^2)$$

$$I_{zdiv8} = 8\left(\frac{1}{12} m_{div}(3a_i^2 + t_{div}^2) + m_{div}R^2\right)$$

$$I_{zdiv8} = \frac{2}{3} m_{div}(3a_i^2 + t_{div}^2 + 12R^2)$$

The systems Ix and Iy MOIs are more complex. Two of the eight disks rotate on their own axis. Two rotate in the perpendicular orientation axis at a distance of R. The other four elliptical disks are on the angled spokes and rotate at an angle of 45 degrees. They rotate at a distance of sin(45) or cos(45) times R from their axes. The x-axis and y-axis MOIs at the angle α for these 8 dividers are:

$$I_{xdiv\alpha} = \frac{1}{12} m_{div}(3a_i^2 \sin^2\alpha + 3c_i^2 + t_{div}^2 \cos^2\alpha) + m_{div}R^2\sin^2\alpha$$

$$I_{ydiv\alpha} = \frac{1}{12} m_{div}(3a_i^2 \cos^2\alpha + 3c_i^2 + t_{div}^2 \sin^2\alpha) + m_{div}R^2\cos^2\alpha$$

To match their position on the 8 spoke positions, the angle α varies from 0 to 315 degrees at increments of 45 degrees. These 8 MOIs are summed to compute the total $I_{xdiv8}$ and $I_{ydiv8}$ in the torus. With the summing of the sine squared and cosine squared terms and simplification, those totals become:

$$I_{xdiv8} = I_{ydiv8} = 8\, I_{xdiv45}$$
$$I_{xdiv45} = \frac{1}{12} m_{div}\left[\left(\frac{3}{2}a_i^2 + 3c_i^2 + \frac{1}{2}t_{div}^2\right) + 12\,\frac{1}{2}R^2\right]$$
$$I_{xdiv45} = \frac{1}{8} m_{div}[a_i^2 + 2c_i^2 + t_{div}^2/3 + 4R^2]$$

The MOI of the 45-degree divider is used to find the total MOI of the 8 dividers around the torus tube:

$$I_{xdiv8} = I_{ydiv8} = 8\,I_{xdiv45} = m_{div}[a_i^2 + 2c_i^2 + t_{div}^2/3 + 4R^2]$$
$$I_{xdiv8} = I_{ydiv8} = \frac{1}{3} m_{div}[3a_i^2 + 6c_i^2 + t_{div}^2 + 12\,R^2]$$

And as a reminder:

$$I_{zdiv8} = \frac{2}{3} m_{div}\left(3a_i^2 + t_{div}^2 + 12R^2\right)$$

## 7.2 Cylinder - Multiple Component Stability

The analysis of the multiple component stability begins with the cylinder geometry. The thin shells and endcaps analysis found that Iz = 1.2 Ix when L=1.3r [Globus et al. 2007]. Thick shells and endcaps consistently produced Iz = 1.2 Ix when L=1.29r over a wide range of radii and wall thicknesses. The length-to-radius ratio increased slightly with very thick shells. This increase implies the station can have a larger L/r but still be stable with thicker shells. The multiple-component stability analysis provides a more realistic estimate of the stability of the large station.

### 7.2.1 Cylinder Components

Figure 7-9 provides a cross-section of the cylinder station. This illustrates the station geometry with labels for components and densities. This includes views on the x-axis (Front) and the z-axis (Side). The drawing shows the major components of the cylinder. These include the outer shell, endcaps, main floor, multiple floors, center spoke, side spoke, and the shuttle bay.

Figure 7-10 also shows a cross-section drawing. This graph shows a specific example of a cylinder station with more details and dimensions. Figure 7-10 shows the polar or axial axis of rotation on the z-axis, and this axis length is labeled c. The vertical axis shows the vertical distance from the axis of rotation (z-axis) and ranges from 0 to 2000 meters. The horizontal axis shows the horizontal distance from the vertical axis (x- or y-axis) and ranges from 0 to 2800 meters. The drawing only shows one-half of the complete cylinder. A full rendering of this example torus would extend from -2000 to 2000 meters on the vertical axis. It would only extend from 0 to 2800 meters on the horizontal axis.

### 7.2.2 Cylinder Mass and Inertia Equations

The 6 components of the hatbox cylinder and their mass equations were presented in *§5.1 Station Component Equations*. These mass equations are used to develop the Moment of Inertia (MOI) equations for each of the major components of the cylinder. Our analysis uses the material densities from Table 5-1. It uses the background and the mass and inertia analysis from *§7.1.3 Component Mass and Inertia Preview*. The following subsections refine the common analysis of the

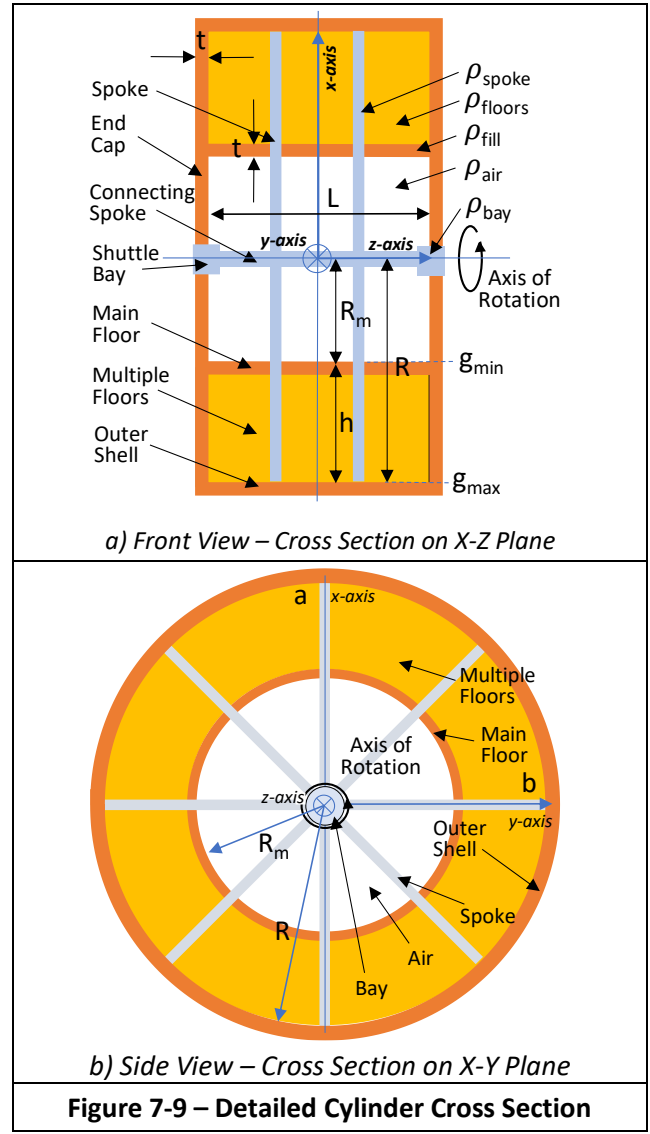

*a) Front View – Cross Section on X-Z Plane*

*b) Side View – Cross Section on X-Y Plane*

**Figure 7-9 – Detailed Cylinder Cross Section**

cylinder station, update those mass and moments of inertia equations, and provide specific examples.

#### 7.2.2.1 Cylinder Outer Shell

The outer shell of the cylinder is comprised of a thick-shelled cylinder and disk endcaps. Both of these components are modeled using the cylinder mass and moment of inertia equations. The analysis typically uses a shell thickness of 20 meters. The thick layer of regolith provides shielding from radiation and debris. A truss framework of the shell provides most of the structural integrity. The truss framework is built with anhydrous rods (or tiles) and panels. It is filled with crushed regolith. The fill density ($\rho_{fill}$) is 1721 kg/m3, and the panels and tiles density ($\rho_{fill}$ and $\rho_{rods}$) is 2790. The fill makes up most of the volume. As such, the fill density from Table 5-1 is used with the shell and endcap volumes to compute their mass.

The thick outer shell mass and MOI use these equations:

$$m_{shell} = \rho_{fill}\, V_{shell} = \rho_{fill}\, pi\, L_i(r_o^2 - r_i^2)$$

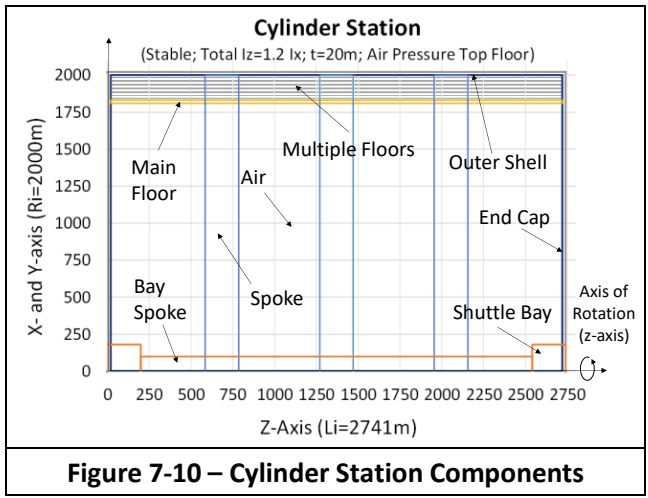

**Figure 7-10 – Cylinder Station Components**

$$I_z = \frac{1}{2}m_{shell}(r_o^2 + r_i^2)$$

$$I_x = I_y = \frac{1}{12}m_{shell}\left(3(r_o^2 + r_i^2) + L_{avg}^2\right)$$

$$I_x = I_y = \frac{1}{12}m_{shell}\left(3(r_o^2 + r_i^2) + \left(\frac{L_o + L_i}{2}\right)^2\right)$$

The following equations are used to compute the mass and MOI of the thick endcaps of the cylinder station:

$$m_{endcap} = \rho_{fill} V_{endcap} = \rho_{fill}\, pi\, t\, r_o^2$$

$$I_z = \frac{1}{2}m_{endcap} r_o^2$$

$$I_x = I_y = \frac{1}{12}m_{endcap}(3r_o^2 + t^2)$$

The endcaps need the parallel axis theorem to adjust their MOI to reflect their distance away from the center of mass:

$$I_x = I_y = \frac{1}{12}m_{endcap}(3r_o^2 + t^2) + m_{endcap}d^2$$

The length of the outer shell is $L_o$, and the length of the inner shell is $L_i$. The average length of the shell is $L_{avg}=(L_o+L_i)/2$, and the endcaps are at a distance of $L_{avg}/2$ from the center. Inertias are added to include the parallel axis theory:

$$I_x = I_y = \frac{1}{12}m_{endcap}(3r_o^2 + t^2) + m_{endcap}\frac{L_{avg}^2}{4}$$

$$I_x = I_y = \frac{1}{12}m_{endcap}(3r_o^2 + t^2 + 3L_{avg}^2)$$

As an example, the outer shell center of mass is at coordinate (1370,0) in Figure 7-10. The inner length is 2701 meters, and the outer length is 2741 meters. The average length is 2720 meters, and the endcap centers of mass are 1360 meters from the station center of mass.

### 7.2.2.2 Cylinder Spokes

The spoke models were presented in *§7.1.3 Component Mass and Inertia Preview*. The spokes are modeled as thick cylinders. For example, in Figure 7-10 the cylinder spokes have an outer radius of 75 meters and an inner radius of 70 meters. The length of the spokes is almost 4000 meters. Only half the spoke is shown in Figure 7-10. The spoke outer shell thickness is 5 meters. The spokes are filled with multiple floors that are 5 meters apart. The spoke total density would be 337.4 kilograms per cubic meter; see Table 5-1. The mass of each cylinder is computed using $m_{spoke} = \rho_{spoke}\, pi\, L\, (r_o^2 - r_i^2)$. The inner and outer cylinders use different densities to provide more generality with changing radius size. The cylinder shell bounds the spoke lengths.

The cylinder station in Figure 7-10 shows the center spoke and two side spokes. The cylinder uses 4 sets of these spokes at equal 45-degree spacing like the ellipsoid station shown in Figure 7-1. The MOIs use equations developed in *§7.1.3* for the four sets of spokes:

$$I_{z4spokes} = \frac{1}{12}m_{4spokes}(3r^2 + L^2)$$

$$I_{x4spokes} = I_{y4spokes} = \frac{1}{12}m_{4spokes}(18r^2 + 2L^2)$$

The spokes require the parallel axis theorem for their MOI. Side spokes are offset from the x-axis and the y-axis by a distance of one-quarter the z-axis length of L. This adds the spoke mass times the distance from their rotation axis squared. This adds the $m_{x4spokes}$ times $(L/4)^2$ to the side spokes Ix and Iy. The mass and MOI values for the 3 sets of spokes are summed to compute the total spoke mass and MOIs. Larger stations use more side spokes, and their mass and inertias are appropriately added to the station MOIs.

### 7.2.2.3 Cylinder Shuttle Bay

Figure 7-10 shows a cylinder example with the shuttle bay cylinders and the connecting shuttle bay spoke. Each shuttle bay cylinder has a radius of 180 meters and a length of 200 meters. The length of the connecting spoke is 2341 meters long. The shuttle bay has a 20-meter thick exterior outer shell and endcaps. Its interior is more like a spoke structure with additional walls. The connecting spoke has a filled outer cylinder that is 5 meters thick, and the interior cylinder is assumed to have multiple floors separated by 5 meters. Each of the shuttle bay cylinders would have a mass of $m_{bay} = \rho_{bay}\, pi\, L\, (r_o^2 - r_i^2)$. The exterior cylinder would have a density of 1721 kilograms per cubic meter, and the interior would have a density of 337.4 kilograms per cubic meter; see Table 5-1. The connecting spoke has the same concentric cylinder structure, and its radii are the same as the other spokes of the station. The outer cylinder density is 337.4, and the inner density is 32.2. The shuttle bay, the connecting spoke, and the endcaps use the same cylinder model. The Iz MOI of the cylinders is computed using $I_z = \frac{1}{2}m(r_o^2 + r_i^2)$ for the z-axis. The cylinder and shuttle bay lengths are used to compute lengths for the parallel axis theorem, which the Ix and Iy MOIs use $I_x = I_y = \frac{1}{12}m(3(r_o^2 + r_i^2) + L^2) + md^2$.

### 7.2.2.4 Cylinder Main Floor

The station's habitable gravity range is designed to be between 0.95g and 1.05g. A cylinder station uses the ratio of the station radius over the main floor height equal to 10.5. The main floor is 190.5 meters above the outer rotating shell, with a minimum gravity of 0.95g. Figure 7-10 shows the cylinder radius is 2000 meters, and the radius of the main floor is 1810 meters. At 1% of the station radius, the thickness would be 20 meters. This matches the thickness limit of 20 meters. With other gravity ranges, shell thickness, and station sizes, the floor thickness ranges from a minimum of 2 meters to a maximum of 20 meters.

Using the cylinder length with the main floor radius and thickness, the mass of the main floor (cylinder) would be:

$$m_{main} = \rho_{fill}\, pi\, L\, ((r_m + t_m)^2 - r_m^2)$$

The main floor is modeled using the thick shell cylinder MOIs from Table 2-3. Those MOI equations are:

$$I_x = I_y = \frac{1}{12} m_{main}(3((r_m + t_m)^2 + r_m^2) + L^2)$$

$$I_z = \frac{1}{2} m_{main}((r_m + t_m)^2 + r_m^2)$$

#### 7.2.2.5 Cylinder Multiple Floors

The cylinder multiple-floors models were presented in *§7.1.3 Component Mass and Inertia Preview*. This portion of the station is modeled with a series of concentric cylinders. The multiple floors are 5 meters apart between the main floor and the outer rim.

Figure 7-10 shows the outer radius of a cylinder at 2000 meters. The radius of the top of the main floor is 1809 meters. Below the topsoil and floor structure of the main floor would be the top of the multiple floors at a height of 168 meters. There would be 34 floors spaced 5 meters apart below the bottom of the main floor. The radii range from the main floor at 1832 meters to the outer rim at 2000 meters.

The multiple-floors component is divided into many more cylinders to more accurately compute the volume, mass, and inertias of the multiple-floors component. Each of the cylinders is analyzed like the main floor. Each cylinder would have a mass of $m_{floor5} = \rho_{floor5}\, pi\, L\, (r_o^2 - r_i^2)$. The length of the cylinder is bound by the cylinder endcaps. In the Figure 7-10 example, the length of the floors is 2701 meters.

The series of concentric cylinders is used to compute the mass and MOIs for the multiple-floors component. The air density is excluded from the multiple-floors analysis because that air density is included in the air analysis. The MOI equation about the z-axis for the cylinder would be $I_z = \frac{1}{2} m(r_o^2 + r_i^2)$ and about the x (or y) axis, the MOI equation would be $I_x = I_y = \frac{1}{12} m(3(r_o^2 + r_i^2) + L^2)$. Those cylinder values are summed to determine the total mass and MOIs of the multiple-floors component.

#### 7.2.2.6 Cylinder Air

The cylinder air component is analyzed using the air mass and inertias of concentric cylinders; see Figure 7-7. These cylinders of air extend from the station rotation center to the outer rim. The mass and MOIs of the concentric cylinders are calculated in the same fashion as the multiple floors. The air component is typically subdivided into many cylinders for modeling. The thickness of the cylinders is set to ensure an accurate summation of the mass and MOIs with the varying air density.

Each cylinder of air would have a mass of $m_{air} = \rho_h\, pi\, L\, (r_o^2 - r_i^2)$. The air density decreases from sea level pressure at the outer rim to a minimum of Denver air pressure. In extremely large cylinders, airtight layers are introduced to provide habitable air pressure on gravity-limited top floors; see *§3.5 Top Floor Limits*. The length of the cylinder is bound by its endcaps. The air cylinder MOI equation about the z-axis for the cylinder would be $I_z = \frac{1}{2} m(r_o^2 + r_i^2)$ and about the x (or y) axis, the MOI equation would be $I_x = I_y = \frac{1}{12} m(3(r_o^2 + r_i^2) + L^2)$. In the Figure 7-10 example, the floor length is 2815 meters and the radii range from 0 to 2000 meters. Those mass and MOI values are summed to determine the air totals.

#### 7.2.2.7 Cylinder Preview

The masses and the MOIs of the components characterize the cylinder station. As a preview, Figure 7-11 shows the distribution of the component masses in two cylinder stations. One of the stations has a radius of 2000 meters, and the other has a radius of 20,000 meters. Both have the same six components. The outer cylindrical shell and endcaps are combined in the shell component of these pie charts. In these stations, the shell has a thickness of 20 meters. Multiple floors extend from the outer rim to the top floor. The floor on the outer rim has a gravity of 1.05g and sea-level air pressure. The top floor is at a height where the gravity is 0.95g or the air pressure is at least Denver air pressure.

Given these station limits, the charts in Figure 7-11 show the shell comprises at least half the station mass. An interesting change from the 2000-meter radius station to the 20,000-meter station is the increase in the mass of the multiple floors. The mass of the floors increases from 6.7% of the total to 40% of the total. The mass of the shell and the main floor both increase as a function of area; the mass of the multiple floors increases as a function of volume. The percentage of air increases from 1.6% to 6.1%. Like with the multiple floors, the size of the station interior is increasing as a function of volume. Other components are more fixed in size and represent a small percentage of the total volume. The spokes and shuttle bay masses increase with the station diameter; however, they are linear or fixed in size and become a smaller percentage of the total with the larger station.

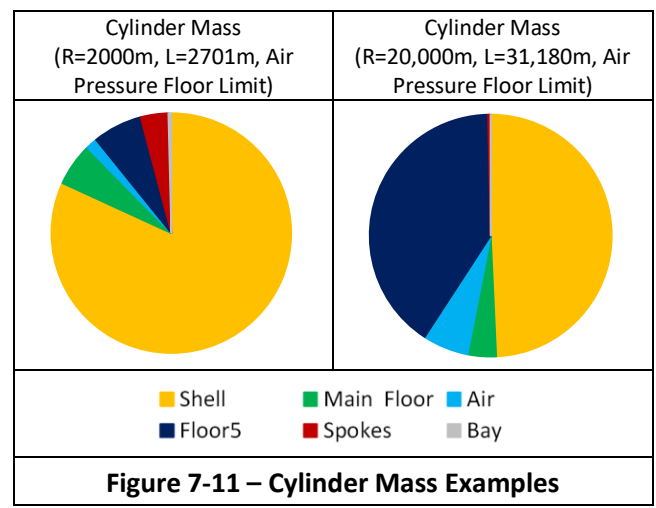

**Figure 7-11 – Cylinder Mass Examples**

### 7.2.3 Cylinder Station Mass Results

As an introduction, Figure 7-12 shows the masses of the components of 6 cylinder stations. The cylinder stations had radii of 2,000, 20,000, and 200,000 meters. Along the x-axis, the charts show the six major components of the cylinder. The y-axes of the charts show the values of the mass and range from 1 to 1e18 kilograms on a logarithmic scale. The component masses change with the increase in the radius. The shell and floor masses represent most of the system mass. Two masses are included for each radius to compare the changes with the top floor limited by gravity only and by air pressure. With the increase in the radius, the mass of the multiple floors increases significantly. Except for the multiple-floors component in the largest radius station, there is minimal change between the masses for the two top-floor limits. This is on a logarithmic scale; with greater detail, the masses of the gravity-limit components tend to be slightly greater than the air pressure top floor limit for nearly all the radii and components. With larger radii, more of the station is available to use with the multiple floors. The graphs show an increase in the amount of air in the station. The spokes and the shuttle bay do not increase significantly with the increasing radii. The relative amount of mass in the spokes and bay decreases significantly with the increase in the station radius.

Figure 7-13 includes two sets of charts. Both sets show the cylinder component masses varying with the cylinder radius. The left two charts show the masses with the top floor limited by gravity, and the right two charts have the top floor limited by air pressure. These cover the same 6 cylinder components. The stations are designed to be rotationally balanced, and the z-axis MOI (rotation axis) is 1.2 times the x-axis MOI. The thickness of the shell is 20 meters.

The graphs' lines of data represent the mass of the cylinder components. The x-axis shows the radius of the cylinder station, and ranges from 100 meters to 50,000 meters. The y-axis shows the mass from 1e6 to 1e16 kilograms. This view of the components clearly shows the relatively small masses of the shuttle bay, spokes, and air. The masses of the shuttle bay and spokes change slowly for the large radius sizes. There is little difference in the data between the gravity-limited and the air-pressure-limited top floor except for the multiple floors. The top-floor air pressure constraint reduces the growth of the multiple-floors mass beyond the radius of 16,000 meters. The charts also show the mass of the multiple floors being constrained by the air pressure limit on large stations.

The stacked bar charts use the same station characteristics and mass data. The stacked bar charts show the mass of the cylinder components as a percentage of the total cylinder station mass. The x-axis shows the radius of the cylinder station, and ranges from 200 meters to 50,000 meters. The stacked bar chart y-axes show mass percentages ranging from 0% to 100%. With the gravity limit on the top floor, the multiple floors become almost 70% of the total mass of the station. With the air pressure limit on the top floors, the multiple floors scale at the same rate as the air, shell, and main floor.

### 7.2.4 Cylinder Station Balance Results

Evaluating the station balance requires the total moments of inertia along the axes of the station. The last subsection provided the equations of the components. Both the mass and moments of inertia can be summed to create a total mass or inertia for the station. The mass is used to compute the moment of inertia (MOI). The MOIs are used to evaluate the rotational balance of the station.

Stability ($I_z \geq 1.2 I_x$) is found by varying the geometry dimensions. In the case of the cylinder, the L/R ratio (the length and radius of the cylinder) is changed. The MOIs for various ratios are computed to find the 1.2 criteria. The Newton-Raphson method (Excel Goal Seek) finds the L/R ratio that provides $I_z/I_x=1.2$. Those station geometries meet the rotational stability criteria.

Figure 7-14 shows the rotating cylinder station stability effect on the cylinder geometry. The chart uses the ratio of the radius and length on the linear y-axis in Figure 7-14 with a range from 0 to 2. The logarithmic x-axis of the chart shows the cylinder radius, and ranges from 100 meters to 500,000 meters. For the multiple component cylinder, the ratio L/R varies from 1.2 to more than 1.9 as the radius increases from R=200 to 500K meters. The extra mass in the multiple-floors component of the gravity-limited stations provides more stabilizing inertia and supports larger L/R ratios. The chart includes the thin and thick shell stability ratios for comparison. It includes a horizontal line at about L/R=1.3 for reference. Figure 7-14 shows that the ratio is smaller for small-radius cylinders and larger for large-radius cylinders. To maintain the passive rotational stability when including the multiple station components, smaller stations must reduce their length-to-radius ratio below the original thin and thick shell guidelines of 1.3. Larger stations can increase their length-to-radius ratio above the 1.3 thin-shell guideline.

Figure 7-15a shows the different changes in the constituent component stabilities. These changes help to understand the

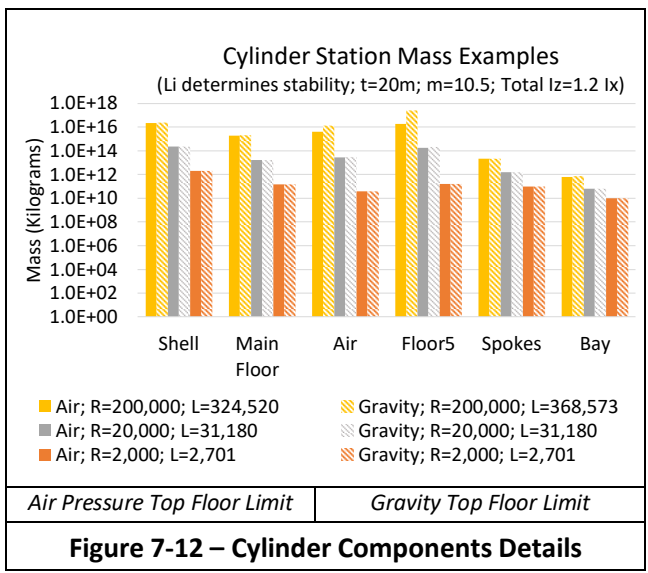

**Figure 7-12 – Cylinder Components Details**

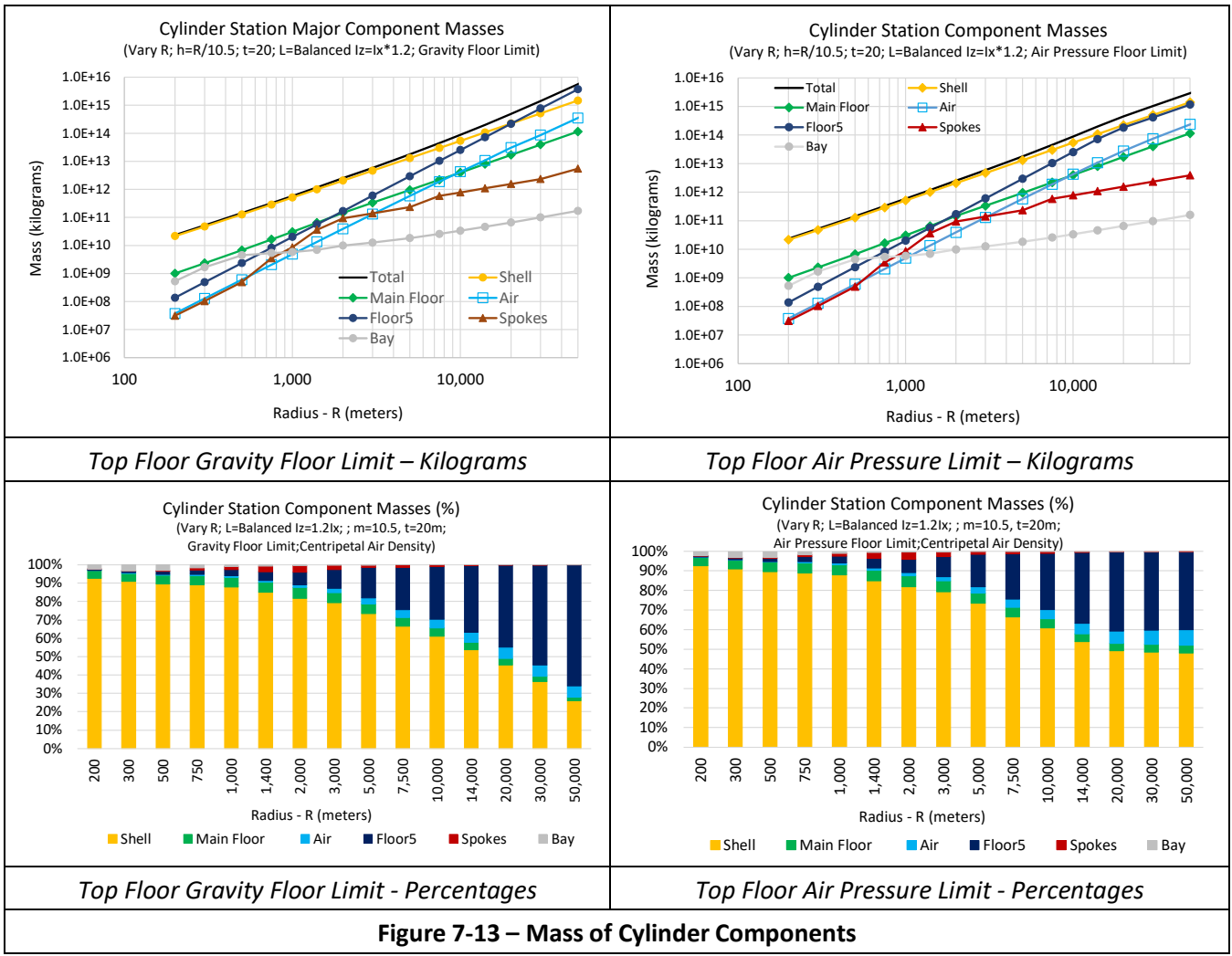

**Figure 7-13 – Mass of Cylinder Components**

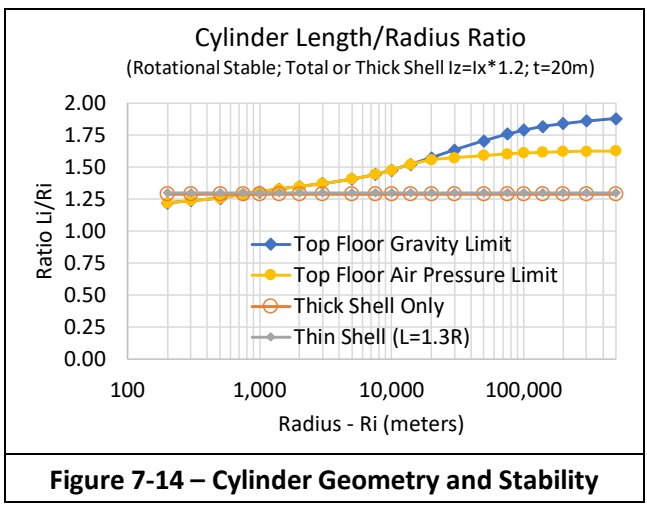

**Figure 7-14 – Cylinder Geometry and Stability**

effects of component changes on the station's stability. The spoke stability ratio drops from 2.0 to 1.5 as the cylinder radius increases from 200 to 1000 meters. The 200-meter cylinder only has a center set of spokes. Larger cylinders have sets of side spokes to provide more structure integrity. These offset spokes increase the Ix inertia with the parallel axis theorem. The spokes ratio Iz/Ix reduces to less than 1.0 with the offset long spokes. Instead of being a stabilizing component, the spokes begin to degrade the stability. Bay component stability ranges from 0.24 to near zero with the increasing radius. The two bays at the end of a long connecting spokes create a large Ix inertia while the Ix inertia remains relatively constant. The shuttle bay is also a destabilizing component. Outer shells greater than 1000 meters have a ratio of less than 1.3 and are also a destabilizing component. The main floor and the multiple-floors components provide stabilization over the entire range of station radii. The mass and geometry of the components determine their inertia. Heavier mass components create larger Ix and Iy. Fortunately, the destabilizing components are some of the lighter mass components.

To illustrate the heavy and light component masses, Figure 7-15b normalizes the component stability using the component mass ($M_{comp}$) and the total station mass ($M_{station}$). The y-axis values represent a form of stability and would be equal to Iz/Ix times $M_{comp}/M_{station}$. These results clearly show the minimal effect that most of the components have on the station stability. The shell stability dominates small station stabilities. The impact from the multiple floors increases and, for large stations, becomes greater than the shell stability. The top floor reaches a maximum height at a station radius of about 16,000 meters. After that height, all the stabilities remain fairly constant. The following paragraphs further investigate these component stability masses and their MOIs.

Figure 7-12 previously showed the masses for 6 cylinder stations. The cylinder stations had radii of 2,000, 20,000, and 200,000 meters. The chart also compared masses with the top floor height limited by air pressure and gravity. Figure 7-16 includes two sets of charts to show the station Iz (z-axis) and Ix (x-axis) MOIs for 3 cylinder station sizes. In the cylinder geometry, Ix is equal to Iy. The top chart in Figure 7-16 shows the MOI values on a logarithmic scale. The bottom three charts show the MOI values on a linear scale.

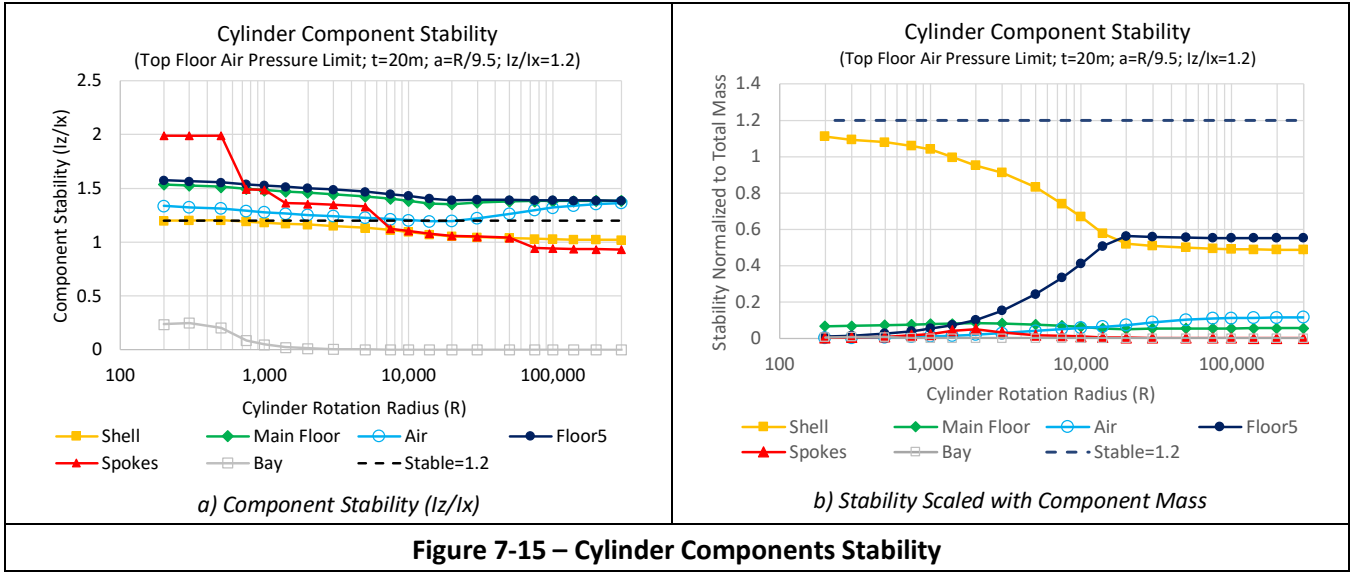

*a) Component Stability (Iz/Ix)*   *b) Stability Scaled with Component Mass*

**Figure 7-15 – Cylinder Components Stability**

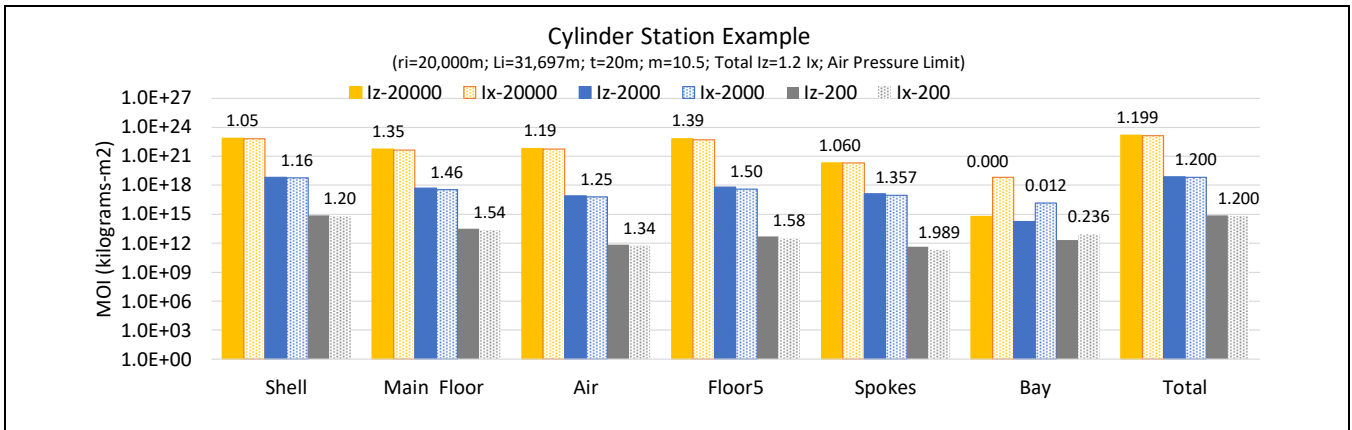

*Logarithmic Scale Comparisons*

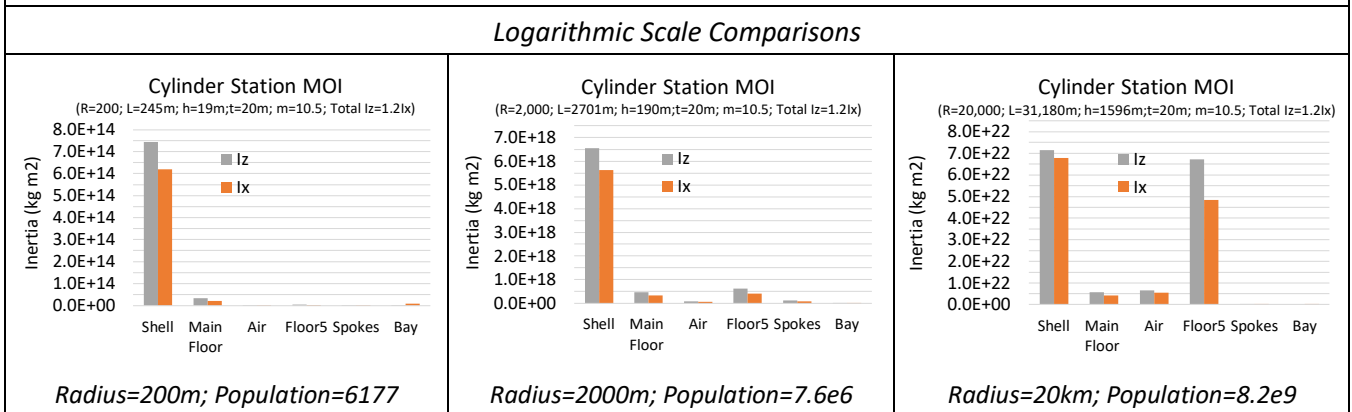

*Radius=200m; Population=6177*   *Radius=2000m; Population=7.6e6*   *Radius=20km; Population=8.2e9*

*Linear Scale Comparisons*

**Figure 7-16 – Cylinder Radius and MOI Stability**

The top column chart shows the Ix and Iz MOIs for the three radii and the six components. The cylinder radius dimensions are shown in the legend with different colored columns. These charts only show the top floor limited by air pressure. These charts help to understand the changing inertias. As expected, the components' moments of inertia (MOIs) vary differently with the changing radius. The charts include the cylinder components along the bottom axes. The y-axis shows the MOIs in kilogram meters squared on a logarithmic axis ranging from 1.0 to 1e27. The logarithmic y-axes support the comparison of the MOIs for all the components.

The bottom three column charts represent the MOIs for the three cylinder radii sizes. The y-axes of the charts vary on the linear scale and range to a maximum of 8e22 kilogram meters squared. The charts have different linear y-axes and show the absolute differences between the Ix and Iz MOIs are shown more clearly. Small MOIs appear near zero, while large MOIs appear near the maximum y-axes value.

Figure 7-16 shows an increase in the multiple-floors MOIs. There are few floors on the 200-meter radius station and many floors on the 20,000-meter radius station. The main floor, air, and multiple floors are all rotationally stable with Iz greater than Ix. The shell Iz and Ix ratio tends to be less than 1.2 and does not improve the station stability. The spokes and air components tend to have the smallest MOIs. The MOI values of the shuttle bay tend to be reversed compared to the other components (Ix greater than Iz) and negatively affect the rotational stability. This tends to unbalance small stations; however, with larger stations, the shuttle bay MOIs are much smaller than the other component MOIs.

Figure 7-17 shows an additional chart to evaluate the stability of the rotating cylinder station. This chart considers the effect of varying the thickness of the cylinder shell. This also influences the thickness of the endcaps, the spokes, and the main floor. The graph shows data for cylinder station radii ranging from 300 meters to 10,000 meters. The x-axes on these charts show the shell thickness ranging from near 0 to 200 meters. The left y-axis shows the ratio of the cylinder length over the cylinder radius, and ranges from 0 to 2. The length is determined to provide a stable system with Iz/Ix=1.2.

The chart includes the thin shell L/R ratio of 1.3 for reference. A length L of the cylinder is found to provide passive balance using Iz=1.2 Ix. For the cylinders, the length L ranges from 0.7R to 1.9R over the range of thickness and cylinder radii. Figure 7-17 also shows the thick shell stability ratio for a 2000-meter radius station for reference. The stability increases from L=1.29R to L=1.40R as the shell thickness increases. This is consistent with earlier results; see Figure 6-2b. The results differ with the multiple components compared to the thin and thick shell stabilities. Other station components bias the stability ratio.

With the thin shell dimensions, the station geometry ratio reflects only the effect of non-shell components. Small radius stations require a length-to-radius ratio of less than one to be rotationally balanced. Larger radius stations can be rotationally balanced with larger length-to-radius ratios. With very thick shells, the geometry ratio begins to decrease. In large stations, the multiple floors are the dominant mass and are on the outer edge of the rotating station. This improves the stability inertia ratios. The shell covers the entire station (from the outer edge to the top poles) and counters that improvement with very large thicknesses. With thinner shells, other components have more influence on the stability and support larger L/R ratios.

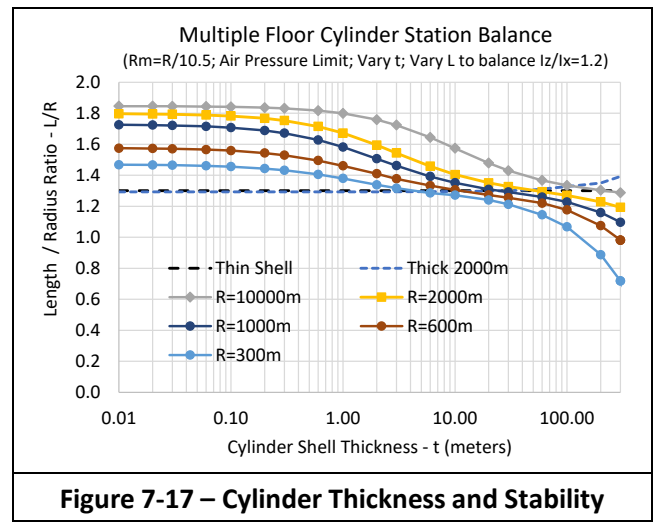

**Figure 7-17 – Cylinder Thickness and Stability**

## 7.3 Ellipsoid - Multiple Component Stability

This section considers multiple component stability for the oblate ellipsoid geometry. In the single floor section, the thin shells were stable (Iz=1.2 Ix) when a>1.225c, where c is the length of the polar or axial axis, and a is the length of the radial axis. Thick shells were stable when $a_o \geq 1.29 c_o$ over a wide range of radii and wall thicknesses. The length-to-radius ratio increased slightly with very thick shells. This increase implies the station can be proportionally larger yet remain stable with thicker shells. In this section, the stability analysis includes other station components. This provides a more realistic estimate of the ellipsoid station stability.

### 7.3.1 Ellipsoid Components

Figure 7-18 shows a cross-section of the elliptical station. This illustrates the station ellipsoid geometry with labels for components and densities. The view is on the x-axis (Front). The side view of the ellipsoid station is identical to the cylinder rendering in Figure 7-9b. The drawings show the major components of the ellipsoid, including the outer shell, main floor, multiple floors, center spokes, side spokes, and shuttle bay.

Figure 7-19 also shows a graph of the station cross-section. It is of a specific example ellipsoid station and shows more details, dimensions, and components. The graph shows the axial rotation axis on the z-axis, and this axis length is labeled c. The radial axes are on the x- and y-axis; their lengths are equal and labeled a and b. The vertical axis shows the vertical distance from the axis of rotation (z-axis) and ranges from 0 to 2200 meters. The horizontal axis shows the horizontal distance from the vertical axis (x- or y-axis) and ranges from 0 to 1800 meters. The drawing only shows one-quarter of the complete ellipsoid. A full rendering of this example ellipsoid would extend from -2000 to 2000 meters on the vertical axis. It would extend from -1654 to 1654 meters on the horizontal axis.

### 7.3.2 *Ellipsoid Mass and Inertia Equations*

The 6 components of the oblate ellipsoid and their mass equations were presented *§5.1 Station Component Equations*. These mass equations are used to develop the Moment of Inertia (MOI) equations for each of the major components of the ellipsoid. This analysis uses the material densities from Table 5-1. This section refines the common analysis from *§7.1.3 Component Mass and Inertia Preview* to apply to the components of the ellipsoid station. The following subsections update those mass and moments of inertia equations and provide specific examples for each of the major components of the ellipsoid.

#### 7.3.2.1 **Ellipsoid Shell**

Thick shell moment of inertia used equations from Table 5-2 in the single-floor stability analysis. That analysis reviewed the ellipsoid shell and its thickness scaled with the axes using a shell with a uniform thickness. The ellipsoid multiple components were evaluated using those equations, and the inertia analysis was validated using the multiple-piece approach.

The outer shell is represented by a series of hollow disks, and it is divided into 1000 disks. Figure 7-19 shows a cross-section of one disk at z=110 and y=2005. Figure 7-19 includes a magnified view of this disk cross-section in the upper right corner. This disk has a width of about 20 meters and a thickness of $c_o$/1000 meters. The mass of one hollow disk would be $m_{disk} = \rho_{fill} \pi t_d (r_o^2 - r_i^2)$. For this example, the disk's volume would be $\pi$ 1.67 ($2015^2-1995^2$), and with the shell density of 1721, the disk's mass would be 7.28e14 kilograms. The approach sums 1000 of these 20-meter-wide by 1.7-meter-thick hollow disks. Their inner radii ranges from 0 to 2000 meters. The summed mass of the shell was 7.73e11 kilograms and matches the mass value using a thick shell equation.

The station and the disks rotate about the z-axis. The moment of inertia for the disks about the z-axis would be $I_z = \frac{1}{2}m(r_o^2 + r_i^2)$ where ri and ro are the inner and outer radii of the ellipsoid shell. The moment of inertia for the disks about the Ix (or Iy) axis would be $I_x = I_y = \frac{1}{12}m(3(r_o^2 + r_i^2) + t^2)$ where t is the thickness of each disk. The parallel axis theorem adds the mass times the distance from the x-axis squared. The 1000 disks were summed and compared the result to the MOIs for the thick ellipsoid shells (see Table 5-2). In this case, the sums match the MOI values using those thick shell equations.

#### 7.3.2.2 **Ellipsoid Spokes**

This subsection uses the spoke models presented in *§7.1.3 Component Mass and Inertia Preview*. The spokes are modeled as thick cylinders. The ellipsoid shell binds the lengths of the spokes. At a distance, z, along the rotation axis, the length of the spoke would be $2a\sqrt{1 - z^2/c^2}$. In Figure 7-19, the length of the center spoke is about 4000 meters. The side spokes are about 3490 meters. The radius of the center spokes is 3% of the ellipsoid diameter on the axial axis. They have a maximum radius of 100 meters and a minimum radius of 10 meters. In this example, the center spoke has an outer radius of 99 meters and an inner radius of 94 meters. The side spokes are only 2% of the axial axis diameter and have an outer radius of 66 and an inner radius of 62 meters. The spokes are filled with multiple floors that are 5 meters apart. The inner floor density is set to 32.2 kilograms per cubic meter, and the outer spoke wall has a density of 337.4 kilograms per cubic meter; see Table 5-1. The mass of each cylinder is computed using $m_{spoke} = \rho \pi L (r_o^2 - r_i^2)$.

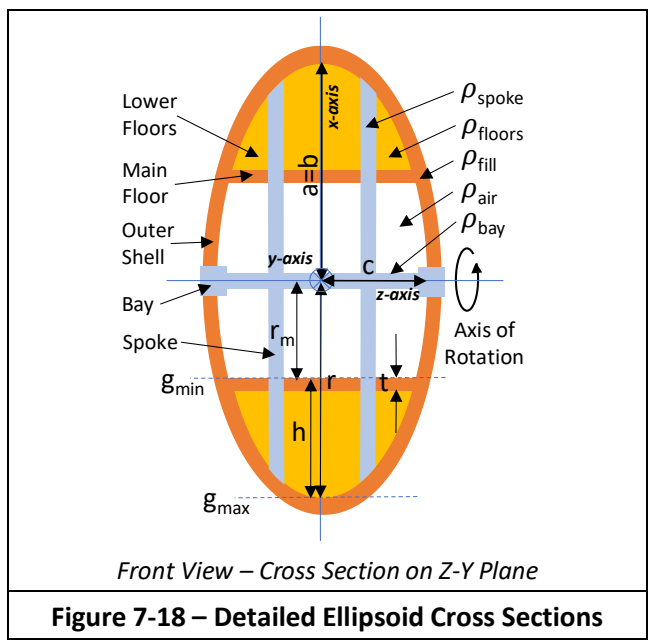

Front View – Cross Section on Z-Y Plane

**Figure 7-18 – Detailed Ellipsoid Cross Sections**

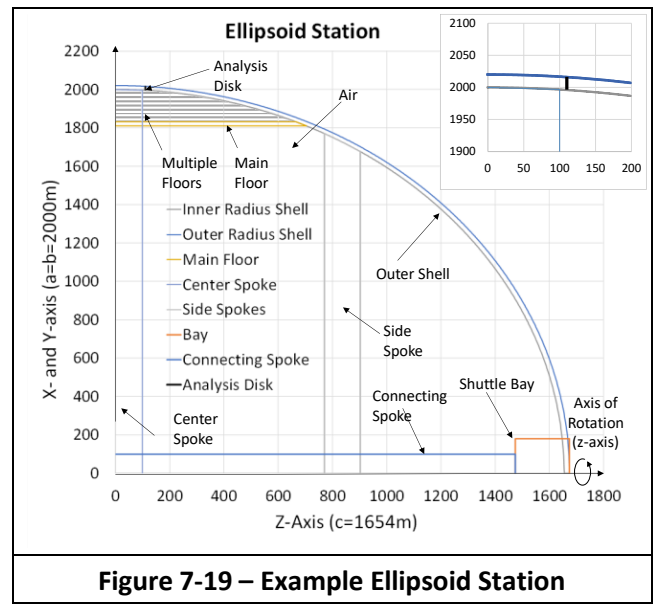

**Figure 7-19 – Example Ellipsoid Station**

The ellipsoid station in Figure 7-18 shows the center spoke and one of the two side spokes. The ellipsoid uses 4 sets of these spokes at equal 45-degree spacing like the cylinder station shown in Figure 7-9b. The MOIs are computed using equations developed in *§7.1.3*. For the four sets of spokes:

$$I_{z4spokes} = 4\left[\frac{1}{12}m\left(3r^2 + L^2\right)\right]$$

$$I_{x4spokes} = I_{y4spokes} = 4I_{x45} = \frac{1}{12}m(18r^2 + 2L^2)$$

The parallel axis theorem is also used for the spoke MOIs. Side spokes are offset from the x-axis and the y-axis by a distance of one-half the z-axis length of c. The equations add the spoke mass times the distance from their rotation axis squared. The $m_{spoke}$ times $(c/2)^2$ is added to the side spokes Ix and Iy. The mass and MOI values for the 3 sets of spokes are added to find the total spoke mass and MOIs. Larger stations use more side spokes, and their mass and inertia add to the station appropriately.

### 7.3.2.3 Ellipsoid Shuttle Bay

The shuttle bay is comprised of two parts: bays on the outer edge of the station, and a connecting spoke between the two bays. A cylinder with endcaps represents the shuttle bay. The thickness of the outer shell of the shuttle bay is the same as the station's outer shell but no larger than 20 meters thick. The endcaps use the same thickness as the shuttle bay shell. For the Figure 7-18 example, each shuttle bay cylinder has a radius of 180 meters and a length of 200 meters. The connecting spoke is modeled like the station center spokes. The length of the connecting spoke is 2948 meters long. Its interior is more like a spoke structure with additional walls. The connecting spoke has a filled outer cylinder that is 5 meters thick, and the interior cylinder is assumed to have multiple floors separated by 5 meters. Each of the shuttle bay cylinders would have a mass of $m_{bay} = \rho_{bay}\,\pi\,L\left(r_o^2 - r_i^2\right)$. The exterior cylinder has a density of 1721 kilograms per cubic meter, and the interior would have a density of 337.4 kilograms per cubic meter; see Table 5-1. The connecting spoke has the same inner and outer cylinder structures. Its radius is the same as the center spokes of the station. The spokes use an outer density of 337.4 and an inner density of 32.2. The mass and MOIs of the cylinders are computed using $I_z = \frac{1}{2}m(r_o^2 + r_i^2)$ for the z-axis and $I_x = I_y = \frac{1}{12}m(3(r_o^2 + r_i^2) + L^2)$ for the x- and y-axis.

### 7.3.2.4 Ellipsoid Main Floor

The main floor is modeled as a cylinder inside the ellipsoid shell. The cylinder rotates about the z-axis. The mass of a floor would be $m_{main} = \rho_{fill}\,pi\,L\left(r_o^2 - r_i^2\right)$. For the Figure 7-19 example, it would have a floor radius, $r_i$, of about 1810 meters. The floor-length would be computed using $2c\sqrt{1 - r^2/a^2}$. This example has a length of about 1369 meters and a thickness of 20 meters with 2 meters of soil. The volume of the main floor would be π 1459 (1832²-1810²) or 1.72e8 cubic meters. With a soil density of 1721 and floor density of 64.4, the mass computes to 7.38e10 kilograms. The MOI equation about the z-axis for the cylinder would be $I_z = \frac{1}{2}m(r_o^2 + r_i^2)$ and Iz computes to 2.43e17 kg-m². About the x (or y) axis, the MOI equation would be $I_x = I_y = \frac{1}{12}m(3(r_o^2 + r_i^2) + L^2)$, and Ix and Iy computes to 1.33e17 kg-m². The stability of the main floor would be 1.83.

### 7.3.2.5 Ellipsoid Multiple Floors

This component of the station is represented with a series of concentric cylinders. Each cylinder is analyzed like the main floor. The multiple floors are 5 meters apart between the main floor and the outer rim. The radii range from the main floor to the outer rim. The lengths of the cylinders are bound by the ellipsoid shell. The floor length is again computed using $2c\sqrt{1 - r^2/a^2}$. The mass and MOIs of each of the series of concentric cylinders are computed like the main floor. The analysis uses many more cylinders than floors to improve the modeling accuracy with the curved shell lengths. The MOI equation about the z-axis for the cylinder would be $I_z = \frac{1}{2}m(r_o^2 + r_i^2)$, and about the x (or y) axis the MOI equation would be $I_x = I_y = \frac{1}{12}m(3(r_o^2 + r_i^2) + L^2)$. The multiple floors are shown at the top of the example in Figure 7-19 The floor lengths range from zero to 1329 meters, and the radii range from 1831 to 2000 meters. The cylinder values are summed to determine this component's total mass and MOIs.

### 7.3.2.6 Ellipsoid Air

At the outer rim, the air density in the station is sea level. It decreases to lower densities with increasing height; see §7.1.3.6. In extremely large ellipsoids, airtight layers are introduced to provide habitable air pressure on gravity-limited top floors; see *§3.5 Top Floor Limits*.

In the same fashion as the multiple floors, the analysis uses concentric cylinders of air. The cylinders start at the center of rotation and continue to the station's outer edge; see Figure 7-7. Each cylinder of air would have a mass of $m_{air} = \rho_h\,\pi\,L\left(r_o^2 - r_i^2\right)$. The length of the cylinder is bound by the ellipsoid shell. Like the floor lengths, the air cylinder lengths are computed using $2c\sqrt{1 - r^2/a^2}$. As before, a and c are the lengths of the ellipsoid axes, and r is the distance to the center.

The mass and MOIs are computed for the series of concentric cylinders. The MOI equation about the z-axis for the cylinder would be $I_z = \frac{1}{2}m(r_o^2 + r_i^2)$ and about the x (or y) axis the MOI equation would be $I_x = I_y = \frac{1}{12}m(3(r_o^2 + r_i^2) + L^2)$. In the Figure 7-18 example, the radius of these air cylinders range from 0 to 2000 meters and their half lengths range from 0 to 1654 meters. The air cylinder mass and MOI values are summed over the entire station to determine the air totals. The air densities are excluded from the multiple floors and the bay to prevent duplication in the computations. The densities of those components would only include the masses of the structures, fill, and panels.

### 7.3.2.7 Ellipsoid Preview

The summed masses and MOIs of the components characterize the ellipsoid station. Figure 7-20 previews the distribution of the component masses in two ellipsoid stations. One of the stations has a rotation radius of 2000 meters, and the other has a radius of 20,000 meters. Both have the same six components. In these stations, the shell has a thickness of 20 meters. Multiple floors extend from the outer rim to the top floor. The top floor is at a height where the top floor gravity is 0.95g, and the top floor air pressure is at least Denver air pressure.

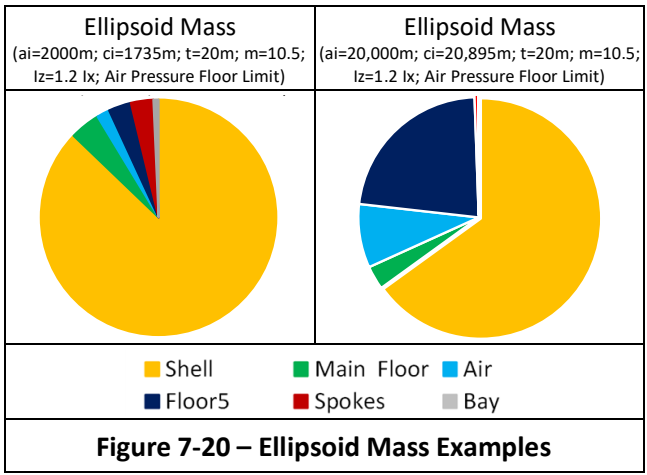

Figure 7-20 – Ellipsoid Mass Examples

Given these station limits, the charts in Figure 7-20 show the shell comprises more than half the station mass. The main floor is 20 meters thick and also represents much of the station mass. An interesting change from the 2000-meter radius station to the 20,000-meter station is the increase in the mass of the multiple floors. The mass of the floors increases from 3% of the total to 23% of the total. The mass of the shell and the main floor both increase as a function of area; the mass of the multiple floors increases as a function of volume. The percentage of air increases from 1.8% to 8.6%. Like with the multiple floors, the size of the station interior is increasing as a function of volume. Other components are more fixed in size and represent a small percentage of the total volume. The mass of the spokes and shuttle bay increases with the station radius; however, the increase is linear or nearly fixed in size. Their masses become even a smaller percentage of the total with the larger station.

### 7.3.3 Ellipsoid Station Mass Results

Figure 7-21 includes the same ellipsoid station constraints, characteristics, and densities as in Figure 7-20. It provides two sets of charts, one with the top floor limited by gravity and the other limited by air pressure. Both include data showing the same six components and the total mass. The graphed line and stacked bar charts present the same component masses. Each illustrates certain characteristics better than the other.

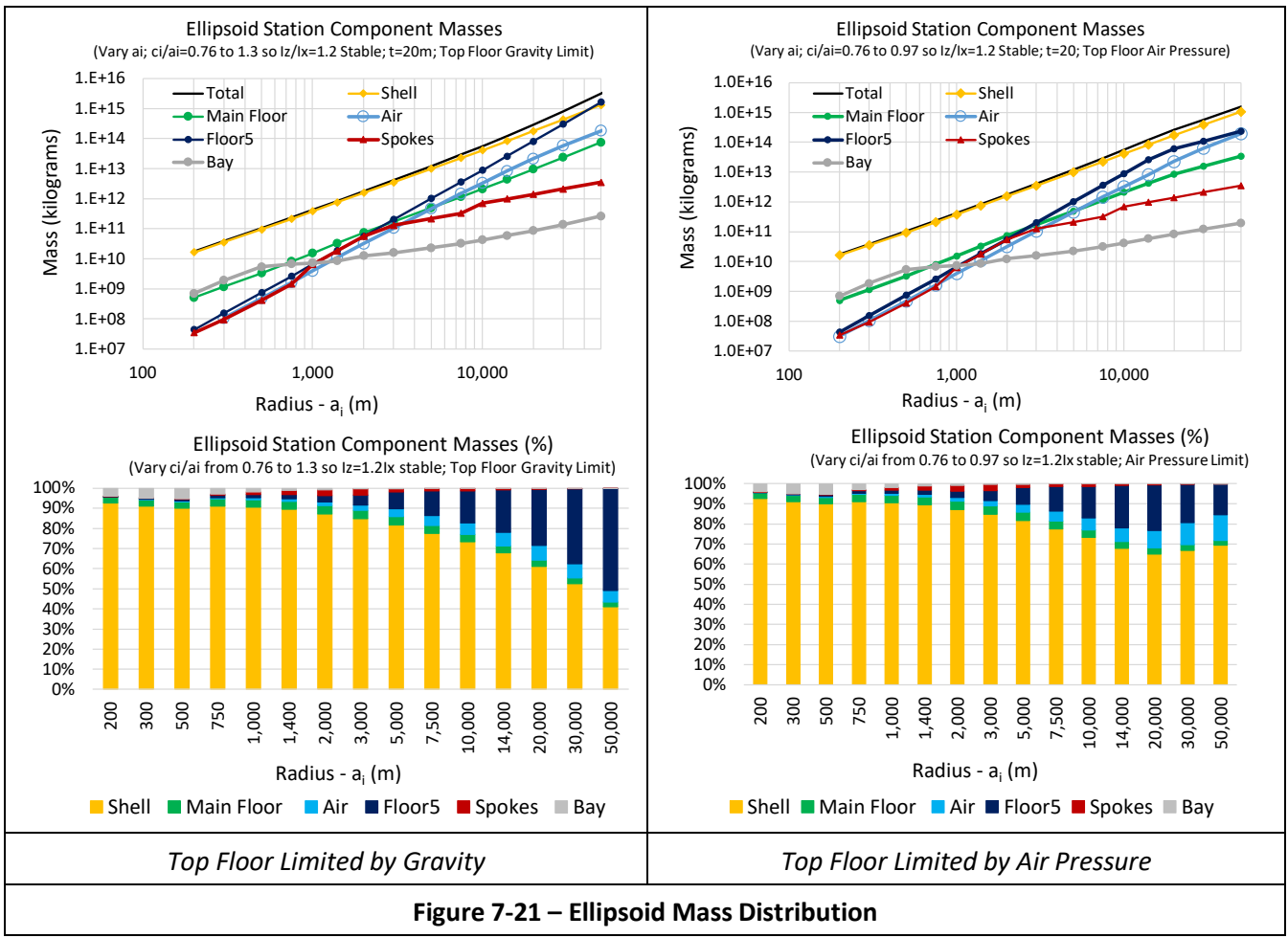

Figure 7-21 – Ellipsoid Mass Distribution

The line graphs show data as the mass of the ellipsoid components. The y-axes shows the mass from 1e7 to 1e16 kilograms on a logarithmic scale. The x-axes shows the radius of the ellipsoid station, and ranges from 100 meters to 50,000 meters. This view of the components clearly shows the slow-changing mass of the shuttle bay and the spokes. The graph also shows the mass of the multiple floors being constrained with the air pressure limit on large stations. With only the gravity limit on the top floor, the multiple floors almost become the total mass of the station. With the air pressure limit on the top floors, the multiple floors component scales at the same rate as the air, shell, and main floor.

Figure 7-21 also includes stacked bar charts as another view of the ellipsoid station component mass. These charts show the mass of the ellipsoid components as a percentage of the total cylinder station mass. The y-axes show the percentages ranging from 0 to 100%. The x-axes show the radius of the ellipsoid station, and range from 200 meters to 50,000 meters. The station is designed to be rotationally balanced, and the MOI on the z-axis (rotation axis) is 1.2 times the MOI on the x-axis. The thickness of the shell is 20 meters. These bar charts again show the decreasing influence of the shuttle bay and spokes. The shell mass dominates smaller stations. With the gravity limit, the mass of the multiple floors dominates large stations, while with the air pressure limit, the shell mass continues to dominate the large stations.

### 7.3.4 Ellipsoid Station Balance Results

Computing the total moments of inertia (MOIs) along the axes of the station is needed to evaluate the station's rotational balance. The equations from the last subsection are used to evaluate the components. Both the mass and moments of inertia of the components can be summed to create a total mass or inertia for the station. Mass is used to compute the MOIs. This section uses the MOIs to evaluate the rotational balance of the station. Stability ($I_z \geq 1.2 I_x$) is found by varying the geometry dimensions. In the case of the ellipsoid, the ratio between the axial and radial axes is varied. The analysis evaluates the MOIs for various ratios to find the 1.2 stability criteria.

Figure 7-22 provides two sets of graphs to investigate the moments of inertia (MOIs) for three ellipsoid stations. The descriptions of the Figure 7-22 ellipsoid charts are identical to the descriptions of the Figure 7-16 cylinder charts. Obviously, the data is different. Again, the top chart in Figure 7-22 shows the MOI values on a logarithmic scale. The bottom three charts show the MOI values on a linear scale.

The bottom three column charts represent the MOIs for the three ellipsoid radii sizes. The charts have different linear y-axes and show the absolute differences between the Ix and Iz MOIs are shown more clearly. Small MOIs appear near zero, while large MOIs appear near the maximum y-axis value.

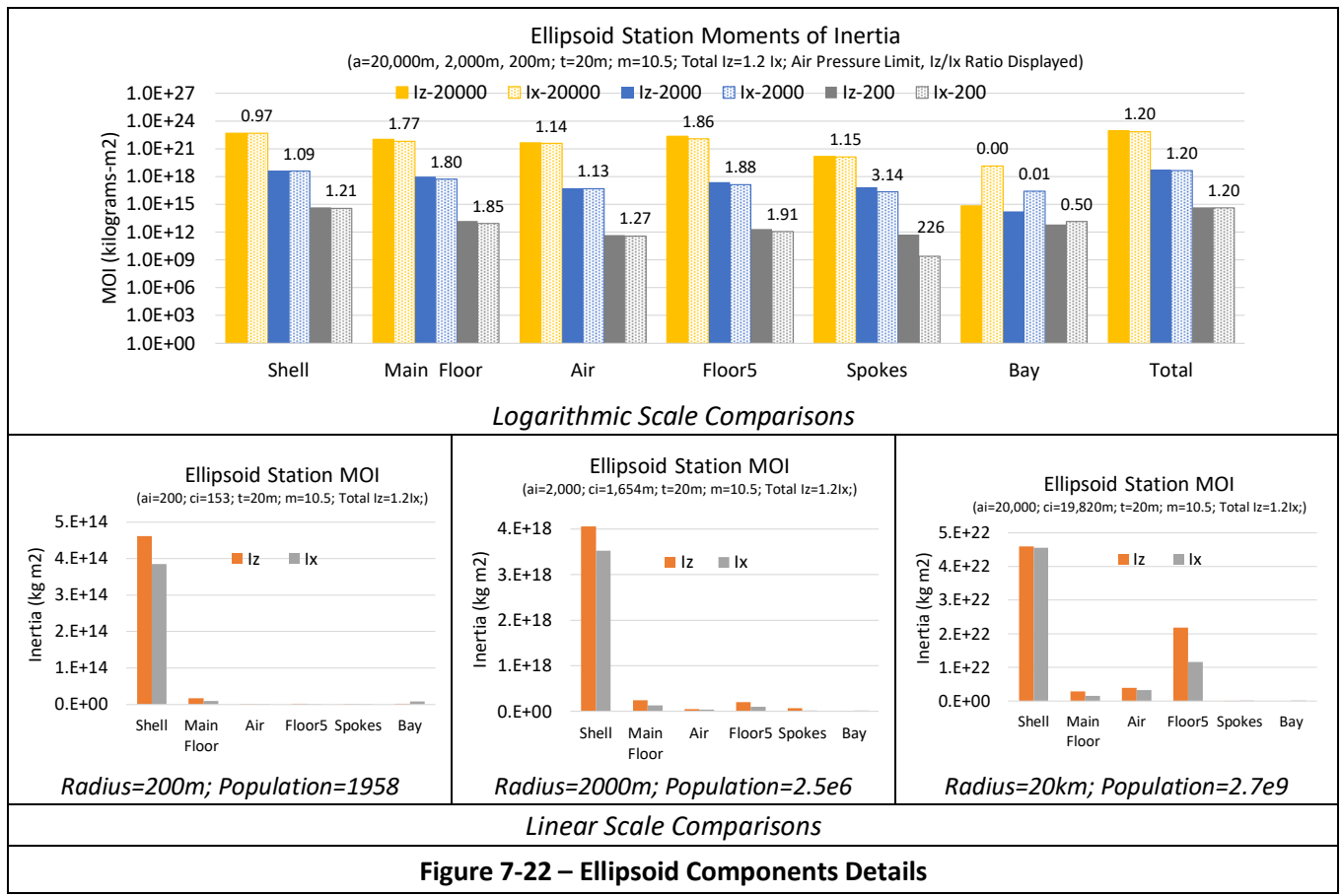

Figure 7-22 – Ellipsoid Components Details

As with the cylinder station, the moments of inertia (MOIs) of the ellipsoid components vary differently with the changing radius. Figure 7-22 shows an increase in the multiple-floors MOIs. There are few floors on the 200-meter radius station and many floors on the 20,000-meter radius station.

The main floor, air, spokes, and multiple floors are all rotationally stable with Iz greater than Ix. Only the small station has a rotationally stable shell. The larger shells are not rotationally stable (Iz<1.2Ix); however, the stability of the floors compensates for the shell instability.

The spokes and air components tend to have the smallest MOIs. The shuttle bay tends to have Ix greater than Iz, which negatively affects the station's rotational stability. This component unbalances small stations and reduces the stability ratio; however, with larger stations, the shuttle bay's MOIs are much smaller than the other component MOIs and do not influence stability.

Figure 7-23 shows the stability of the rotating ellipsoid station. The linear y-axis in Figure 7-23 shows the ratio of two ellipsoid axes lengths and ranges from 0 to 2.25. This geometry ratio changes with the radius to create a stable station. The logarithmic x-axis of the chart shows the ellipsoid radius a, and ranges from 100 meters to 500,000 meters. For the multiple component ellipsoid using the air pressure limit, the ratio c/a varies from 0.75 to 1.1 as the radius increases from R=200 to 500K meters. With the gravity top floor limit, the stability ratio varies from 0.75 to more than 2.25. Additional floors and mass on the outer rim improve the stability. The graph includes the thin and thick shell stability ratios for comparison to the entire station with components. Earlier analysis found the thin ellipsoid shell stability ratio was c/a=0.8165, and the thick shell stability was approximately 0.775. Figure 7-23 shows the stability ratio with multiple components is smaller for small-radius ellipsoids and larger for large-radius ellipsoids. To maintain the passive rotational stability when including the multiple station components, smaller stations must reduce their axes length ratio ($c_i/a_i$) below the thin or thick shell guidelines. Larger stations can increase the ratio of their axial axis (c) length to the radial axis (a) length above the thick or thin shell guidelines. It is intriguing that a spherical station can be rotationally stable with its internal components. Figure 7-23 shows that the ellipsoid can become a rotationally stable sphere at about a radius of 20,000 meters. Beyond that radius, a spherical station would have a stability ratio (Iz/Ix) greater than 1.2 and be rotationally stable. The large ratios imply proportionately greater surface area and population.

Figure 7-24 shows an additional chart to better understand the stability of the rotating ellipsoid station. This case considers the effect of varying the thickness of the ellipsoid shell. The station thickness also influences the thickness of the spokes, the main floor, and the shuttle bay. The graph shows data from a 2000-meter radius station. The x-axis shows the shell thickness ranging from near 0 to 200 meters. The left y-axis shows the ratio of the axial axis length (ci) over the radial axis length (ai) and ranges from 0 to 2. The polar axis length c is set to provide a stable system with Iz/Ix=1.2. The chart includes the thin and thick shell c/a ratios thin thick of 0.8165 and 0.775 for reference. The shell mass dominates the stability as the shell thickness increases. The chart includes the shell percentage of the total mass for reference. With thinner shells, other components have more influence on the stability. With our typical thickness of 20 meters, the a/c ratio is 0.827, and the 2000-meter radius station has a polar axis length of 1654 meters.

Thinner shells increase this ratio and the length of the polar axis. Thinner shells improve the stability of the shell and allow spherical stations with c=a and Iz/Ix>1.2. With a shell thickness of 1 meter, the c/a is 1.23, and the 2000-meter radius station has a polar axis of length 2,455 meters. This increases the volume and population. The population would increase from 2.5 million to 4.3 million. The thinner shell would use less material, decreasing the station mass from 1.77e12 to 3.85e11 kilograms. The thicker shells reduce this ratio and the length of the polar axis. This would have the opposite effect, decreasing the population and using more material.

## 7.4 Torus - Multiple Component Stability

The multiple-component stability analysis continues by considering the elliptical torus. The single-floor analysis found elliptical tori were stable when R>=6.33a and c=2a or c=3a. That analysis considered only the torus shell, and other components were not included. This subsection defines and analyzes the seven major components of the rotating torus space station. This section covers equations used to model the components' mass and inertia. Those equations are used to compute the station mass, the MOIs, and the station stability. This subsection concludes by presenting the station mass and stability balance results.

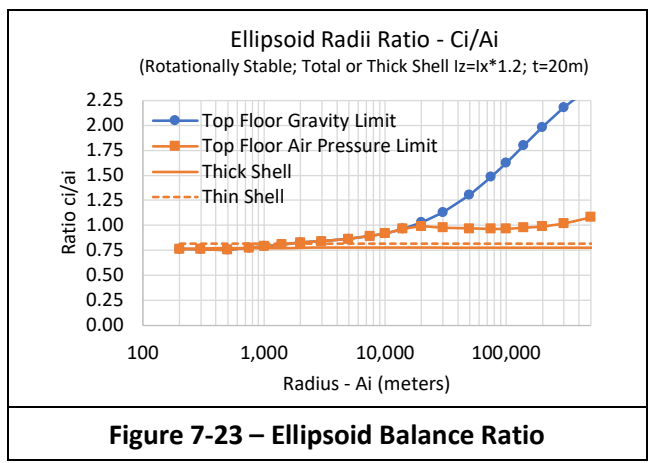

**Figure 7-23 – Ellipsoid Balance Ratio**

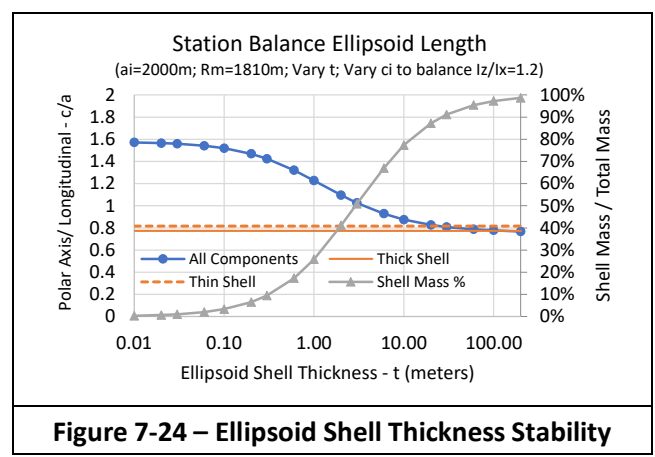

**Figure 7-24 – Ellipsoid Shell Thickness Stability**

*7.4.1   Torus Station Components*

Figure 7-25 includes a rendering using a cross-section of the elliptical station. This illustrates the station elliptical torus geometry with labels for components and densities. The view is along the y-axis. The drawing includes the outer shell, main floor, lower multiple floors, center spoke, side spokes, and shuttle bay. The torus station also includes eight elliptical-shaped cross-section dividers aligned with the spokes of the torus. These dividers would limit severe impact and air loss damage to 1/8 of the station. The drawing does not illustrate nor label the divider walls that separate the torus tube into multiple sections. They are omitted because they would obscure several components.

Figure 7-26 shows a cross-section drawing of a specific elliptical torus station example. The vertical axis (the station x- or y-axis) shows the distance from the axis of rotation (z-axis) and ranges from 0 to 2600 meters. The horizontal axis shows the distance from the vertical axis and ranges from 0 to 1000 meters. The drawing only shows one-quarter of the complete torus. A full rendering of this example torus would extend from -2300 to 2300 meters on the vertical axis. It would extend from -900 to 900 meters on the horizontal axis. The drawing labels the major components of the torus.

*7.4.2   Torus Station Mass and Inertia Equations*

The 7 components of the torus and their mass equations were presented in *§5.1 Station Component Equations*. These mass equations were used to develop the Moment of Inertia (MOI) equations for each of the major components of the elliptical torus. These components include the masses and volumes of the air, structures, fill, and panels. This analysis uses the material densities from Table 5-1. Summing the results from these individual components produces the total station mass and rotational moments of inertia. A ratio of the MOIs serves as the stability of the station. The following subsections update and refine the common analysis from *§7.1.3 Component Mass and Inertia Preview* for the components of the elliptical torus station.

#### 7.4.2.1   Elliptical Torus Shell

The station shell uses a uniform thickness. We could use torus volume, mass, and inertia equations to model the shell. As previous geometry subsections showed, the outer shell is modeled with a series of hollow disks. These disks are circular about the torus axis of rotation. The torus shell is divided into thousands of slices along that axis of rotation. Each slice has a thickness of $t_d$ equal to $c_o/1000$. Because this is a torus, two disks are used: one disk is above the major axis distance R, and one is below. The mass of the outer disk would be $m_{disk} = \rho_{fill}\, pi\, t_d\, ((R + r_o)^2 - (R + r_i)^2)$, where R is the major radius of the torus and $r_o$ and $r_i$ are the inner and outer edges of the disk. The mass of the inner disk would be $m_{disk} = \rho_{fill}\, pi\, t_d\, ((R - r_i)^2 - (R - r_o)^2)$. The summed mass of the two sets of disks was within 0.1% of the thick elliptical torus shell mass equation value.

The same hollow disks are used to determine the elliptical torus station inertia. The station and the disks rotate about the z-axis. Table 5-2 provides the MOIs for thick shell disks. The moment of inertia about the z-axis for the disks beyond the major radius R would be $I_z = \frac{1}{2}m((R + r_o)^2 + (R + r_i)^2)$. The moment of inertia for the disks about the Ix (or Iy) axis would be $I_x = I_y = \frac{1}{12}m\big(3((R + r_o)^2 + (R + r_i)^2) + t_d^2\big)$ plus the mass times the distance $R + (r_o + r_i)/2$ from the x-axis squared (the parallel axis theorem). The second set of disks inside the major radius would be $I_z = \frac{1}{2}m((R - r_o)^2 + (R - r_i)^2)$. Similarly, the MOI about the Ix (or Iy) axis would be $I_x = I_y = \frac{1}{12}m\big(3((R - r_o)^2 + (R - r_i)^2) + t_d^2\big)$ plus the mass times the distance $R - (r_o + r_i)/2$ from the x-axis squared (the parallel axis theorem). The MOIs from the thousands of pairs of hollow disks were summed. The sum was compared to the result of the MOI equations for the thick elliptical torus shells. The two results nearly exactly matched (within 0.1%).

#### 7.4.2.2   Elliptical Torus Spokes

The spokes are modeled as thick cylinders that extend from one outer edge of the torus tube, through the center of rotation, and to the other outer edge of the torus tube. For example, in Figure 7-26, the length of the center spoke is 4600 meters. Only half the spoke is shown in the drawing. The lengths of the spokes are bound by the elliptical torus shell. At a distance, z, along the rotation axis, the length of the side spoke would be $2\left(R + a\sqrt{1 - z^2/c^2}\right)$. The side spoke is located at 460 meters and is 4516 meters in length. The spokes are designed with an outer radius proportional (2%) to the station radius. For large stations, the spokes have a maximum outer radius of 100 meters. Smaller stations have spokes with a minimum radius of 10 meters. The spoke thickness is 5% of the spoke radius. The thickness has a maximum thickness of 50 meters.

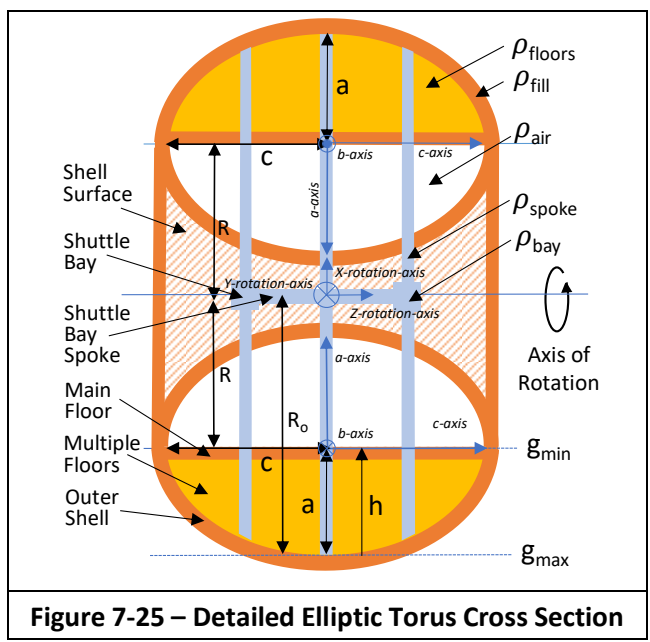

**Figure 7-25 – Detailed Elliptic Torus Cross Section**

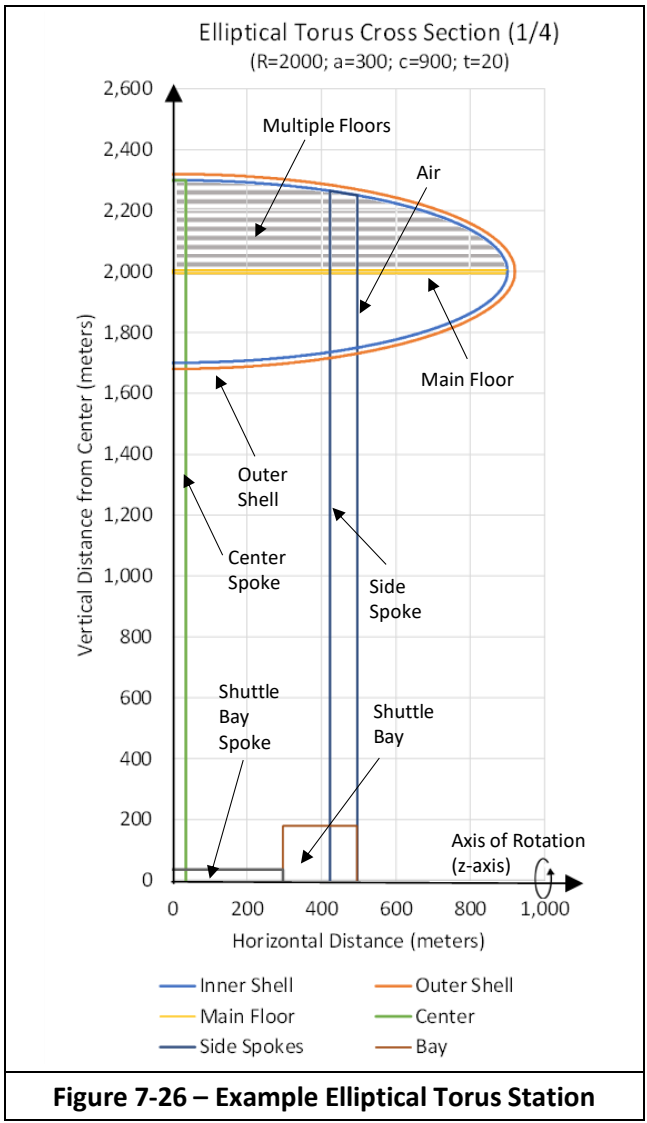

**Figure 7-26 – Example Elliptical Torus Station**

The spokes are modeled using two concentric cylinders. The spokes are filled with multiple floors that are 5 meters apart. The inner cylinder has a dense structure density of 64.4 kilograms per cubic meter, and the outer cylinder has a density of 337.4 kilograms per cubic meter; see Table 5-1. The spokes are not inhabited and do not require a thick-filled shell. The spokes provide structural strength and use a space truss that doubles the floor density. Each spoke cylinder would have a mass of $m_{spoke} = \pi L \left( \rho_{spoke} \left( r_o^2 - r_i^2 \right) + \rho_{structure}\, r_i^2 \right)$.

The elliptical torus station in Figure 7-25 shows the center spoke and two side spokes. Like the cylinder, it uses 4 sets of spokes at equal 45-degree spacing, as shown in Figure 7-1. The MOIs are computed from equations developed in *§7.1.3* for the four sets of spokes:

$$I_{z4spokes} = 4 \left[ \frac{1}{12} m \left( 3r^2 + L^2 \right) \right]$$

$$I_{x4spokes} = I_{y4spokes} = 4 I_{x45} = \frac{1}{12} m (18r^2 + 2 L^2)$$

The spokes also require the parallel axis theorem for their MOI. Side spokes are offset from the x-axis and the y-axis by a distance of one-half the z-axis semi-major axis length c. The spoke mass times the distance from their rotation axis squared is added to the MOIs. The mass $m_{spoke}$ times $(c/2)^2$ is added to the side spokes inertias. The mass and MOI values are summed for the 3 sets of spokes to compute the total spoke mass and MOIs.

The masses of the spokes in the Figure 7-26 example sum to 7.0e9 kilograms. The Iz MOI sums to 1.23e16, and the Ix MOI sums to 6.17e15. The spokes stability balance metric of this example is 1.76 and much better than the minimum 1.2 stability metric.

7.4.2.3 **Elliptical Torus Shuttle Bay**

As with all our geometries, the torus station includes a shuttle bay along its axis of rotation. We considered extending the entire shuttle bay along the rotation axis to the width of the torus (2c in Figure 7-25). This low-gravity region is not suitable for long-term habitation. The extended shuttle bay would not increase the population and would not increase the accessibility for shuttles. Such long bays require significant amounts of material, and that mass is better used to increase the torus radii and population. Instead of a full radius size bay across the entire torus, the design uses a spoke to connect two shuttle bays at opposite sides of the torus; see Figure 7-3 or Figure 7-25.

Each of the shuttle bays has a diameter of 360 meters and a length of 200 meters; see Figure 7-26. The distance between the two bays increases with large stations. As an example, our example has a 593-meter-long connecting spoke. The mass and inertia equations are from §7.1.3.3. The masses of the shuttle bay and connecting spoke in the Figure 7-26 example sum to 3.46e10 kilograms. Their Iz MOI sums to 5.52e14, and Ix MOI sums to 5.99e15. The stability balance metric is 0.1 and is worse than our minimum of 1.2 stability metric. This geometry and inertia of the shuttle bay decrease the desired station stability; however, in most stations, the other components more than compensate for that negative impact.

7.4.2.4 **Elliptical Torus Main Floor**

The main floor is modeled as a cylinder inside the elliptical torus shell. The torus is designed to provide a habitable gravity range between the torus major radius and the outer rim; see *§3.1 Gravity Limits*. The main floor is located at the major radius, at the center of the torus tube.

The main floor is again a cylinder. The main floor mass in the elliptical torus station would be $m_{main} = \rho_{fill}\, pi\, 2c\, ((R + t)^2 - R^2)$. For the Figure 7-26 example, it would have a radius of 2000 meters, a length of 1800 meters, and a floor thickness of 5 meters. The volume of the main floors would be $\pi\, 1800\, (2005^2 - 2000^2)$ or 5.66e7 cubic meters. With a floor density of 64.4 and a soil density of 1791, the mass would be 8.23e10 kilograms.

The MOI equation about the z-axis for the cylinder representing the floor would be $I_z = \frac{1}{2}m(r_o^2 + r_i^2)$ and the MOI Iz would be 3.30e17 kg-m$^2$. About the x (or y) axis the MOI equation would be $I_x = I_y = \frac{1}{12}m(3(r_o^2 + r_i^2) + 4c^2)$, and the MOIs Ix and Iy would be 1.57e17 kg-m$^2$. Considering only the floor, the stability metric would be Iz/Ix or 1.993. The floor by itself is rotationally stable given the 1.2 metric from [Brown 2002] and [Globus et al. 2007].

### 7.4.2.5 Elliptical Torus Multiple Floors

Like the previous geometries, the floors are modeled with a series of concentric cylinders about the rotation z-axis. The multiple floors are 5 meters apart between the main floor and the outer rim. Unlike the cylinder geometry example, the lengths of the cylinders in the torus vary with their radius. They are analyzed more like the multiple floors in the ellipsoid station. The cylinder lengths are bound by the elliptic torus shell, and the radii range from the top floor to the outer rim. The floor length is again computed using $2c\sqrt{1 - h^2/a^2}$, where c is the axial axis length, a is the radial semi-minor axis length, and h is the height below the semi-major radial axis R. In the Figure 7-26 example, the lengths range from zero to about 1800 meters, the radii range from 2000 to 2300 meters, and the heights range from 0 to 300 meters.

The mass and MOIs are computed for the series of concentric cylinders. Each cylinder would have a mass of $m_{floor5} = \rho_{floor5} \, pi \, L \, (r_o^2 - r_i^2)$. The density of the multiple floors is from Table 5-1. The mass from the height-varying air density is excluded from the floor density because it is included in the air component analysis. The MOI equation about the z-axis for each of the cylinders would be $I_z = \frac{1}{2}m(r_o^2 + r_i^2)$ and about the x (or y) axis the MOI equation would be $I_x = I_y = \frac{1}{12}m(3(r_o^2 + r_i^2) + L^2)$. The r$_o$ and r$_i$ are varying from R to R+a. Those values are summed to determine the total mass and MOIs of the multiple-floors component. The total mass of the concentric cylinders is within 0.1% of the mass using a closed outer half elliptic torus formula ($V_{outer\ half} = \pi^2 Rac + \frac{4}{3}\pi a^2 c$). In the example Figure 7-26 torus, the multiple cylinders' MOIs about the x-axis sum to 7.9e17 kg-m$^2$, and about the z-axis they sum to 7.77e17 kg-m$^2$. The stability of the multiple-floors Iz/Ix would be equal to 1.84 and greater than the minimum stability limit of 1.2.

### 7.4.2.6 Elliptical Torus Air

The density of the air is pressurized to sea level at the outer rim. It decreases to lower densities with increasing height. The air pressure equations were described in *§3.2 Air Pressure Limits*. The outer rim typically has a sea-level air density of 1.225 kg/m3 and a maximum gravity of 1.05g. In extremely large toruses, airtight layers are introduced to provide habitable air pressure on gravity-limited top floors; see *§3.5 Top Floor Limits*.

The mass and MOIs are computed for the series of concentric cylinders of air inside the torus tube. Each air cylinder would have a mass of $m_{air} = \rho_h \, pi \, L \, (r_o^2 - r_i^2)$. The length of the cylinder is bound by the elliptical torus shell. In the same fashion as the multiple floors, the analysis uses concentric cylinders of air. The cylinders of air start at the outer edge and continue to the inner edge of the torus shell; see Figure 7-26. Their radii range from R-a to R+a. The masses of the concentric cylinders of air sum to 1.26e10 kilograms in the example torus in Figure 7-26. This is less than 1/100 the mass of the outer shell with 20 meter thickness.

The MOI equation about the z-axis for the air cylinders would be $I_z = \frac{1}{2}m(r_o^2 + r_i^2)$ and about the x (or y) axis the MOI equation would be $I_x = I_y = \frac{1}{12}m(3(r_o^2 + r_i^2) + L^2)$. In the Figure 7-26 example, the half lengths range from 0 to 900 meters, and the radii range from 1700 to 2300 meters. Those mass and MOI values are summed to determine the air totals. For our example station in Figure 7-26, the Iz MOI sums to 5.15e16 kg-m$^2$ for the air in the torus. The Ix MOI sums to 2.83e16. The Iz/Ix stability ratio is 1.82 and more stable than our minimum of 1.2 metric.

### 7.4.2.7 Elliptical Torus Divider

Figure 7-8 shows a diagram of torus dividers. One of the eight dividers would be across the elliptical torus tube in Figure 7-25. The divider is not shown because it would obscure the other details. The dividers rotate about the z-axis at a distance of R. The divider provides an airtight seal between sections of the torus. The dividers enclose the spokes; as such, their thickness is about the diameter of the spokes.

A divider in the torus is modeled as two filled elliptical disks with a structure of trusses between them. It also includes multiple floors on the dividers overlooking the open space of the torus; see Figure 2-2. These divider structures are represented as disks with different densities and thicknesses. The mass of the interior divider structure, m$_{div}$, would be the area of the divider ellipse times the structure thickness, t$_{div}$ (twice the radius of the spoke $r_{i\text{-}spoke}$). Conceptually, the mass $m_{div} = \rho \, t_{div} \, \pi \, a_i \, c_i$.

The divider MOI equations were developed in *§7.1.3.7*. The MOIs of each divider are:

$$I_{x\alpha} = \frac{1}{12}m_{div}(3a_i^2 \sin^2 \alpha + 3c_i^2 + t_{div}^2 \cos^2 \alpha) + m_{div}R^2\sin^2(\alpha)$$

$$I_{y\alpha} = \frac{1}{12}m_{div}(3a_i^2 \cos^2 \alpha + 3c_i^2 + t_{div}^2 \sin^2 \alpha) + m_{div}R^2\cos^2(\alpha)$$

$$I_{zdiv} = \frac{1}{12}m_{div}(3a_i^2 + t_{div}^2) + m_{div}R^2$$

The position of the 8 dividers match the spokes in the torus. To match their positions, the angle α varies from 0 to 315 degrees at increments of 45 degrees. The MOI equations for the eight dividers are:

$$I_{xdiv8} = I_{ydiv8} = \frac{1}{3}m_{div}[3a_i^2 + 6c_i^2 + t_{div}^2 + 12\,R^2]$$

$$I_{zdiv8} = \frac{2}{3}m_{div}[3a_i^2 + t_{div}^2 + 12R^2]$$

In the example torus of Figure 7-26, the mass of the eight dividers is 5.86e10 kilograms. The Iz MOI sums to 2.36e17

kg-m$^2$, and the Ix MOI sums to 1.30e17 kg-m$^2$. The divider stability balance metric is 1.8. This is more stable than our minimum of 1.2 stability metric.

#### 7.4.2.8 Elliptical Torus Preview

The masses and the MOIs of the components are summed to characterize the elliptical torus station. As a preview, Figure 7-27 shows three pie charts illustrating the distribution of the component masses in three elliptical torus stations. The stations have radii of 200, 2000, and 20000 meters. All three have the same seven components. The torus station includes the divider component unlike the other geometries. In these stations, the shell has a thickness of 20 meters. Multiple floors extend from the outer rim to the top floor. The floor on the outer rim has a gravity of 1.05g and sea-level air pressure. The top floor is at a height where the top floor gravity is at least 0.95g, and the air pressure is higher or equal to Denver air pressure.

Given these station limits, the charts in Figure 7-27 show the shell comprises more than half of those station masses. An interesting change from the 2000-meter radius station to the 20,000-meter station is the increase in the mass of the multiple floors. With the increasing radius, the mass of the multiple floors increases faster than the shell. The mass of the floors increases from 5.4% of the total to 28% of the total. The mass of the shell and the main floor both increase as a function of area; the mass of the multiple floors increases as a function of volume. The percentage of air increases from 0.4% to 2.4%. Like with the multiple floors, the size of the station interior is increasing as a function of volume. The masses of the air, spokes, and dividers are much smaller than the shell. Those components have less effect on the construction mass and the overall station stability. In the smallest station, the bay represents nearly ¼ of the station mass. In the larger stations, the bay mass becomes insignificant. The spokes and shuttle bay masses increase with the station diameter; however, their dimensions change linearly or are fixed in size. Their masses become a smaller percentage of the total with the larger station. The following subsections delve into more detail and explanation.

### 7.4.3 Torus Station Mass Results

Figure 7-28 is included to better illustrate the change in the component masses. The radius increases by a factor of 10 for the 2000 and 20000 meter radius stations. Obviously, the larger torus station has larger and heavier components. The x-axis of the chart shows the major components of the elliptical torus, and the two chart columns represent the 2,000-meter radius station and the 20,000-meter radius station. The left y-axis shows the mass of the components, and ranges from 1 to 1e18 kilograms on a logarithmic scale. The right y-axis shows the ratio of the component masses and ranges from 1e-4 to 1e5.

Figure 7-28 also presents the ratio of the station masses using the components from the 2,000-meter station over the 20,000-meter station. The shuttle bays in both stations are the same size and mass, but the 20,000-meter station has a much longer connecting spoke; as such, the ratio is 1.2. The shell and main floor both increase in size by about 100. Their mass is a quadratic (area) relationship to the radius. The air and multiple floors (floor5) increase by factors of 811 and 700. Their masses have a cubic relationship (volume) with the radius. The spoke's length and radius both increase linearly with the station radius. Their scaling is quadratic, and their ratio is 242. The mass of the total station with the gravity limit is 134.7 times the mass with the air pressure top floor limit.

Figure 7-29 reviews the component mass changes for a wider range of station radii. Figure 7-29a has a stacked bar chart showing the torus station component masses. The same seven components are included. This chart shows the mass of the torus components as a percentage of the total torus station mass. The y-axis shows the percentages ranging from 0 to 100%. The x-axis shows the major radius of the torus

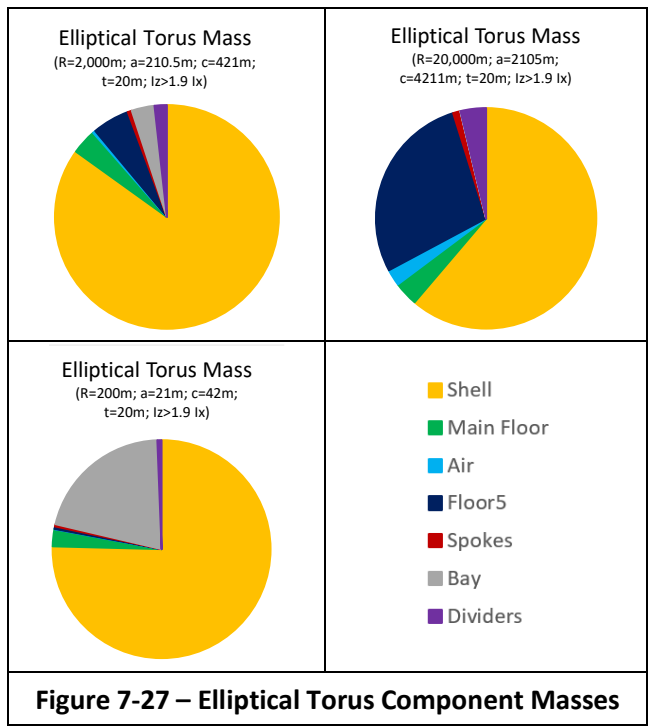

**Figure 7-27 – Elliptical Torus Component Masses**

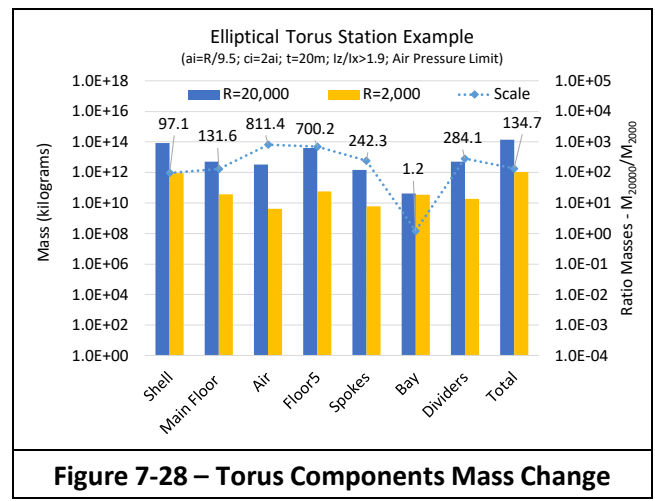

**Figure 7-28 – Torus Components Mass Change**

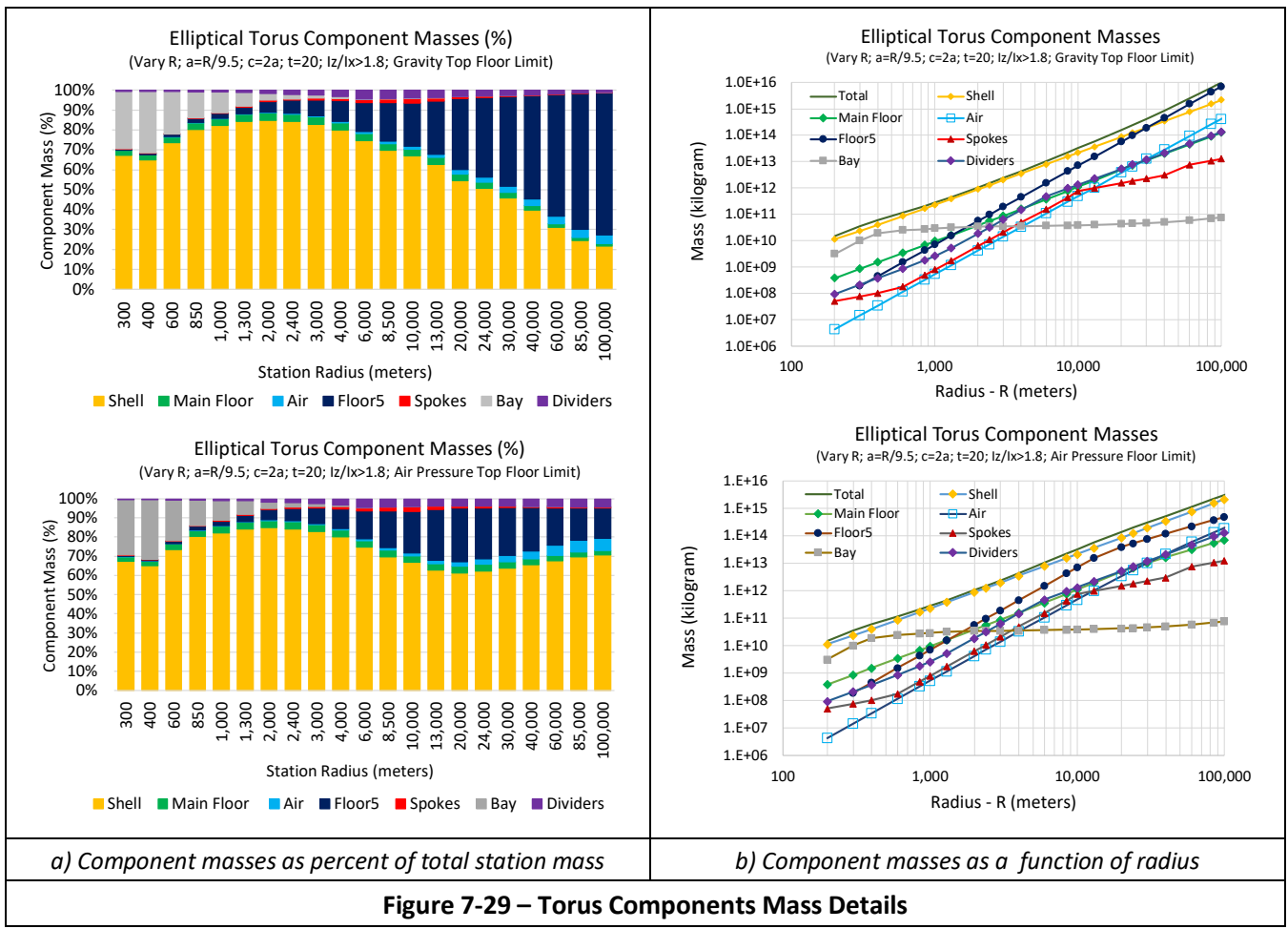

*a) Component masses as percent of total station mass*

*b) Component masses as a function of radius*

**Figure 7-29 – Torus Components Mass Details**

station, and ranges from 300 meters to 100 thousand meters. The station is designed to be rotationally balanced, and the MOI on the z-axis (rotation axis) is 1.2 times the MOI on the x-axis. The thickness of the shell is 20 meters. This chart shows that for small stations, the fixed size of the shuttle bay is a large percentage of the total mass. As the radius increases, the shell and main floor both increase in size and mass to become much more of the station mass. The interior of the torus becomes larger with the larger radius. The volume available for the multiple floors increases, and the chart shows those floors become much more of the station mass. The mass of the spokes becomes negligible with the increasing station radius. The small density of the air and its centripetal gathering along the outer rim reduces the air mass growth. The air mass is barely shown on the bar chart in Figure 7-29a.

An alternative view of the same mass data is shown in Figure 7-29b. This figure contains two graphs showing the mass of the components as a function of the torus radius. The y-axes of the charts show mass ranging from 1e6 to 1e16 kilograms on a logarithmic scale. The x-axes of the charts show the elliptical torus major radius ranging from 100 to 100 thousand meters on a logarithmic scale. Both graphs show data from tori with multiple floors, and the main floor height is limited by gravity or air pressure. The graphs have lines representing the total mass and the individual masses of the seven major components of the elliptical torus. Again, this is the same data as presented in Figure 7-29a. The lower graph uses the air pressure top floor limit constraint. The total mass of the largest station is about 3 times greater with the top floor at the gravity limit (9.8e15kg over 3.0e15kg). The shell mass remains the same between the two charts. On larger stations, the top floor drops from the center of the torus tube to a lower height where the air pressure is habitable. On the largest station with a radius of 100,000 meters, the center of the tube has an air density of 0.35 kg/m3, and the inner rim of the tube has only 0.115 kg/m3. This highlights the need for airtight layers. The top floor is at a height of 1543 meters and has the Denver air pressure (83,728 Pascals) and air density (1.01 kg/m3). The top floor height (and the number of floors) is almost 7 times greater without the air pressure constraint. Similarly, the multiple-floors mass on the largest station is 14.5 times greater without the air pressure constraint. With the air pressure limit, the top main floor has a greater radius and a larger circumference, but the top floor width is smaller in the elliptical cross-section. Overall, the main floor area (and mass) is about 1.8 times greater without the air pressure limit. There are many interesting observations from these charts. Overall, it shows that the air pressure limit has minimal effect on the design and mass until the station radius is greater than 15,000 meters. With larger stations, the air pressure limit significantly reduces the number of multiple floors and the station mass.

Figure 7-30 is included for completeness and shows the effect of the shell thickness on the station's mass. It shows the stacked bar chart with the torus station's percentage of component masses. It also includes a line chart with the component masses. The same seven torus components are included.

The stacked bar chart shows the mass of the torus components as a percentage of the total torus station mass. The torus has a major radius of R=20,000 meters, and the minor axes are a=2105 meters and c=4211 meters (a=R/9.5 and c=2a). The y-axis shows the percentages ranging from 0 to 100%. The x-axis shows the station shell thickness, and ranges from 0.01 meters to 500 meters. The mass values are included in the right chart of Figure 7-30. That graph shows the mass in kilograms ranging from 1e9 to 1e16 on the logarithmic y-axis. The x-axis shows the shell thickness on a logarithmic scale ranging from 0.01 to 1000 meters.

As one would expect, the mass of the shell increases with the thicker shell. For very thin shells, the multiple floors dominate the station's mass. The volume and mass of the air (torus interior), the multiple floors, and the shuttle bay do not change with the changing shell thickness. Their percent contribution to the total mass decreases with the increasing shell thickness. The thickness of the main floor, the dividers, and the spokes increase with the shell thickness (between a defined minimum and maximum thickness) to provide more structural support. The spokes and dividers percentages are much less than the other components and are barely visible in Figure 7-30We advocate a fairly thick shell to protect the station from space debris. For thicknesses greater than 10 meters, the shell mass dominates the total station mass.

The right chart in Figure 7-30 shows the component masses for the same range of station shell thicknesses. This chart provides details on the scaling of the main floor, spokes, and dividers with the changing shell thickness. This scaling of the structural design needs additional material strength analysis and finite element analysis. We base our current strength and scaling estimates on earlier work [O'Neill et al. 1979] and engineering estimates.

### 7.4.4 Torus Station Balance Results

The results from the mass and inertia equations are combined to evaluate the elliptical torus's rotational stability. Multiple station parameters could affect the stability, including the radius, the minor axes ratio, and the shell thickness. The moments of inertia were evaluated to determine the station's rotational stability.

The single-floor analysis showed the elliptic torus was rotationally stable for a broad range of torus axes lengths and ratios; see Figure 6-8. The radius length has much less effect on the stability than the ratio of the elliptic cross-section axes. The perpendicular axis (c) can be more than 7 times the length of the rotation axis (a); see Table 6-1. The single-floor stability results were found with a straightforward equation analysis. The multiple-floors stability results were found by combining the moments of inertia of the station components.

Figure 7-31 shows four sets of data in bar charts. They show results for elliptical torus stations with radii of 200, 2000, 20K, and 200K meters. These charts show the rotational moments of inertia (MOIs) for the seven major components of the station. The seven components are along the x-axis. The logarithmic y-axis shows the MOI in kilograms-meter squared ranging from 1 to 1e27.

Above each pair of MOIs ($I_z$ and $I_x$) is the stability metric ($I_z/I_x$). All the stability metrics in Figure 7-31 are close to 1.9, except for the shuttle bay component. All the components are rotationally stable with $I_z/I_x$ greater than 1.2. Torus stations are very stable with $I_z/I_x$ much greater than 1.2. The MOIs of the shell and multiple-floors components are the greatest of all the components and provide much stability to the station.

The spokes and air components tend to have the smallest MOIs. The shuttle bay has $I_x$ greater than $I_z$ and negatively affects the station's rotational stability. This component unbalances small stations and slightly reduces their stability ratio; however, with larger stations, the shuttle bay MOIs are much smaller than the other component MOIs and do not influence the stability.

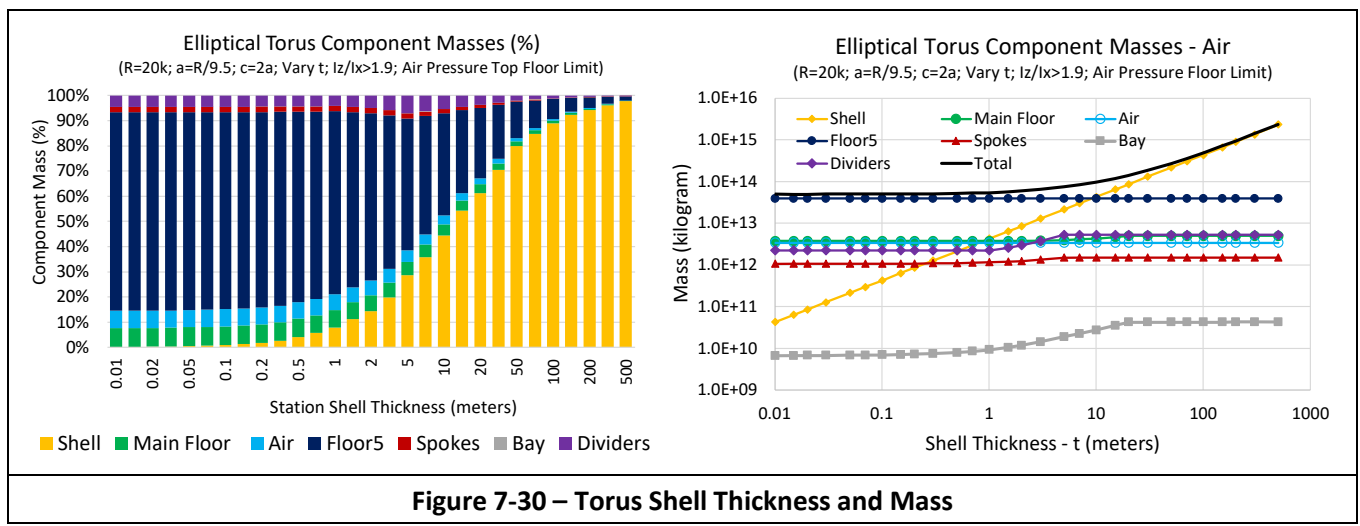

**Figure 7-30 – Torus Shell Thickness and Mass**

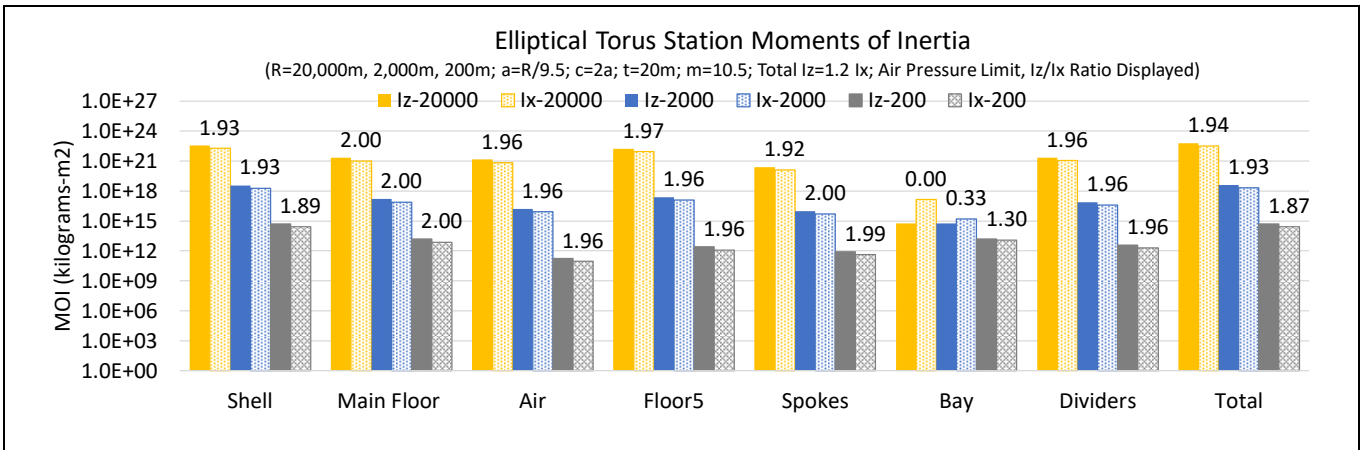

**Figure 7-31 – Torus Components MOI Details**

Figure 7-32 provides the elliptical torus multiple-floors stability for a broad range of station radii. This graph shows the ratio of the MOIs Iz/Ix as a stability metric. The station is considered rotationally stable with a ratio metric of 1.2, as described in [Brown 2002] and [Globus et al. 2007]. The y-axis of the graph shows the stability (Iz/Ix) ratio ranging from 1.8 to 2.0. The x-axis shows the torus major radius ranging from 100 to 800 thousand meters on a logarithmic scale. The small range on the y-axis overemphasizes the variation with radius size. This torus has a gravity range of 0.95g to 1.05g, resulting in a scale of 9.5. The radial minor axis a has a length of the major radius R over 9.5, which provides the habitable gravity range from the center of the tube (at radius R) to the outer rim (at radius R+a). The perpendicular minor axis (c) has a length of twice the radial minor axis (c=2a). The shell is a regolith-filled structure that is 20 meters thick.

Figure 7-32 shows the stability of the station with the gravity limit using multiple floors from the outer rim to the center of the torus tube. It also shows the stability of the station with the air pressure limit, where the top floor has an air pressure of at least Denver's air pressure. The stabilities range from 1.86 to 1.96. The gravity and air pressure stabilities are nearly identical for most of the graph. A small difference in the stability appears with very large torus radii. The small difference is accentuated with the small range on the y-axis. The elliptical torus stabilities shown in Figure 7-32 are greater (more stable) than our minimum stability ratio metric of 1.2. For reference, the graph also includes the stability ratio of 1.92 for the thin shell single-floor elliptical torus using the same major and minor radii.

Figure 7-33 is included to better understand the component stabilities in the system. The x-axes of the charts in Figure 7-33a show the rotation radius on a logarithmic scale ranging from 100 to 400,000 meters. The left y-axis in Figure 7-33a show the components' stability ratios (Iz/Ix), and ranges from 1.75 to 2.00. The right y-axis shows the ratios for the total stability and for the bay component and ranges from 0.0 to 3.5 The two scales of the y-axes provide more detail for the different components.

Each component has a different mass, so the total station stability ratio is not simply the sum of these individual component stabilities. The shuttle bay component stability quickly drops as the two 200-meter length bays are separated with a connecting spoke. Their combined length matches the distance to the outer edge of the torus spokes. The bay spoke is approximately the minor axis length, c. The spoke dimensions increase with increasing station radius to provide more structural strength. The number of spokes increases at station radii of 8500 and 85,000 meters. The increase in the number of spokes causes the step functions in the spoke stability. The station stability for the air pressure limit remains fairly constant at about 1.95 beyond the 1,000-meter radius. The main floor, the air, and the multiple floors all have a stability Iz/Ix of about 1.95 until radii greater than 10,000 meters. The stability of the multiple-floors component increases past the 10,000-meter radius. With the larger stations and the air pressure top floor limit, the stability of the floors is larger. The mass increases slower with the top floor constrained. Even with the less mass, the greater distance from the rotation center increases the stability. The air component has a similar increase as more air is centripetally pushed to the outer edge.

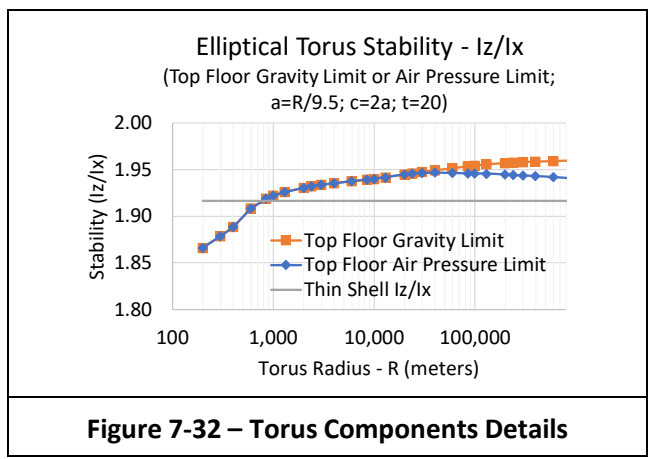

**Figure 7-32 – Torus Components Details**

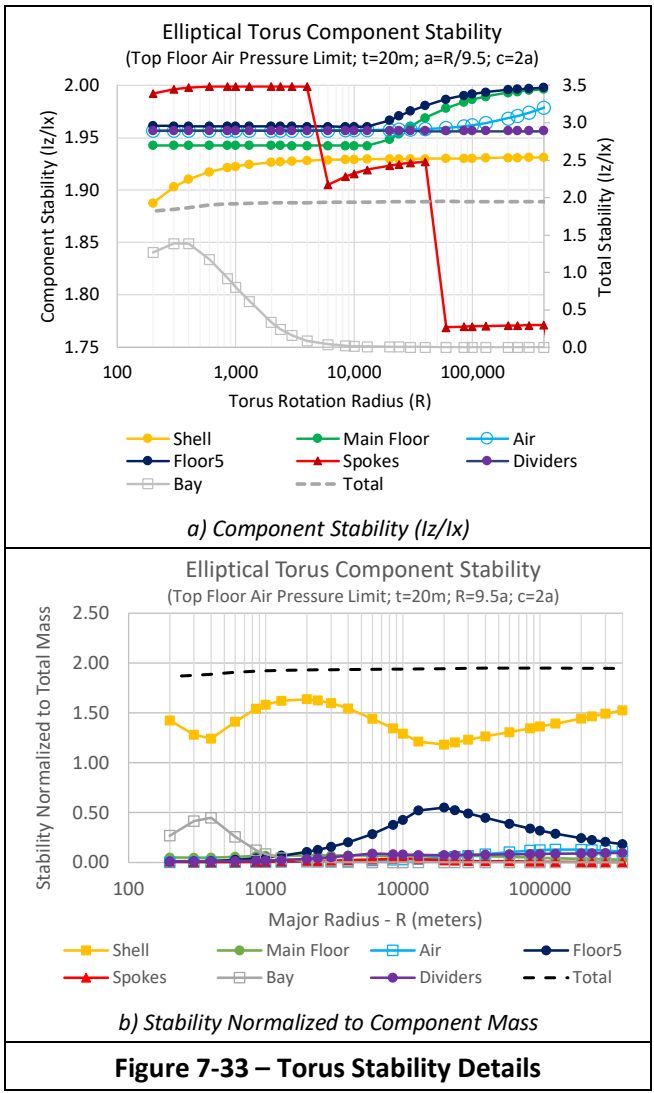

*a) Component Stability (Iz/Ix)*

*b) Stability Normalized to Component Mass*

**Figure 7-33 – Torus Stability Details**

For completeness, we include the same stability results in Figure 7-33b except we normalize each stability using the component mass ($M_{comp}$) and the total station mass ($M_{station}$). The y-axis values represent a form of stability and would be equal to Iz/Ix times $M_{comp}/M_{station}$. These results clearly show the minimal effect that most of the components have on the station stability. The shell stability dominates all the station stabilities. The shuttle bay stability, the main floor stability, and the multiple-floors stability assert some effect on the station stability, but much less than the shell effect. The following paragraphs further investigate these component stability masses and their MOIs.

The effect of the station geometry is used to further investigate the stability of the elliptical torus. Figure 7-34 shows more details on this stability with the two charts. Those charts vary the station elliptical cross-section ratio (c/a) and the gravity range ratio (R/a).

The station is stable (Iz/Ix=1.2) for all the data in Figure 7-34a. The elliptical torus is designed with the minor axis a set to the radius over 6.33 or 9.5. This produces a gravity range from 0.95g on the center main floor to 1.1g or 1.05g on the outer rim. The x-axis of the chart shows the station radius and is logarithmic from 100 to 40,000 meters. The y-axis shows the ratio of the two minor elliptic axes (c/a). The single-floor example found the station stable across a broad range of major radii. With R=6.33a, the station is stable with the axes ratio (c/a) from 6.0 to 8.0 as the radius R ranges from 200 to 400,000 meters. With R=9.5a, the torus would remain stable even with the torus tube cross-section being very elongated, with c/a ranging between 9 to 12. The station is typically designed using c=2a or c=3a, and the chart includes those ratios for reference. With these design ratios, the station would be stable with a significant margin for safety. The chart also shows the increase in elliptic ratio for large stations using the top floor air pressure constraint. This increase is from the increase in stability with the smaller floor height limit. The thinner cylinder of multiple floors is more stable than the thicker cylinder of multiple floors.

Figure 7-34b shows the chart with an alternative view of this stability data. Again, the x-axis shows the radius. The y-axis shows the stability ratio of the station (Iz/Ix). The chart shows the minimum stability ratio of Iz=1.2Ix for reference.

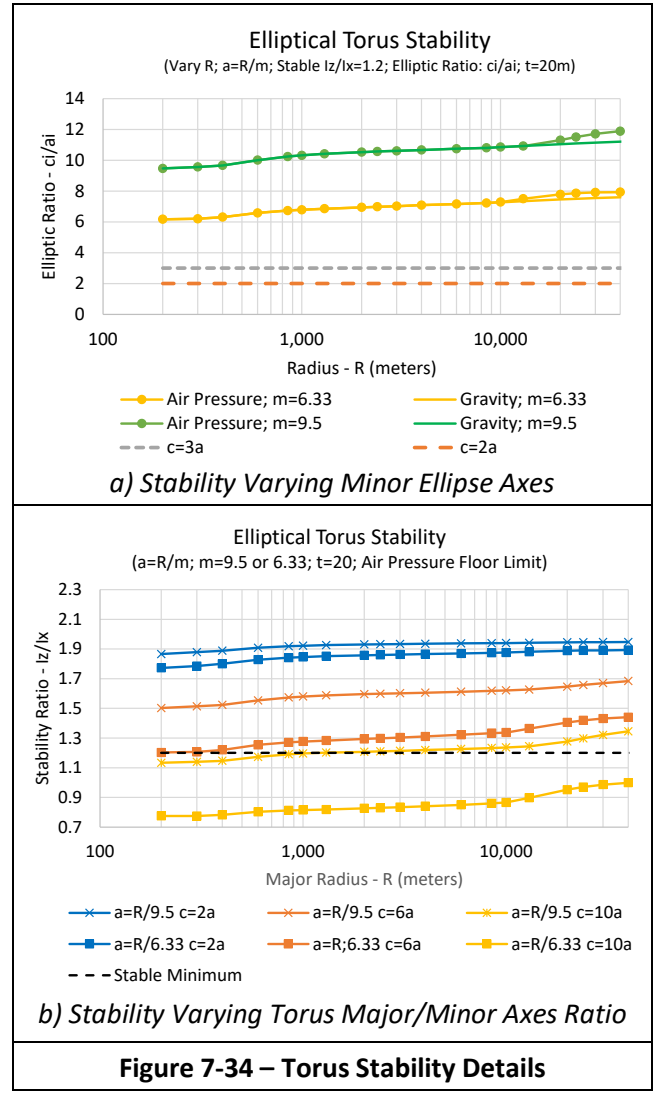

*a) Stability Varying Minor Ellipse Axes*

*b) Stability Varying Torus Major/Minor Axes Ratio*

**Figure 7-34 – Torus Stability Details**

The three sets of ratios show the effect of varying the axes ratio (c/a). Consistent with the previous chart in Figure 7-32, most of the ratio data is greater than the minimum stability value. This shows the station is stable for c=2a and c=6a. The station is not rotationally stable when c=10a and R=6.33a. There are limits on the torus geometry ratio.

For completeness, Figure 7-35 is included to illustrate the effect of the shell thickness on the stability of the multiple component elliptical torus. The x-axis provides the same range of thicknesses as in our earlier thickness chart. The y-axis only presents a narrow range of stability (Iz/Ix) from 1.9 to 2.0. The increasing thickness reduces the stability. Recall that a ratio of 1.2 provides passive stability, and all thicknesses are stable for the 20,000-meter radius elliptical torus station.

## 7.5 Dumbbell - Multiple Component Stability

This study continues with the multiple component stability analysis for dumbbells. The dumbbell is unique compared to the earlier three geometries. In particular, the dumbbell rotational axis MOI is nearly identical to one of the other axes' MOIs. This implies that the station would not be stable based on our previous three stability rules; see §6The dumbbell single-floor stability subsection covered this instability. The station is expected to settle into a limit cycle, causing a wobble on its rotational axis [Fitzpatrick 2023]. For the most part, this paper ignores this wobbling and continues the analysis of the multiple-floors dumbbell station.

This section defines and analyzes the six major components of the rotating dumbbell space station. It covers equations used to represent the components' mass and inertia. It presents the location and limits on the top floor of the multiple floors of the dumbbell station. It presents the station mass and stability balance results. This section concludes by introducing a double dumbbell design. This design can address the single dumbbell wobbling. That subsection includes a brief analysis of its mass and balance.

### 7.5.1 Dumbbell Station Components

Figure 7-36 shows a cross-section rendering of a full dumbbell station. It illustrates the station dumbbell geometry with labels for components and densities. This view is on the y-axis. It also includes a three-dimensional line drawing of the geometry. The drawings include the outer shell, main floor, multiple floors, center spoke, side spokes, and shuttle bay.

Figure 7-37 shows another drawing of an example dumbbell station with ellipsoid nodes. The vertical axis shows the vertical distance from the axis of rotation (z-axis) and ranges from 0 to 2400 meters. The horizontal axis shows the horizontal distance from the vertical axis (x- or y-axis) and ranges from 0 to 1000 meters. The drawing only shows one-quarter of the complete dumbbell. A full drawing of this example dumbbell would extend from -2300 to 2300 meters on the vertical axis. It would extend from 900 to -900 meters on the horizontal axis. The drawing labels the dumbbell's major components including the outer shell, main floor, multiple floors, center spoke, side spoke, and shuttle bay.

### 7.5.2 Dumbbell Mass and Inertia Equations

The 6 components of the dumbbell and their mass equations were presented in *§5.1 Station Component Equations*. These mass equations are extended to provide the Moment of Inertia (MOI) equations for each of the major components of the dumbbell. These components include the masses and volumes of the structures, fill, and panels. Summing the results from these individual components produces the total station mass and rotational moments of inertia. The mass equations use the material densities in Table 5-1. The rotational moments of inertia (MOIs) equations use the mass. The stability of the station is measured by using the ratio of the MOIs. The dumbbell rotational axis MOI (Iz) is nearly identical to one of its other axes' MOIs (Iy) and creates a rotationally unbalanced system.

#### 7.5.2.1 Dumbbell Shell

The dumbbells of this paper use ellipsoid nodes. The single-floor stability analysis used the thick shell moment of inertia equations from Table 5-2. The ellipsoid equation characteristics were presented in *§5.2 Ellipsoid Analysis*. This station node design uses the uniform shell thickness model. As shown in previous subsections, the outer shell is modeled with a series of hollow disks. The mass and inertias of the rings are summed to produce the ellipsoid node values. The ellipsoid node inertia is computed, and then the parallel axis theorem is applied to calculate the inertia of the two rotating dumbbell nodes. These disks are circular about the center node axis. The dumbbell node shell is divided into thousands of slices along that axis of rotation. Each slice has a thickness of $t_d$ equal to $c_o/1000$. Because this is a prolate ellipsoid node, hollow-cylinder disks can be used to compute the mass and MOIs. The disks are stacked along the z-axis, ranging from $-c_o$ to $c_o$. The mass of a disk would be $m_{disk} = \rho_{fill}\, \pi\, t_d\, (r_o^2 - r_i^2)$ where $r_o$ and $r_i$ are the inner and outer edges of each disk. These radii are computed using the ellipse cross-section along the z and y axis. The summed mass of the disks was within 0.1% of the thick ellipsoid shell mass equation value.

A similar approach is used for the dumbbell node inertias. The moments of inertia are computed using the same hollow disks, but the Iz and Iy MOIs for the nodes require the parallel axis theorem. This adds the node mass times the distance to the node center of gravity from the z-axis squared. The station and the disks rotate about the z-axis. We use the MOIs for thick shell disks from Table 5-2. The moment of inertia about the z-axis for the disks would be $I_z = \frac{1}{2} m_{disk}(r_o^2 + r_i^2)$, where ro and ri are the disk radii along the node ellipse. The moment of inertia for the disks about the Ix (or Iy) axis would be $I_x = I_y = \frac{1}{12} m(3(r_o^2 + r_i^2) + t_d^2) + m d_x^2$ where $d_x$ is the distance from the x-axis. The thousands of disks are summed and compared to the result of the MOIs for the thick ellipsoid shells in the dumbbell. The summed MOIs of the disks are within 0.1% of the thick shell inertia equation results.

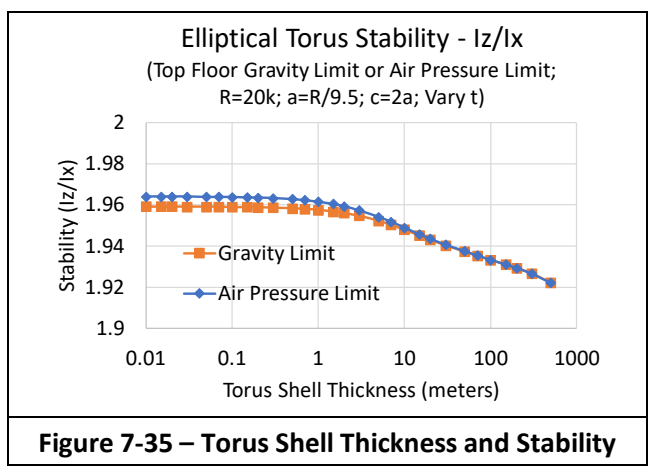

**Figure 7-35 – Torus Shell Thickness and Stability**

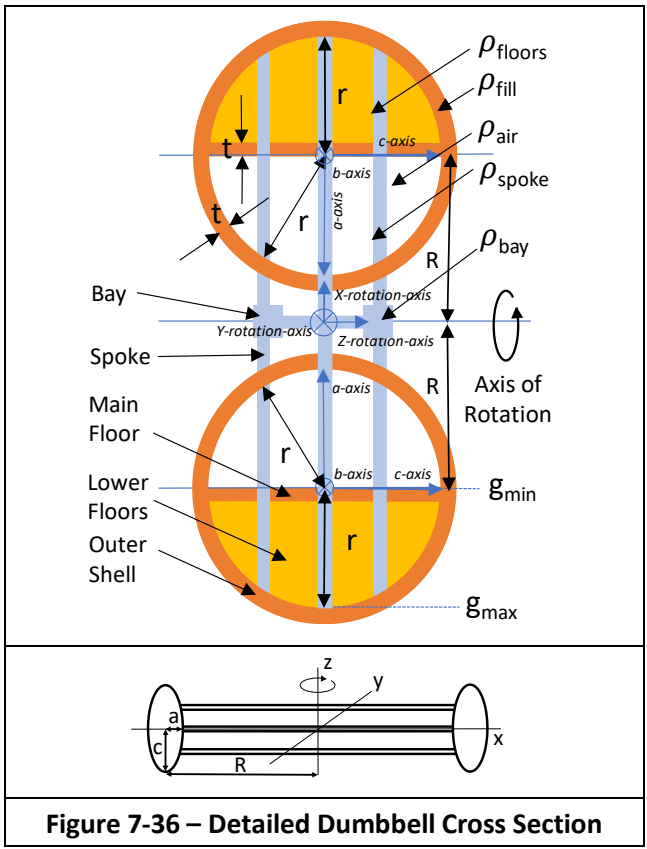

Figure 7-36 – Detailed Dumbbell Cross Section

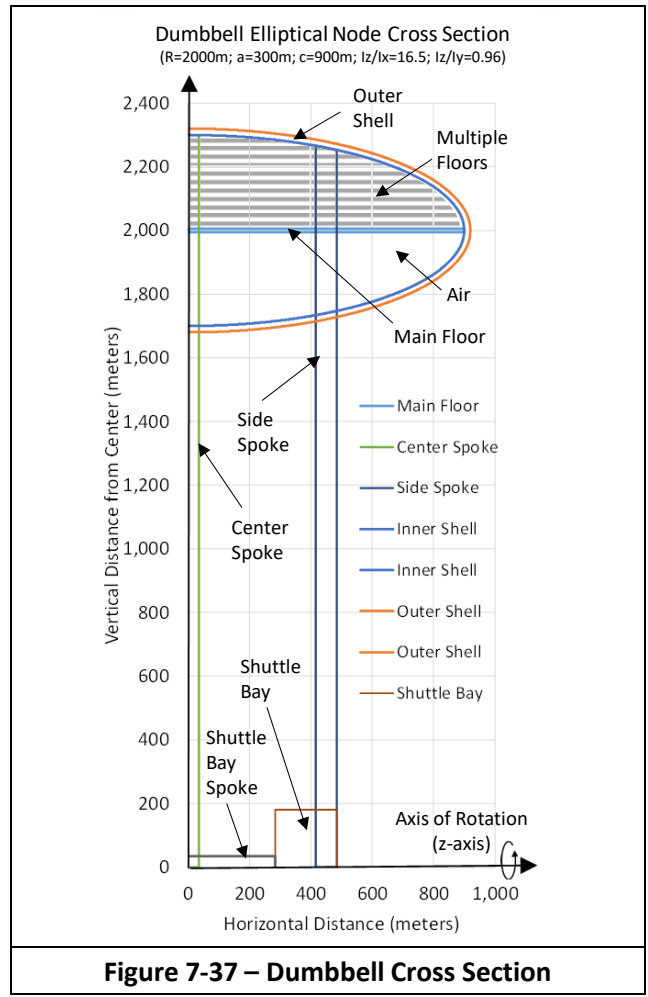

Figure 7-37 – Dumbbell Cross Section

There are two nodes, and we double the mass of the dumbbell shell components. We double the node x-axis MOI to compute the Ix for the dumbbell shell component $I_{xshell} = 2\ I_x$. The node y-axis and z-axis use the parallel axis theorem with their MOIs at a distance of R. Those MOIs are doubled because of the two nodes in the dumbbell and results in $I_{zshell}=2\ (I_z+m\ R^2)$ and $I_{yshell}=2\ (I_y+ m\ R^2)$.

### 7.5.2.2 Dumbbell Spokes

The spokes are again modeled as thick shell cylinders. The dumbbell only has one set of spokes instead of the four shown with the other geometries. These spokes must be strong enough to support the entire rotating node mass. The stress on the spokes can be estimated using the mass of the node and the area of the spokes. The effective radius of the spokes is set to be 3% of the dumbbell's major radius. Multiple spokes are used and are designed to provide the same combined area as a spoke with the effective area. The wall thickness of the spokes is 10% of the radius and a maximum shell thickness of 50 meters. The length of the center spoke is the dumbbell's major radius plus the minor elliptical node radius. The offset side spoke is shorter than the center spoke and is constrained by the curve of the elliptical node. In Figure 7-36 the side spokes are positioned halfway across the minor axis c. The spokes extend from the outer rim of one dumbbell node to the outer rim of the other dumbbell node. The mass and MOIs of the spokes are computed using the same approach as used in the previous cylinder geometry.

### 7.5.2.3 Dumbbell Shuttle Bay

The shuttle bay is along the rotation axis of the dumbbell station. The geometry and analysis are nearly identical to the other geometry shuttle bays. The shuttle bay has a maximum radius of 180 meters and a maximum length of 400 meters. For larger stations, the bay is split in half. The halves are positioned at the outer edges of the two side spokes; see Figure 7-37. A shuttle bay spoke connects the two bays to provide structural integrity. The shuttle bay mass and MOIs are computed using the equations shown with previous geometries.

### 7.5.2.4 Dumbbell Main Floor

The mass and MOIs of the main floor are computed using a flat elliptical disk. It may be more appropriate to curve the floor in the rotation direction. Centripetal gravity would be directed perpendicular to the curved plate. One would experience a lean when walking on the rotating flat disk; see *§5.3 Dumbbell Modeling Details*. For an engineering estimate, the volumes and areas of flat and curved disks are close to equal for large-radius stations. The main floor is built and modeled as an ellipse disk, and the elliptical disk equations from Table 5-2 are used to compute the main floor MOIs. That ellipse disk is in Figure 7-36 on the node axis b-c plane and rotates about the z-center axis at a distance R in the x direction. The mass of the floor is m and is equal to $m_{floor} = \rho_{fill} \, t_f \, \pi \, a \, c$. The variables a and c are the elliptical node axes length. The variable $t_f$ is the thickness of the main floor or the soil on top of the floor. The floor density is twice the 5-meter floor density (64.4 kg/m3), and the soil density is $\rho_{fill}$ (1791 kg/m3). The elliptical disk equations in Table 5-2 are rotated to match the orientation of the dumbbell's main floor. The analysis uses the node inertias $I_x = \frac{1}{4}m(a^2 + c^2)$ for the x-axis, $I_z = \frac{1}{12}m(3a^2 + t_f^2)$ for the z-axis, and $I_y = \frac{1}{12}m(3c^2 + t_f^2)$ for the y-axis. The elliptical disk rotates about the z-axis and is offset to the major radius. The same would be true if rotating about the y-axis. The $I_z$ and $I_y$ MOIs for the main floor in the dumbbell require the parallel axis theorem and add the floor mass times the distance R from the z-axis squared.

### 7.5.2.5 Dumbbell Multiple Floors

This component of the station is represented with a series of ellipse disks. These are analyzed like the main floor. In this analysis, the multiple floors are 5 meters apart, and their radii range between the main floor and the outer rim. The minor length of the floor at height h uses the ellipse equation $a_h = a_i \sqrt{1 - d^2/a_i^2}$ where $a_i$ is the node minor axis and d is the depth below the center main floor (d=$a_i$-h). The minor axes' lengths are typically scaled by 2 or 3 to compute the major lengths, $c_i$ and $c_h$. The density of the multiple floors is from Table 5-1; however, the air density is excluded from the floors and included in the air analysis. The curved elliptical node shell binds the size of the ellipse floor layer. The multiple floors analysis uses a layer thickness $t_f$ much smaller than 5 meters. The thin layer improves the accuracy of the analysis within the curved shell. The mass and MOIs are computed for the series of floor layers. Each layer would have a mass of $m_{layer} = \rho_{floor5} \, t_l \, \pi \, a_h \, c_h$, where $t_l$ is the layer thickness and $a_h$ and $c_h$ are the dimensions of the ellipse layer floor at height h. The layer MOIs are computed on the three axes. The elliptical disk equations in Table 5-2 are rotated to match the orientation of the dumbbell floors. They become $I_x = \frac{1}{4}m_{layer}(a_h^2 + c_h^2)$ for the x-axis, $I_z = \frac{1}{12}m_{layer}(3a_h^2 + t_l^2)$ for the z-axis and $I_y = \frac{1}{12}m_{layer}(3c_h^2 + t_l^2)$ for the y-axis. The $I_z$ and $I_y$ MOIs for the floors in the dumbbell require the parallel axis theorem and the location of the multiple-floors center of gravity (CoG) from the z-axis. The MOIs add the floor mass times the CoG distance squared.

In the Figure 7-37 example, the floor lengths range from zero to about 1800 meters as the radii range from 2300 to 2000 meters. The multiple-layer values are summed to determine the total mass and MOIs of the multiple-floors component. The total mass of the layers is less than 1% different than the mass using an ellipsoid volume formula: $V = \frac{4}{3}\pi a^2 c$. In the Figure 7-26 torus example, the MOIs of the multiple floors about the x-axis sum to 6.20e14 kg-m$^2$, about the y-axis sum to 6.21e16, and about the z-axis sum to 6.17e16 kg-m$^2$. The stability Iz/Ix of the multiple floors equals 99.7 and is much greater than the minimum stability limit of 1.2 [Brown 2002] [Globus et al. 2007]. The other stability would be Iz/Iy, and equals 0.99 and is less than the minimum stability ratio. This corroborates the unstable wobble issue discussed earlier in this section.

### 7.5.2.6 Dumbbell Air

The air mass and inertia analysis is very similar to the analysis shown for the elliptical torus. The air density in the station is sea level at the outer rim. It decreases to lower densities with increasing height. The formulae and examples were introduced in *§3.2 Air Pressure Limits*. It is assumed the air behaves like a low-density solid. The masses of each layer of air would be analyzed like those in the multiple floors except with varying densities. Each layer would have a mass of $m_{layer} = \rho_{air} \, t_l \, \pi \, a_i \, c_i$. The layers of air start at the outer node shell and continue to the inner node shell; see Figure 7-36. The MOIs of the layers of air use the same approach as the multiple-floors layers. In the Figure 7-37 example, the half lengths range from 0 to approximately 900 meters, and the radii range from about 1700 to 2300 meters. The analysis uses 2000 layers; the thickness $t_l$ equals ai/2000 or 0.3 meters. The volume of those summed layers matches the result using an ellipsoid volume equation. The mass and MOI values are summed to determine the air totals. The Iz and Iy use the parallel axis theorem and the center of gravity (CoG) of the ellipsoid air.

### 7.5.2.7 Dumbbell Preview

The masses and the MOIs of the components characterize the dumbbell station. As a preview, Figure 7-38 shows pie charts illustrating the distribution of the component masses in three dumbbell stations. One of the stations has a radius of 200 meters, another has 2000 meters, and the third has 20,000 meters. All have the same six components. In these stations, the shell has a thickness of 20 meters. Multiple floors extend from the outer rim to the top floor. The top floor is at a height where the top floor gravity is 0.95g, and the air pressure is higher or equal to the Denver air pressure.

Given these station limits, the mass distributions change between the different size stations and are quite different than those of previous geometries. In the other geometries, the spokes are typically insignificant compared to their other components. The shells of the cylinder, ellipsoid, and torus extend around the rotation axis at the major radius R. The dumbbell node does not extend around the rotation axis and has a much smaller radius equal to R/m. The spokes in all four designs have a length nearly the twice the radius R. With the smallest dumbbell in Figure 7-38 the shuttle bay dominates the system mass. The shuttle bay is fairly fixed in size and becomes a smaller percentage of the total mass as the station size increases. With the increasing station size, the outer shell remains a significant contribution to the total mass. With increasing radius size and the increase in the node size, the mass of the multiple floors begins to increase faster than the shell. The mass of the floors increases from 0.1% to 2% and to 11% of the total with the three example dumbbells in Figure 7-38. With increasing floor and node mass, the spokes increase in radius and mass to support the increasing stress. The mass of the spokes increases from near 0.5% to 5% and to 40% of the total. As with the other geometries, the mass of the shell and the main floor both increase as a function of area; the mass of the multiple floors increases as a function of volume. The following subsections delve into more details and explanations for the dumbbell.

### 7.5.3 Dumbbells - Small and Large Comparison

Figure 7-39 illustrates design changes between a small and large dumbbell. The left graph shows a dumbbell with a radius of 2000 meters, and the right shows a dumbbell with a radius of 200 meters. The axes and scales of the charts are different to better illustrate details in the stations. Both use a shell with a thickness of 20 meters. The shell appears quite thin in the 2000-meter radius station and excessively thick in the 200-meter radius station. The internal space appears much more constrained in the small station. The small station would only hold 4 floors spaced 5 meters apart, and the large station would hold 40 floors. Both stations provide 0.95g on the main floor and 1.05g at the outer rim of the station. The small station would rotate at 2 rpm, and the large station would rotate at 0.65 rpm. For most people, neither would cause nausea from rotation effects on the inner ear. Both stations would be rotationally unstable (Iz/Iy<1.2) and would have wobble that may cause problems for the inhabitants and equipment. Table 7-1 compares and summarizes a set of metrics for the two stations.

### 7.5.4 Dumbbell Station Mass Results

Figure 7-40 provide a view of the dumbbell station component masses. It includes the same six components as reviewed with other geometries. The thickness of the outer shell is 20 meters. The station has stability metrics of Iz/Ix>1.2 and Iz/Iy<1.2, and would be rotationally unstable. This instability would manifest as a periodic wobble [Fitzpatrick 2023].

Figure 7-40a contains two stacked bar charts as one view of the dumbbell station component mass. The stacked bar chart shows the masses of the dumbbell components as percentages of the total dumbbell station mass. The y-axes show the

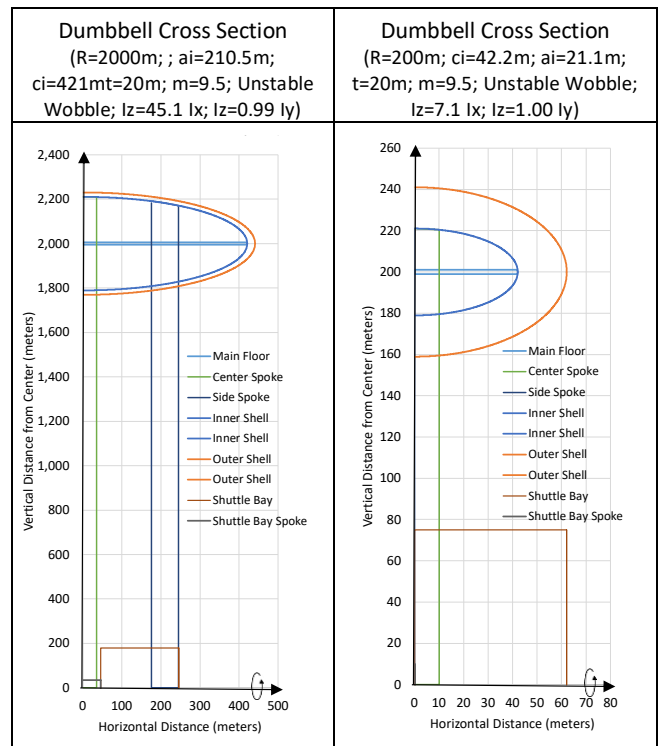

**Figure 7-39 – Large and Small Dumbbell**

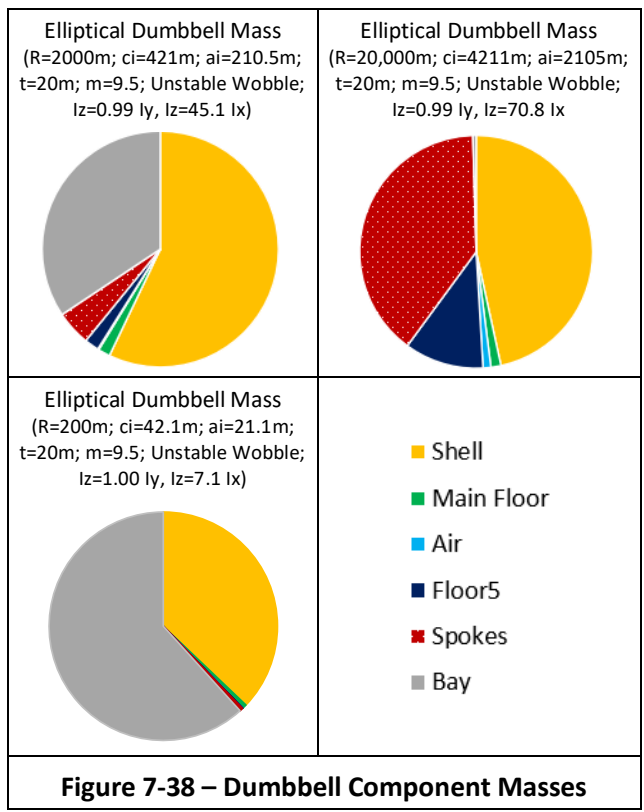

**Figure 7-38 – Dumbbell Component Masses**

**Table 7-1 – Dumbbell Size Comparison**

| Size | Large Dumbbell | Small Dumbbell |
|---|---|---|
| Radius | 2000 meters | 200 meters |
| Minor A | 210.5 meters | 21.1 meters |
| Minor C | 421.1 meters | 42.1 meters |
| Shell Thickness | 20 meters | 20 meters |
| Spokes | 3 | 1 |
| Spoke Thickness | 3.5 meters | 1.0 meters |
| Population | 107,605 people | 101 people |
| Mass | 1.2e11 kilograms | 2.6e9 kilograms |
| Half Bay | 90R x 200L meters | 62R x 124L meters |
| Rotation | 0.65 rpm | 2.06 rpm |

percentages ranging from 0 to 100%. The x-axes show the major radius of the dumbbell station, and range from 200 meters to 40 thousand meters.

Figure 7-40b shows an alternative view of the same mass data. This figure contains two line graphs showing the mass of the components as a function of the torus radius. The y-axes of the charts show mass ranging from 1e6 to 1e16 kilograms on a logarithmic scale. The x-axes of the charts show the elliptical torus major radius ranging from 100 to 40 thousand meters on a logarithmic scale. Both graphs show data from dumbbells with multiple components and a gravity range of 0.95g to 1.05g from the node center to the outer rim. The graphs have lines representing the total mass and the individual masses of the six major components of the elliptical dumbbell. The lower graph includes the air pressure limit constraint.

Both charts show that the shuttle bays in small-radius stations dominate the total station mass. They are fixed in size to accommodate arriving supplies and personnel. The shuttle bay size is scaled smaller for the smallest stations. The shuttle bay reaches its maximum size at a station radius of about 1000 meters. There is small growth with larger stations because of the connecting spoke between two halves of the large shuttle bay cylinders.

The largest station of the charts in Figure 7-40 has a radius of 40 kilometers, a semi-minor radius of 4211 meters, and a semi-major radius of 8421 meters. The total mass of the largest station is about 30% greater with top-floor gravity limit. Much of that increase is from the multiple-floors component. The shell mass remains the same between the two charts. The air pressure is habitable from the outer rim to 1560 meters above the outer rim. The top floor is at this height of 1560 meters and has the Denver air pressure (83,728 Pascals) and air density (1.01 kg/m3). The top floor with the gravity limit would be at 4211 meters above the outer rim. The air pressure at this height would be 0.74 kg/m3 and not habitable. This is a case where airtight layers would be introduced to provide habitable air pressure on gravity-limited top floors; see *§3.5 Top Floor Limits*.

With a radius of 40K meters, the top floor height (and the number of floors) is 2.7 times greater with the gravity limit. Similarly, with more floors, the mass of the multiple-floors on the largest station is 5.6 times greater without the air pressure constraint. The top main floor at 4211 meters is a larger

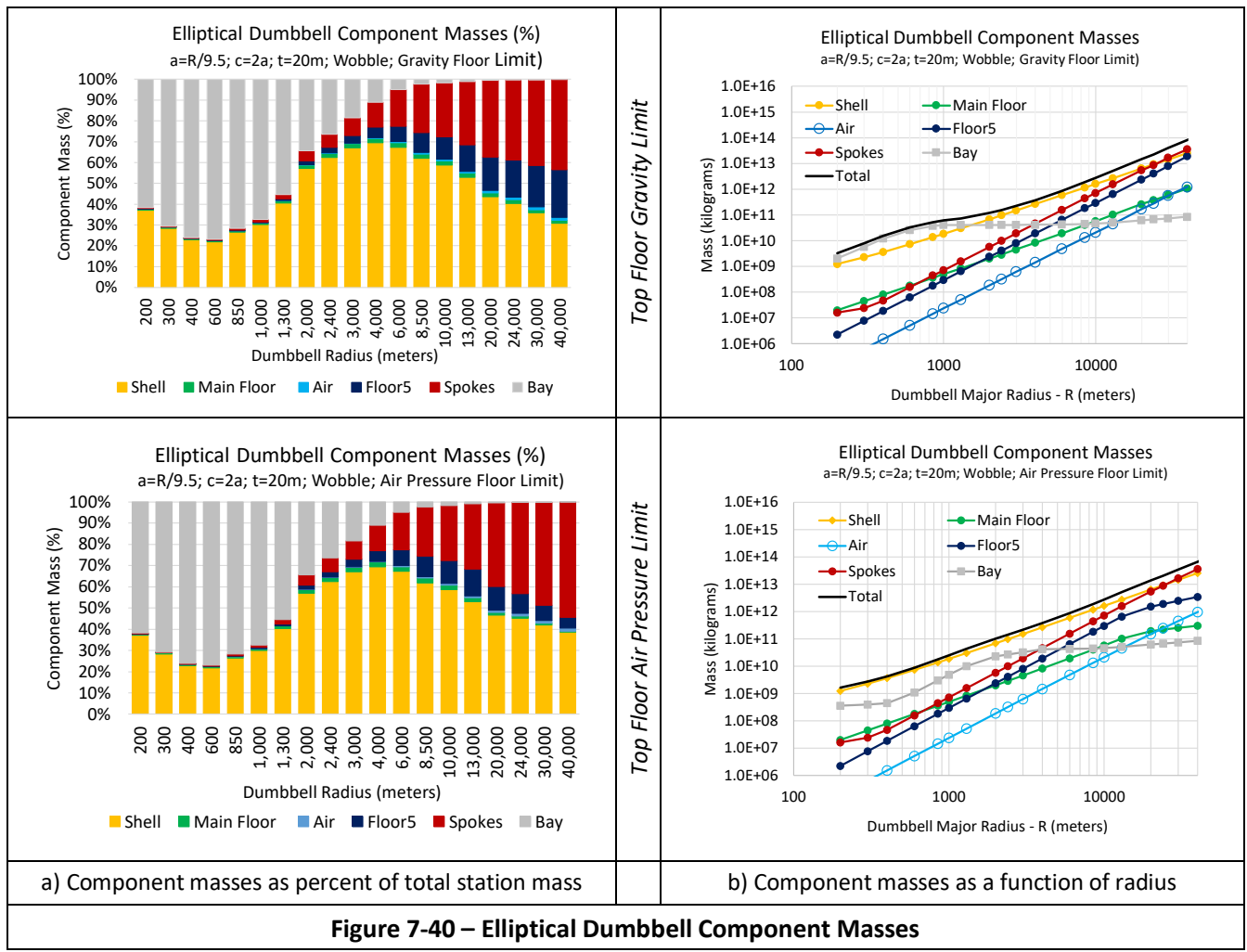

**Figure 7-40 – Elliptical Dumbbell Component Masses**

elliptic disk and is about 3.5 times greater than the disk at 1560 meters.

There are many interesting observations from these charts. Overall, the charts show that the air pressure limit has minimal effect on the design and mass until the station radius is greater than 15,000 meters. Placing the top floor in larger stations at the air pressure limit significantly reduces the number of multiple floors and the station mass. Fortunately, airtight layers can overcome this air pressure limit and support extremely large populations.

Figure 7-41 provides charts with another view of the dumbbell station component mass. It includes the same six components. Those charts study a dumbbell station with a 2,000-meter radius. The dumbbell shell thickness is varied to see the effect on the station components. The left chart shows the mass of the dumbbell components as a percentage of the total dumbbell station mass. The y-axis shows the percentages ranging from 0 to 100%. The right chart shows the component masses in kilograms on the y-axis. The y-axis is logarithmic and ranges from 10 million to 100 billion kilograms. In both charts, the x-axis shows the thickness of the dumbbell station shell, and ranges from 0.01 meters to 500 meters to 1000 meters.

As one would expect, the mass of the shell increases with the thicker shell. With large shell thicknesses, the shell mass dominates the station's mass. For very thin shells, the multiple floors dominate the station's mass. The volume and mass of the air (torus interior) and the multiple floors do not change with the shell thickness. The shuttle bay and the spokes barely increase with large shell thicknesses. Only the shell, multiple floors, and shuttle bay represent a significant percentage of the total mass. The thickness of the main floor increases with the shell thickness but is limited to a minimum and a maximum. In general, we advocate a fairly thick shell to provide protection from space debris. For thicknesses greater than 10 meters, the shell mass dominates the total station mass. The second chart in Figure 7-41 shows the component masses for the same range of station shell thickness. This chart provides details on the scaling of the main floor, spokes, and dividers with the changing shell thickness. This scaling analysis needs additional material strength analysis (perhaps using finite element analysis). The current scaling estimates are based on earlier work [O'Neill et al. 1979] and engineering spoke stress estimates.

Figure 7-42 illustrates the effect of the gravity range ratios on the mass of the dumbbell station. The x-axis of the chart shows the dumbbell radius on a logarithmic scale ranging from 100 to 40,000 meters. The left y-axis shows the station mass on a logarithmic scale ranging from 1e9 to 1e16 kilograms. The right y-axis shows the height of the top floor above the outer side of the dumbbell node. The data is shown for two gravity range ratios using the major radius R over the node radius equal to 6.33 and 9.5. These two values produce gravity ranges from 0.95g on the center floor to 1.05g or 1.1g on the outer edge of the node.

The chart shows that the smaller gravity range (m=9.5) has a smaller station mass than the larger gravity range (m=6.33). For a given radius, the node dimensions are smaller with the larger scaling factor m=9.5. With smaller dumbbell radii, there are more floors with the larger gravity range (m=6.33) and larger node dimensions. The top floor is at a greater height with the larger node dimensions (m=6.33) until the air pressure effects begin to limit the floor height. The longer radius to the outer rim and slower rotation affect the air pressure and increase the top floor height. Even with the lower mass, an additional 14 floors are available in very large stations with the smaller gravity range (m=9.5).

### 7.5.5 Dumbbell Spoke Stress Consideration

The stress distribution in cylinders, ellipsoids, and tori dominates as hoop stress. The dumbbell stress must be modeled differently. The stress in the dumbbell spokes is modeled as force over area. The spoke cross-section area varies with the rotation radius. The spokes are modeled with concentric outer and inner cylinders. The spoke cross-section area uses the cross-section area of the outer cylinder plus a fraction of the inner cylinder cross-section area. The force is the mass rotating about the center of the station. The mass is the summation of the masses of the entire shell, main floor, air, and multiple floors. This force includes the 1g acceleration from centripetal rotation. The total force also includes the mass of the spoke accelerating at an average of one-half of 1g.

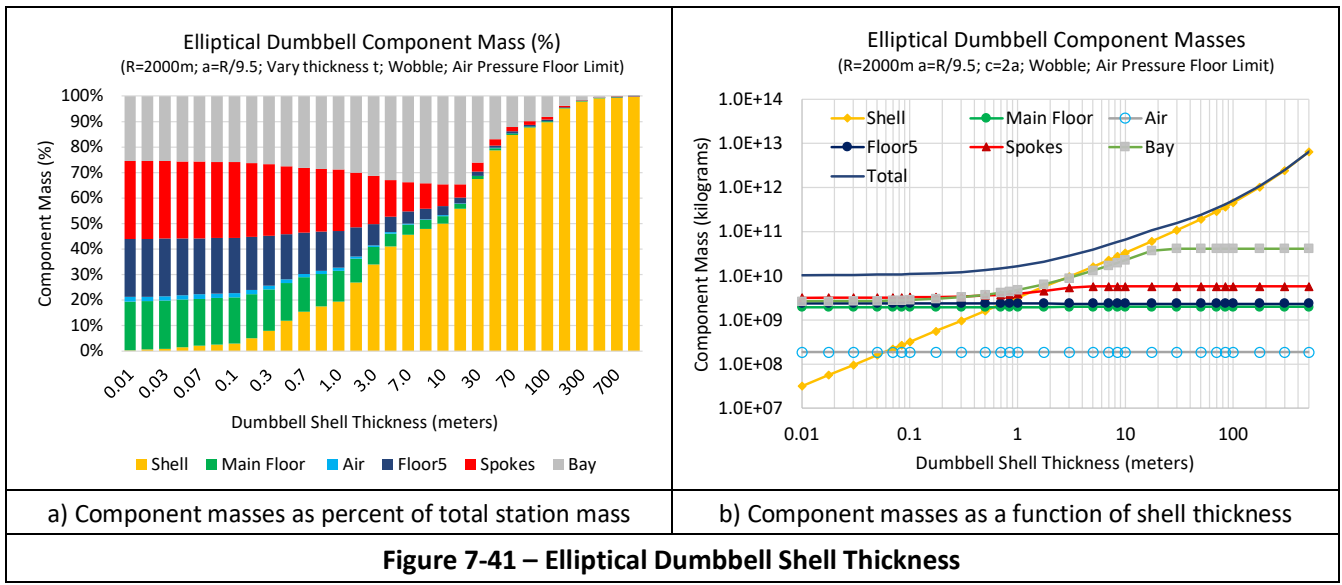

**Figure 7-41 – Elliptical Dumbbell Shell Thickness**

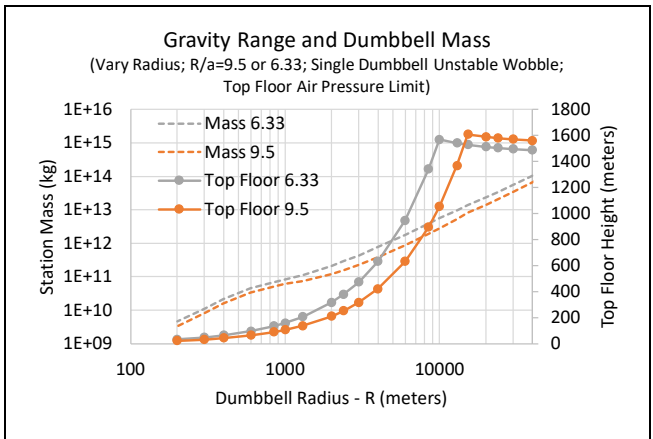

**Figure 7-42 – Dumbbell Mass and Gravity Range**

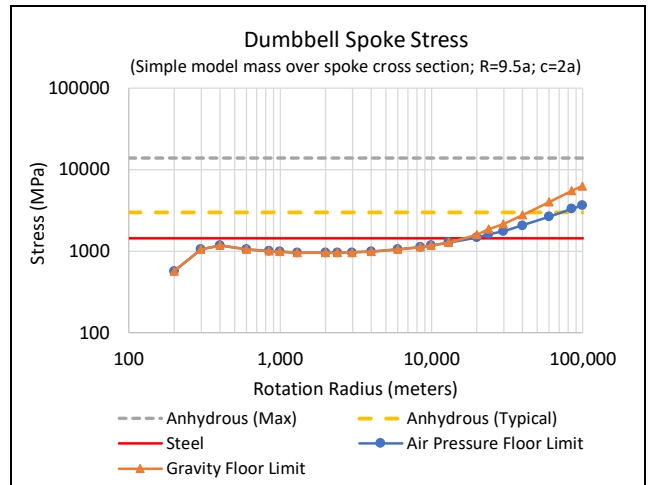

**Figure 7-43 – Dumbbell Spoke Stress**

Figure 7-43 shows the effect of this simple stress analysis on the dumbbell tether or spoke. The x-axis shows the rotation radius on a logarithmic scale ranging from 100 to 100,000 meters. The y-axis shows the stress in megapascals (MPa). Material strengths from [Jensen 2023] are used; specifically, Anhydrous Glass (max) is 13,800 MPa, Anhydrous Glass is 3,000 MPa, and Steel is 1,240 MPa. The data shows the computed spoke stress for the dumbbell geometry. The spoke radius increases with the station radius. To support the stress, no maximum radius size is defined for the dumbbell spoke. There are reasonable limits on the spoke radius, such as exceeding the size of the nodes or using all the available construction mass. Figure 7-43 uses the results for stations where the top floor is limited by the gravity range and limited by the habitable air pressure. The node has more floors and mass, with the top floor being limited by the gravity range. The graph shows greater stress from that increased mass in the large stations. With the small station sizes, the graph shows reduced stress from the scaling of components and their reduced mass.

The chart and simple analysis show that even steel structures could support the dumbbell spoke stress in large stations. With structures made from the ideal laminated anhydrous glass, the stations could be greater than 100,000 meters. Tests on the strength of laminated anhydrous glass and structural analysis using finite element analysis tools are still required to fully validate such results. The results of this simple analysis provide confidence to continue with large space station design examples.

### 7.5.6 Dumbbell Station Balance Results

The results from the mass and inertia equations are combined to further evaluate the dumbbell's rotational stability. Multiple station parameters could affect stability. These parameters include the radius, the minor axes ratio, and the shell thickness. The moments of inertia are evaluated to determine the station's rotational stability. As previously covered, the single dumbbell station does not meet the standard criteria for stability [Brown 2002] [Fitzpatrick 2023].

The single-floor analysis showed that the dumbbell was not rotationally stable. The single-floor stability results were found with straightforward equation analysis. The multiple-floors stability results were found using the moment of inertia of the components. We had hoped that the additional components might improve the stability.

Figure 7-44 introduces the dumbbell's multiple-floors stability. The top graph shows the moments of inertia for an elliptic node dumbbell as a function of the dumbbell radius. This represents the summation of the inertia of the 6 components. The bottom graph shows the MOIs as a function of the station shell thickness. The inertia about the z-axis and the y-axis are almost identical. The MOI about the x-axis is almost 100 times less than the MOIs about the other two axes. As a reminder, these axes were shown in the bottom half of Figure 7-36. Section *6.5.1 Are dumbbell stations rotationally stable?* determined that single dumbbells are rotationally unstable. The ratio $I_z/I_y$ is approximately 1.0 and represents an unstable rotation. The ratio $I_z/I_x$ is much greater than the 1.2 ratio and would represent a stable system.

Figure 7-45 further investigates the elliptical dumbbell stability. This graph shows the ratio of the MOIs $I_z/I_x$ and $I_z/I_y$ as stability ratio metrics. Typically, the station is considered rotationally stable with a stability metric of 1.2, as described in [Brown 2002] and [Globus et al. 2007]. In the case of the dumbbell, the MOIs of the rotation axis and another principal axis are nearly equal. The y-axis of the graph shows the stability ratio ranging from 0.1 to 200. The x-axis shows the dumbbell's major radius ranging from 100 to 40,000 meters on a logarithmic scale. This dumbbell has a gravity range of 0.95g to 1.05g, resulting in a design scale of 9.5. The radial node a-axis length is the major axis length R over 9.5. This provides the habitable gravity range from the center of the tube (at radius R) to the outer rim (at radius R+a). The

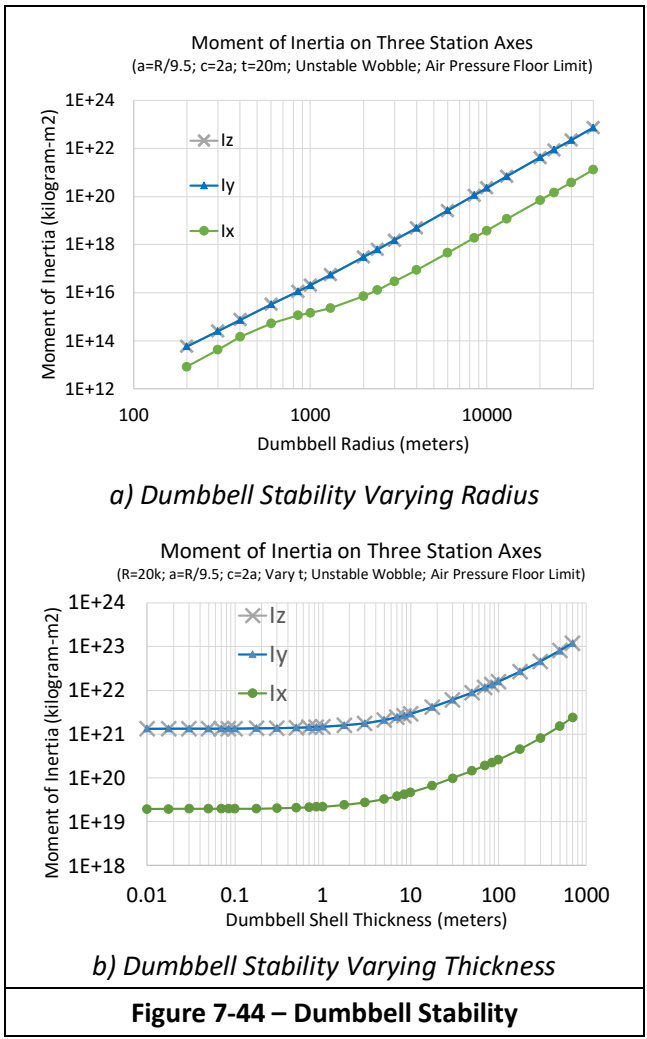

*a) Dumbbell Stability Varying Radius*

*b) Dumbbell Stability Varying Thickness*

**Figure 7-44 – Dumbbell Stability**

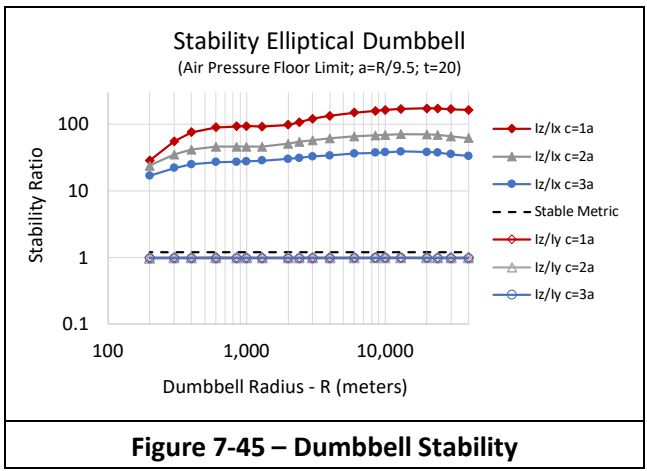

**Figure 7-45 – Dumbbell Stability**

perpendicular minor c-axis length is a multiple of the radial minor axis. The shell is a regolith-filled 20-meter-thick structure. As previously described, the Iz/Iy stability ratio is close to 1.0 for these elliptical dumbbells. The other stability ratio, Iz/Ix, is much greater than the minimum stability ratio of 1.2. The ellipsoid node axes are varied for this analysis. The node radial axis, a, remains at R/9.5 to provide habitable gravity over the floors. The perpendicular axis, c, is varied from 1x, 2x, and 3x the radial axis length. The more prolate ellipsoid tends to reduce the stability Iz/Ix ratio; however, the design ratio c/a appears to have minimal effect on the Iz/Iy stability ratio.

The analysis of both the single-floor and multiple-component stability found the dumbbell was not rotationally stable. The rotation axis and one of the other principal axes are nearly equal. This breaks one of the stability rules presented earlier [Fitzpatrick 2011] [Globus et al. 2007] [Brown 2002]. Fortunately, the instability grows algebraically and not exponentially. The station would not flip between the two equal principal axes; instead, a periodic wobble would be introduced [Fitzpatrick 2023]. For this paper, the wobble and its effect on the station will remain an open issue for the single dumbbell system.

### 7.5.7 Double Dumbbell Alternative

Alternative dumbbell architectures can resolve the stability issues of previous dumbbell subsections. Multiple variations of the dumbbell are discussed in [Johnson and Holbrow 1977] and [NSF 2014]. This subsection introduces one of those alternatives and includes a brief overview of the design, mass, stability, and population.

**Design:** Figure 7-46 shows a line drawing of the double dumbbell concept. This design is essentially two dumbbells oriented perpendicular about a common rotation center. The double dumbbell architecture has the same components described in the previous subsections on the single dumbbell station geometry. Figure 7-46 shows the prolate ellipsoid nodes, the spokes, and the central shuttle bay. The main floor, the multiple floors, and the air are internal to the nodes. These components can be used to compute the station's mass, required building material, construction schedules, and moments of inertia (MOI).

**Mass:** Figure 7-47 shows a comparison of the single and double dumbbell masses as a function of their radius. The chart includes data from two gravity ranges where the top floor is at 0.95g, and the outer rim floor is at 1.05g or 1.1g. The node radial axis, a, is set to R/9.5 or R/6.33 to provide these habitable gravity ranges over the floors. The air pressure and top floor analysis remain the same for both the single and double dumbbell geometries. As expected, the mass of the double dumbbell is approximately double the single dumbbell. The mass of the nodes and spokes double while only one shuttle bay is needed at the center.

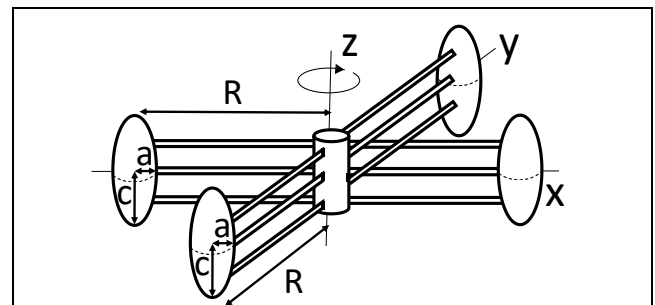

**Figure 7-46 – Double Dumbbell Concept**

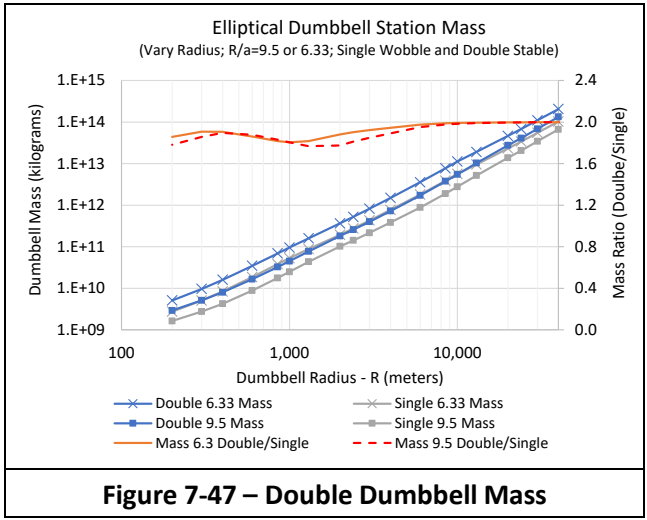

**Figure 7-47 – Double Dumbbell Mass**

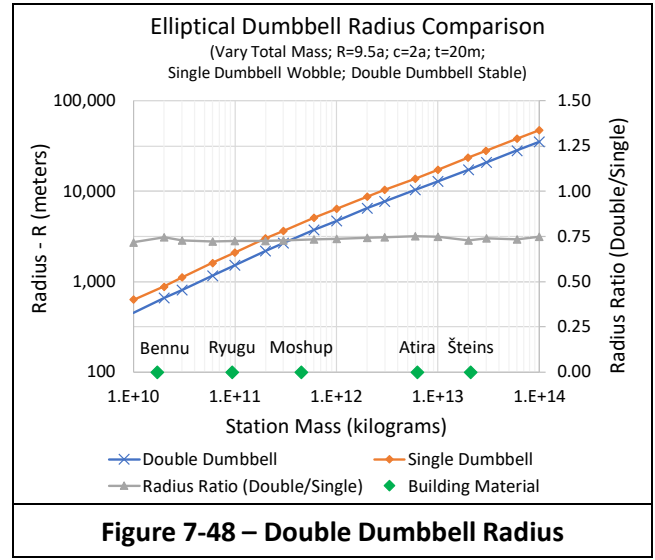

**Figure 7-48 – Double Dumbbell Radius**

The Figure 7-47 x-axis ranges from 100 to 40,000 meters on a logarithmic scale. The left y-axis ranges from 1e9 to 1e15 kilograms on a logarithmic scale. The right y-axis shows the ratio of the double dumbbell mass over the single dumbbell mass. That ratio axis ranges from 0 to 2.4 on a linear scale. The chart includes data from single and double dumbbell geometries with two gravity ranges.

The masses of the stations increase with their radius. Earlier subsections showed that the mass of the shuttle bay dominates the small station mass. This explains the slower rate of increase in the small stations. At sizes larger than 1000 meters in radius the rate of increase is nearly linear between the radius and the mass. The ratio data in Figure 7-47 shows that the mass of a large double dumbbell station is twice that of a single dumbbell station. The shuttle bay does not get doubled in the double dumbbell station. Its dimensions are reduced for small radii. These two reductions reduce the ratio data in smaller stations. Figure 7-47 also shows the masses of the stations with larger gravity ranges (R/a=6.33) are larger than the smaller gravity ranges (R/a= 9.5). The associated smaller nodes explain this difference.

Figure 7-48 shows a comparison of the single and double dumbbell radii as a function of their construction mass. The chart shows data from a gravity range where the top floor is at 0.95g, and the outer rim floor is at 1.05g. The node radial axis, a, is set to R/9.5 to provide this habitable gravity range. The air pressure top floor limit is used for both of the geometries.

The x-axis ranges from 1e10 to 1e14 kilograms. The left y-axis ranges from 100 to 100,000 meters on a logarithmic scale. A right y-axis shows the ratio of the double dumbbell station radius over the single dumbbell radius. The masses of 5 asteroids are marked along the x-axis for reference. It is assumed that 1/3 of the bulk material of these asteroids is used to create the station [Jensen 2023].

The radius ratio data in Figure 7-48 shows that a given mass asteroid can consistently be converted to a double dumbbell station with about 0.75 the radius of the single dumbbell station radius. The smaller radius produces a smaller node and a smaller total population for the double dumbbell. It may be a smaller population, but we will find that the double dumbbell is rotationally stable while the single dumbbell has a rotational wobble.

**Stability:** The single dumbbell stability is unique from the other geometries used in this paper. In particular, this section identified that the dumbbell rotational axis MOI is nearly identical to the MOI of one of the other principle axes. This resulted in an unstable design with a wobble in the rotation.

An analytic thin shell analysis is used to overview the MOIs of the double dumbbell. The MOIs for the double dumbbell nodes are as follows:

$$Ixdd = 2\,Ix + 2(Iy + m_{node}\,R^2)$$
$$Iydd = 2\,Iy + 2(Ix + m_{node}\,R^2)$$
$$Izdd = 4(Iz + m_{node}\,R^2)$$

Where Ixdd, Iydd, and Izdd are the MOIs for the nodes on the three double dumbbell axes. Ix, Iy, and Iz are MOIs for the prolate ellipsoid nodes. The variable $m_{node}$ is the mass of the node, and R is the radius from the station center to the node center. To simplify this introduction, the focus is only on spherical nodes in the double dumbbell. The spoke masses are assumed to be significantly less than the nodes and are ignored. A thin shell MOI is considered for the spherical nodes. The system MOIs become:

$$Ixdd = Iydd = 2m_{node}\left(\frac{4}{3}a^2 + R^2\right)$$
$$Izdd = 4m_{node}\left(\frac{2}{3}a^2 + R^2\right)$$

The moment of inertia about the rotation axis z (Izdd) is almost double that of the other principal axes (Ixdd and Iydd). The Iz/Ix and the Iz/Iy would be identical for the simple model and equal to:

$$\frac{Izdd}{Ixdd} = 2\,\frac{\left(\frac{2}{3}a^2 + R^2\right)}{\left(\frac{4}{3}a^2 + R^2\right)} = \frac{4a^2 + 6R^2}{4a^2 + 3R^2} = 1 + \frac{3R^2}{4a^2 + 3R^2}$$

When a=R/9.5, the stabilities Izdd/Ixdd and Izdd/Iydd are 1.9855. When a=R/6.33, both are 1.9678. In both cases, the stability ratio of the simple model is much greater than our minimum of 1.2. Unlike the single dumbbell station, this station is quite stable!

Thick shell spherical nodes can also be used to evaluate the MOIs. The MOIs for the system simplify to:

$$Ixdd = Iydd = 2\left(\frac{4}{5}\left(\frac{r_o^5 - r_i^5}{r_o^3 - r_i^3}\right) + R^2\right)$$

$$Izdd = 4\left(\frac{2}{5}\left(\frac{r_o^5 - r_i^5}{r_o^3 - r_i^3}\right) + R^2\right)$$

The Iz/Ix and the Iz/Iy would be identical for the simple model and equal to:

$$\frac{Izdd}{Ixdd} = 2\frac{\left(\frac{2}{5}\left(\frac{r_o^5 - r_i^5}{r_o^3 - r_i^3}\right) + R^2\right)}{\left(\frac{4}{5}\left(\frac{r_o^5 - r_i^5}{r_o^3 - r_i^3}\right) + R^2\right)} = \frac{4\left(\frac{r_o^5 - r_i^5}{r_o^3 - r_i^3}\right) + 10R^2}{4\left(\frac{r_o^5 - r_i^5}{r_o^3 - r_i^3}\right) + 5R^2}$$

$$\frac{Izdd}{Ixdd} = 1 + \frac{5R^2}{4\left(\frac{r_o^5 - r_i^5}{r_o^3 - r_i^3}\right) + 5R^2}$$

For a shell thickness of t, $r_o = r_i + t$, the equation becomes:

$$(r_o^5 - r_i^5)/(r_o^3 - r_i^3) = r_i^2 + \frac{2r_i^4 + r_i^3 t}{3r_i^2 + 3r_i + t^2} + 2r_i t + t^2$$

For a small t compared to ri, this equation approaches $5r_i^2/3$, and the thick shell stability approaches the thin shell stability Iz/Ix:

$$\frac{Izdd}{Ixdd} = 1 + \frac{5R^2}{4(5r_i^2/3) + 5R^2} = 1 + \frac{3R^2}{4r_i^2 + 3R^2}$$

When ri=R/9.5 and for a range of thicknesses t and major radii R, the stabilities Izdd/Ixdd and Izdd/Iydd are an average of 1.982. When a=R/6.33, both ratios are an average of 1.963. In both cases, the stability ratio of the simple model is much greater than our minimum of 1.2. Again, the thick shell analysis shows this double dumbbell station is quite stable!

Elliptical nodes and all the station components can also be used to evaluate the rotational stability of the double dumbbell station. Again, the shell, multiple floors, main floor, air, spokes, and shuttle bay components are included. The single dumbbell cross-section in Figure 7-36 can also represents one arm of the double dumbbell.

Figure 7-49 is included to help understand the elliptical dumbbell station stability. This figure includes a chart for the single dumbbell and the double dumbbell stability ratios. The y-axes of the graphs show the stability ratio (Iz/Ix). Both include the minimum stability ratio of 1.2 for reference. The x-axes show the dumbbell radius (R), and ranges from 100 meters to 100 thousand meters on a logarithmic scale. The dumbbell nodes in this analysis have a minor axis with a length of R/9.5 and R/6.33. These provide the gravity range of 0.95g at the center floor. The 9.5 gravity scaling uses 1.05g on the outer edge of the node, and the 6.33 gravity scaling uses 1.1g on the outer edge. The stability ratios were generated using the same six components.

The single dumbbell results are on the top chart in Figure 7-49. The stability ratio Iz/Ix is much greater than 1.2 (on the logarithmic y-axis). The stability ratio Iz/Iy is about 1.0, and the station would not be stable and would have a wobble.

The double dumbbell results are on the lower chart in Figure 7-49. The stability ratios Iz/Ix and Iz/Iy range from 1.62 to 1.95 with these designs. The ratios of the MOI about the Iz axis over the MOIs about the Ix and Iy axes are both more than 1.2; as such, the double dumbbell is rotationally stable. These results were for dumbbells with minor axes c=2a. Their stability ratios averaged 1.86 and were better than the 1.2 stability metric. For comparison, with spherical nodes (c=a), the stability ratios were better and averaged 1.91. For c=3a, the stability ratios averaged 1.79. For c=10a, the stability ratios were even smaller and averaged 1.24, barely above the 1.2 stability metric. The double dumbbell stations with multiple components are rotationally stable for the designs being considered.

**Population:** Converting an asteroid into a torus was introduced in [Jensen 2023]. Another paper covering limits on large stations shows charts and data from single and double dumbbell geometries with two gravity ranges [Jensen 2024d]. The population increases with radius. The node sizes for a specific radius and gravity range are identical for the single and double dumbbell geometries. There are twice the number of nodes in the double dumbbell design; as such, the double dumbbell design supports twice the population at a specific radius. At a specific mass, the double dumbbell population is consistently about 0.75 times the single dumbbell population. This is consistent with the 0.75 ratio found for the previous radius analysis shown in Figure 7-48. Population impacts the dumbbell or double dumbbell geometry selection. For a given mass, the double dumbbell population may be smaller than the single dumbbell; however, the double dumbbell does not have the instability wobble. To avoid active stability control techniques, the double dumbbell appears to be the better choice.

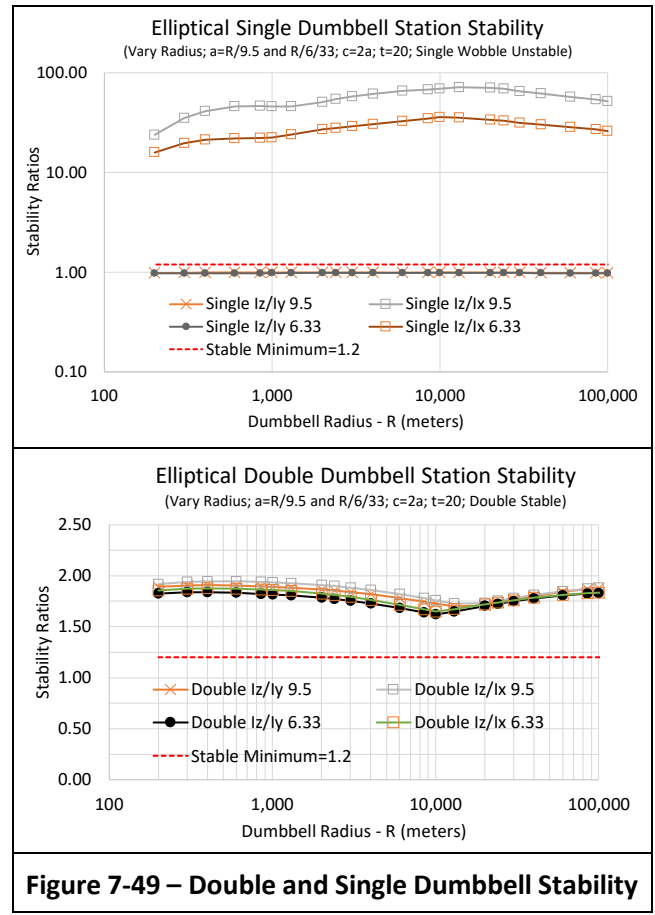

**Figure 7-49 – Double and Single Dumbbell Stability**

This subsection introduced an alternative dumbbell design. This double dumbbell would be stable and avoid the wobble expected with the single dumbbell. Other dumbbell designs with 3 or more arms would be rotationally stable and could be considered in the future.

# 8   Station Stability Summary

This section summarizes the station details and limitations identified in this paper. It reviews the space station limitations including gravity and air pressure. The paper's analysis included single floors and multiple component designs.

## 8.1   Limitations and Constraints

Limitations and constraints were not the focus of this paper. They define the station geometry and, as such, the station's stability. This subsection summarizes the multiple constraints limiting the size and geometry of a rotating space station. This includes single-floor and multiple-component stations. It also includes limits from gravity, air pressure, materials, top-floor limits, multiple floors, and population.

This paper reviewed geometry limits for the station design. It provided details on the station mass and air pressure for those geometries. Rotating the space station provides centripetal gravity. It would be best to provide one Earth gravity (1g), and the station rotation radius should be greater than 200 meters to avoid issues with rapid rotation. Larger stations support multiple floors and increase the available floor space. The centripetal gravity pushes the station atmosphere towards the outer rotating rim. This produces varying air density and pressure somewhat similar to Earth. The air pressure at the outer rim is sea level. The air density on a rotating station becomes similar to Denver at a height of about 1600 meters. This limits the top floor height and the maximum number of floors in large stations. To overcome that air pressure limit, airtight layers can provide habitable air pressure on the gravity-limited top floors in large stations.

This paper reviewed the limitations of the materials used in the space station. It covered the availability, strength, and desired characteristics. A simple analysis of the strength of materials showed that large stations can be built. Anhydrous glass from asteroid oxide is one choice for constructing large stations [Jensen 2023]. The strength of this material could support a station of 10 to 20 kilometers in radius.

This paper provided details on passive rotational stability concepts. There is a risk for rotation instabilities with asymmetric geometries. This instability can cause abrupt changes in orientation between two rotational states. This abrupt change would not be desirable in a space station. One must design the geometry of the space station to avoid this behavior. Four space station geometries were considered. Experience with spin-stabilized spacecraft suggests that the desired axis of rotation should have an angular moment of inertia at least 1.2 times greater than any other axis [Brown 2002]. Globus and his co-authors designed a cylinder station that would not have this risk [Globus et al. 2007]. This paper extended that single-floor analysis to all four geometries and to multiple-component stations.

Given these limitations and constraints, the space station geometries were refined. Spherical stations became ellipsoidal stations; long cylindrical stations became short hatbox stations; circular cross-section torus stations became elliptical cross-section torus stations; and dumbbells were doubled.

## 8.2   Stability Limitations

This paper focuses on the rotational stability of space stations. Moments of inertia were used to compute the stability, which requires knowing the station's mass and major components. This paper documented the geometries and densities of the major components. The following two subsections review this paper's single-floor and multiple-component stability results.

### 8.2.1   Single Floors Stability Summary

These results match and extend the cylinder results from [Globus et al. 2007]. Table 6-2 contains a summary of our current stability equations for the four station geometry types. These equations represent moments of inertia for thin-shell hollow geometries. This table was adapted from [Jensen 2023] with minor description changes. The analysis includes thick-shell geometries. The thick shell results are typically close to the thin shell results when the thickness is much less than the rotation radius. Spherical stations are not stable with single-floor analysis; however, ellipsoid geometry stations can be rotationally stable. Torus designs would be rotationally stable for typical (and desired) station dimensions. This analysis also found that dumbbell stations would not be rotationally stable. They would have a wobble that might impact the station residents. The analysis also found that a double dumbbell station is rotationally stable.

### 8.2.2   Multiple Component Stability Summary

This paper covered the multiple-component station rotational stability for the station geometries. Each geometry had a geometry design ratio that could be modified to produce a station with passive rotational stability. This summary describes each of those ratios and the desired range to create stable stations. Table 8-1 summarizes those stability ratios.

**Cylinder:** The stability of the cylinder is varied by changing the cylinder length (L) and the cylinder radius (R). For the thin shell cylinder, the analysis duplicated the Globus ratio L<1.29R [Globus et al. 2007]. For thick shells, the ratio increases from L<1.29R to L<1.35R as the shell thickness increases. Shells with thicknesses up to 1% of the radius require a length-to-radius ratio (L/R) of 1.29, like the thin shell stability ratio. Shells with thicknesses greater than 1% of the radius permit longer-length cylinder sides. A shell with a thickness of 10% of the radius permits a ratio of L/R up to

| Table 8-1 – Typical Stability Ratios | | | | |
|---|---|---|---|---|
| Geometry Analysis | Cylinder Iz/Ix=1.2 | Ellipsoid Iz/Ix=1.2 | Elliptical Torus R/a=9.5 | Double Dumbbell R/a=9.5 |
| Single Floor Thin Shell | L/R=1.29 | a/c=1.225 | Iz/Ix=1.92 c=2a | Iz/Ix=1.95 c=a |
| Single Floor Thick Shell | L/R=1.29 to 1.35 | a/c=1.29 | Iz/Ix=1.93 c=2a | Iz/Ix=1.98 c=a |
| Multiple Components | L/R=1.2 to 1.9 | a/c=0.75 to 2.25 | Iz/Ix= 1.86 to 1.96 c=2a | Iz/Ix=1.7 to 1.95 c=2a |

1.35. With the multiple components, the length-to-radius ratio can vary between 1.2 and 1.9 compared to the 1.3 ratio for thin or thick shells. Smaller stations (R<1000 meters) must reduce their length-to-radius ratio below the thin and thick shell guidelines of 1.3. With larger stations (R>4000 meters), the length can be 1.9 times the radius. Figure 7-14 illustrates these ratios. In cylinder stations with thinner shells, other components have more influence on the stability and support larger L/R ratios.

**Ellipsoid:** The oblate ellipsoid has two major axes. Two are the rotation radius with lengths (a=b), and the third is the polar minor axis with a length (c). The stability is controlled by changing the length of the polar axis (c) or the rotation radius length (a). The thin shell analysis found when c<0.8165a, the station would be stable with $I_z>1.2 I_x$. The thick shell analysis found when $c_o<0.775a_o$, the station would be stable with $I_z>1.2 I_x$. With the multiple components and multiple floors, varying the c/a ratio can still control the station's rotational stability. Depending on the rotation radius, the ratio can vary between 0.75 to 2.25; see Figure 7-23. Smaller stations require smaller polar axes for a stable station. Larger stations can have larger polar axes. In fact, our multiple component analysis found radii where the ellipsoid station can be spherical (a=c) and rotationally stable.

**Elliptical Torus:** The torus has a major radius of R and an elliptical cross-section with a minor radial axis a=R/m and a perpendicular minor axis c=2a or 3a. The m scaling value defines the gravity range in the torus tube, and typically, 9.5 or 6.33 is used; see *§3.1 Gravity Limits*. The stability is controlled by changing the length of the radial minor axis or the gravity scaling ratio. The analysis found that the torus is always rotationally stable (Iz/Ix>1.2) in our design space. The thin shell elliptical torus has Iz/Ix=1.916 for R=9.5a and c=2a. The thick shell elliptical torus has Iz/Ix=1.931 for R=9.5a and c=2a. With R=6.33a and c=2a, the stability Iz/Ix equals 1.852. All have a stability ratio greater than the minimum of 1.2. For the multiple component torus, analysis finds the station can be stable with Iz/Ix>1.2 as long as ci/ai is less than 8.0 for nearly any station radius.

**Dumbbell:** Dumbbell stations are not rotationally stable. The analysis found that Iz/Iy will be approximately 1.0 and produce an unstable rotation. The analysis found that Iz/Ix is much greater than 1.2 and, by itself, would represent a stable system. The Iz/Ix stability ratio ranges from 1.83 to 1.88 as the gravity range scaling is a=R/6.33 and a=R/9.5.

Fortunately, an expert states that *"the instability with two equal moments of inertia is algebraic rather than exponential"* and that this unstable rotation *"will settle into a limit cycle in which the rotational axis wobbles slightly [Fitzpatrick 2023]"*. For our analysis, a wobble will remain an open issue for the single dumbbell system.

**Double Dumbbell:** The double dumbbell is rotationally stable. This paper analyzed the stability of thin-shell spherical nodes in a double dumbbell station. The stability ratios of Iz/Ix and Iz/Iy were identical, and both were greater than 1.8. This is significantly greater than the recommended stability limit of 1.2 [Brown 2002]. For this thin shell model, when a=R/9.5, the stability was 1.88, and when a=R/6.33, it was 1.83. The multiple component double dumbbell is always rotationally stable with Iz/Ix and Iz/Iy both greater than 1.7 for our design space.

## 8.3 Stability Evaluation Conclusion

Stations can be made rotationally stable by varying their geometry. Spherical stations become ellipsoidal stations; long cylindrical stations become short hatbox stations; and dumbbells are doubled. Nearly all torus stations are rotationally stable. Different geometry ratio design results are found with the thin, thick, and multiple components modeling. Each model produces a more realistic result.

For the most accurate model in this paper, the station geometries were decomposed into components including the outer shell, end caps, air, floors, main floor, dividers, spokes, and shuttle bay. Each component was modeled and included in the stability analysis. This was a refinement of the thin and thick shell stability analysis.

This study provided insights and new limitations for future space station designs. The results of this analysis suggest that large, passive rotationally stable stations can be built. Obviously, refinement of these multiple-component models would be appropriate in the future. We note that additional research on material strengths and station stress models would be valuable.

# 9 References

[Baker and Haynes 2024] Engineering Statics: Open and Interactive, Daniel W. Baker and William Haynes, 28 March 2024, Libretext Engineering, Colorado State University and Massachusetts Maritime Academy, https://engineeringstatics.org/parallel-axis-theorem-section.html, https://batch.libretexts.org/print/Letter/Finished/eng-70197/Full.pdf

[Barr 2018] The High Life? On the Psychological Impacts of Highrise Living, Jason M. Barr, 31 January 2018, https://buildingtheskyline.org/highrise-living/

[Bell and Hines 2012] Space Structures and Support Systems, Larry Bell and Gerald D. Hines, 2012, MS-Space Architecture, SICSA Space Architecture Seminar Lecture Series, Sasakawa International Center for Space Architecture (SICSA), Cullen College of Engineering, University of Houston, http://sicsa.egr.uh.edu/sites/sicsa/files/files/lectures/space-structures-and-support-systems.pdf

[Birse 2000] 4.8 Stability of free rotation, Mike Birse, 17 May 2000, PC1672 Advanced Dynamics, Rigid-body motion, University of Manchester, https://www.theory.physics.manchester.ac.uk/~mikeb/lecture/pc167/rigidbody/stability.html

[Birse 2000p] 4.7 The symmetric top, Mike Birse, 17 May 2000, PC1672 Advanced Dynamics, Rigid-body motion, University of Manchester, https://www.theory.physics.manchester.ac.uk/~mikeb/lecture/pc167/rigidbody/top.html

[Bock, Lambrou, and Simon 1979] Effect of Environmental Parameters on Habitat Structural Weight and Cost, Edward Bock, Fred Lambrou, Jr., and Michael Simon, 1979, NASA SP-428 Space Resources and Space Settlements, pp. 33-60, [O'Neill et al. 1979], https://ntrs.nasa.gov/api/citations/19790024054/downloads/19790024054.pdf

[Brody 2013] Not 'Elysium,' But Better 'Ringworld' Settlements Could Return Our Future to Its Past (Commentary), David Sky Brody, 10 August 2013, https://www.space.com/22326-elysium-movie-space-colonies-future.html

[Brown 2002] Elements of Spacecraft Design, Charles D. Brown, AIAA Education Series, Reston, VA, 2002. https://www.worldcat.org/title/elements-of-spacecraft-design/oclc/850628450&referer=brief_results

[Diaz, Herrera, and Martinez 2005] Using symmetries and generating functions to calculate and minimize moments of inertia, Rodolfo A. Diaz, William J. Herrera, R. Martinez, 27 Jul 2005, Version 2 submission to arXiv, https://arxiv.org/pdf/physics/0404005.pdf

[Dunbar 2019] Why Space Radiation Matters, NASA Official: Brian Dunbar, 8 October 2019, National Aeronautics and Space Administration, Analog Missions, https://www.nasa.gov/analogs/nsrl/why-space-radiation-matters

[Fitzpatrick 2011] Newtonian Dynamics, Richard Fitzpatrick, 2011 March 31, Rigid Body Rotation, Rotational Stability, The University of Texas at Austin, https://farside.ph.utexas.edu/teaching/336k/Newton/node71.html

[Fitzpatrick 2023] Dumbbell Space Stability, Richard Fitzpatrick, 2023 June 28, Personal Correspondence with Professor University of Texas

[Garrett 2019] A cinema, a pool, a bar: inside the post-apocalyptic underground future, Bradley L Garrett, 16 December 2019, The Guardian, https://www.theguardian.com/cities/2019/dec/16/a-cinema-a-pool-a-bar-inside-the-post-apocalyptic-underground-future

[Globus et al. 2007] The Kalpana One Orbital Space Settlement Revised, Al Globus, Nitin Arora, Ankur Bajoria, and Joe Strout, 2007, American Institute of Aeronautics and Astronautics, http://alglobus.net/NASAwork/papers/2007KalpanaOne.pdf

[Graem 2006] Visions 2200 - A Perspective on the Future Space Habitat, H Graem, 2006, http://visions2200.com/SpaceHabitat.html, Wayback Machine Access: https://web.archive.org/web/20080209111914/http://visions2200.com/SpaceHabitat.html

[Hall 1991] The Architecture of Artificial Gravity: Mathematical Musings on Designing for Life and Motion in a Centripetally Accelerated Environment, Theodore W. Hall, November 1991, Space Manufacturing 8 Energy and Materials from Space, Proceedings of the Tenth Princeton/AIAA/SSI Conference, 15-18 May 1991, https://citeseerx.ist.psu.edu/viewdoc/download?doi=10.1.1.551.3694&rep=rep1&type=pdf

[Hall 1999] Inhabiting artificial gravity, Theodore Hall, September 1999, AIAA Space Technology Conference, 28-30 September 1999, Albuquerque, New Mexico 10.2514/6.1999-4524, https://www.researchgate.net/publication/269226838_Inhabiting_artificial_gravity

[Hand and Finch 1998] Analytical Mechanics, Louis N. Hand and Janet D. Finch, 1998, Cambridge University Press. p. 267. ISBN 978-0-521-57572-0, https://www.iaa.csic.es/~dani/ebooks/Mechanics/Analytical%20mechanics%20-%20Hand,%20Finch.pdf

[Holloway 2011] The Effect of Footway Crossfall Gradient on Wheelchair Accessibility, Catherine Holloway, April 2011, University College London, Department of Civil, Environmental & Geomatic Engineering, https://discovery.ucl.ac.uk/id/eprint/1310252/1/1310252.pdf

[Jensen 2023] Autonomous Restructuring of Asteroids into Rotating Space Stations, David W. Jensen, 23 February 2023, https://arxiv.org/abs/2302.12353

[Jensen 2024d] Design Limits on Large Space Stations, David W. Jensen, 22 July 2024, Submitted to arXiv.org

[Johnson and Holbrow 1977] NASA SP-413: Space Settlements: A Design Study, Eds: Richard D. Johnson and Charles Holbrow, 1977, Scientific and Technical Information Office, https://ntrs.nasa.gov/citations/19770014162

[Keeter 2020] Long-Term Challenges to Human Space Exploration, Bill Keeter, 4 September 2020, National Aeronautics and Space Administration, https://www.nasa.gov/centers/hq/library/find/bibliographies/Long-Term_Challenges_to_Human_Space_Exploration

[Lente and Ősz 2020] Barometric formulas: various derivations and comparisons to environmentally relevant observations, Gábor Lente and Katalin Ősz, 4 April 2020, ChemTexts 6, 13 (2020). https://doi.org/10.1007/s40828-020-0111-6, https://link.springer.com/article/10.1007/s40828-020-0111-6

[Lipsett 2005] Edward Lipsett's 2303AD - Japan's Battle for the Stars, Edward Lip-sett, Retrieved 14 September 2020 from web.archive.org, 22 December 2005, https://web.archive.org/web/20051222104439/http://www.kuroto-kage.org/2300AD/Shintenchi/outline.html

[McGraw-Hill 2023] Rotational Stability, Accessed: 27 December 2023, McGraw-Hill Dictionary of Scientific & Technical Terms, 6E, Copyright © 2003 by The McGraw-Hill Companies, Inc. https://encyclopedia2.thefreedictionary.com/rotational+stability

[McKendree 1995] Implications of Molecular Nanotechnology Technical Performance Parameters on Previously Defined Space System Architectures, Thomas Lawrence McKendree, 9-11 November 1995, The Fourth Foresight Conference on Molecular Nanotechnology, Palo Alto, California, http://www.zyvex.com/nanotech/nano4/mckendreePaper.html

[Misra 2010] The "Tesla" Orbital Space Settlement, Gaurav Misra, 2010, Barcelona, Spain, 40th International Conference on Environmental Systems, American Institute of Aeronautics and Astronautics, Inc, AIAA 2010-6133, https://doi.org/10.2514/6.2010-6133, http://spacearchitect.org/pubs/AIAA-2010-6133.pdf

[Myers 1962] Handbook of Equations for Mass and Area Properties of Various Geometrical Shapes; Compiled by: Jack A. Myers; Weapons Development Department; Ad274936. NAVWEPS Report 7827, NOTS TP 2838, Copy 313, U.S. Naval Ordnance Test Station, China Lake, California, April 1962, https://apps.dtic.mil/sti/pdfs/AD0274936.pdf

[NSF 2014] Topic: Realistic, near-term, rotating Space Station, Member: Roy_H, 16 February 2014, NASA Space Flight Forum, Discussion 2014 through 2019, https://forum.nasaspaceflight.com/index.php?topic=34036.0

[O'Neill 1974] The Colonization of Space, Gerard K. O'Neill, September 1974, Physics Today. 27 (9): 32-40. Bibcode:1974PhT....27i..32O. doi:10.1063/1.3128863, https://physicstoday.scitation.org/doi/pdf/10.1063/1.3128863

[O'Neill 1976] The High Frontier: Human Colonies in Space, Gerard K. O'Neill, Morrow New York 1976, ISBN-13:978-1896522678; https://archive.org/details/highfrontierhuma0000onei_y9i8/mode/1up

[O'Neill 2008] Building a Moon Base: Part 1 - Challenges and Hazards, Ian O'Neill, 7 February 2008, Universe Today, Space and Astronomy News, https://www.universetoday.com/12726/building-a-base-on-the-moon-challenges-and-hazards/ [CC BY-4.0]

[O'Neill et al. 1979] NASA SP-428, Space Resources and Space Settlements, Gerard O'Neill, Eds: John Billingham, William Gilbreath, and Brian O'Leary, 1 January 1979, NASA, Technical papers derived from the 1977 Summer Study at NASA Ames Research Center, Moffett Field, California, https://ntrs.nasa.gov/api/citations/19790024054/downloads/19790024054.pdf

[O'Reilly et al. 2021] Rotations: A tumbling T-handle in space: the Dzhanibekov effect, Daniel T. Kawano, Alyssa Novelia, and Oliver M. O'Reilly, Retrieved: 10 March 2022, Rose-Hulman Institute of Technology, Berkeley University of California, https://rotations.berkeley.edu/a-tumbling-t-handle-in-space/

[Queijo et al. 1988] Some Operational Aspects of a Rotating Advanced-Technology Space Station for the Year 2025, M.J. Queijo, A.J. Butterfield, W.F. Cuddihy, C.B. King, R.W. Stone, J.R. Wrobel, and P.A. Garn, June 1988, The Bionetics Corporation, NASA Contractor Report 181617, Report 19880017013, National Aeronautics and Space Administration, Langley Research Center, https://ntrs.nasa.gov/api/citations/19880017013/downloads/19880017013.pdf

[Routh 1877] An elementary treatise on the dynamics of a system of rigid bodies; Edward John Routh; 1877; Macmillan and Company, London; https://archive.org/details/elementarytreati00rout/page/20/mode/2up

[Tatum 2017] 2.20: Ellipses and Ellipsoids, Jeremy Tatum, 17 September 2017, Classical Mechanics, LibreTexts, Physics, University of Victoria, https://phys.libretexts.org/Bookshelves/Classical_Mechanics/Classical_Mechanics_(Tatum)/02%3A_Moments_of_Inertia/2.20%3A_Ellipses_and_Ellipsoids

[Temples 2015] Solution Set Six, Dylan J. Temples, 11 November 2015, Classical Mechanics, Classical Mechanics, http://www.dylanjtemples.com:82/solutions/GoldsteinSolution6.pdf

[Tessmer 2014] Making 'Impossible' ADA Slopes Possible: Predictable concrete flatness techniques, Jean Tessmer, 28 April 2014, The Construction Specifier, Construction Specifications Institute (CSI), https://www.constructionspecifier.com/making-impossible-ada-slopes-possible-predictable-concrete-flatness-techniques/

[Turner and Kunkel 2017] Radiation Environment inside a Lunar Lava Tube, Ronald E. Turner and Robert Kunkel, 47th International Conference on Environmental Systems, 16-20 July 2017, Charleston, South Carolina, https://ttu-ir.tdl.org/ttu-ir/bitstream/handle/2346/72863/ICES_2017_15.pdf

[Young and Budynas 2001] Roark's Formulas for Stress and Strain, Warren C. Young and Richard G. Budynas, January 2001, Seventh Edition, McGraw-Hill Professional, http://materiales.azc.uam.mx/gjl/Clases/MA10_I/Roark's%20formulas%20for%20stress%20and%20strain.pdf

# 10 License Types

This document is Copyright © 2024 to David W. Jensen. All self-produced material is licensed under a Creative Common License CC BY-SA 4.0 [CC BY-SA 4.0]. Licensing for materials from other sources is referenced in the text and is described in this section.

[CC BY-4.0] Attribution 4.0 International (CC BY 4.0), Creative Commons, You are free to share and adapt for any purpose, even commercially. You can copy and redistribute the material in any medium or format. You can remix, transform, and build upon the material. You must give appropriate credit, provide a link to the license, and indicate if changes were made; https://creativecommons.org/licenses/by/4.0

[CC BY-NC 4.0] Attribution-NonCommercial 4.0 International (CC BY-NC 4.0), Creative Commons, You are free to copy and redistribute the material. You may remix, transform, and build on the material. You must give appropriate credit, provide a link to the license, and indicate if changes were made. You may not use for commercial purposes; https://creativecommons.org/licenses/by-nc/4.0/

[CC BY-SA 4.0] Attribution-ShareAlike 4.0 International (CC BY-SA 4.0), Creative Commons, Free to Share and Adapt; You must give appropriate credit, provide a link to the license, and indicate if changes were made; You may adapt the material for any purpose, even commercially; ShareAlike: You must distribute your contributions under the same license as the original, https://creativecommons.org/licenses/by-sa/4.0/

[Facts] Facts and data are generally not eligible for copyright. "Charts, graphs, and tables are not subject to copyright protection because they do not meet the first requirement for copyright protection, that is, they are not "original works of authorship," under the definitions of 17 U.S.C. § 102(a). Essentially, that means that a graph, chart, or table that expresses data is treated the same as the underlying data. Facts, data, and the representations of those facts and data are excellent examples of things that require much "sweat of the brow" to create, but yet still do not receive copyright protection." https://deepblue.lib.umich.edu/bitstream/handle/2027.42/83329/copyrightability_of_tables_charts_and_graphs.pdf

[NASA Image Public Domain] National Aeronautics and Space Administration, Accessed 21 November 2021, NASA content is generally not subject to copyright in the United States. You may use this material for educational or informational purposes, including photo collections, textbooks, public exhibits, computer graphical simulations, and Internet Web pages, https://images.nasa.gov/, https://www.nasa.gov/multimedia/guidelines/index.html .

# 11 Contents – Space Station Stability